\documentclass[twoside,11pt]{article}

\usepackage[preprint]{jmlr2e}
\hypersetup{hypertexnames=false}   
\usepackage[T1]{fontenc}
\usepackage{times}
\usepackage{amsmath,bm}
\usepackage{booktabs}
\usepackage{array}
\newcolumntype{L}[1]{>{\raggedright\arraybackslash}p{#1}}
\usepackage{algorithm}
\usepackage[noend]{algpseudocode}
\usepackage{lastpage}
\usepackage{placeins}
\usepackage{pdfpages}
\def\repourl{https://github.com/hhuang45/DeepNB-STAR}
\hypersetup{hidelinks}

\graphicspath{{figures/}}
\newtheorem{assumption}{Assumption}
\providecommand{\qed}{\hfill\BlackBox}
\newcommand{\E}{\mathbb{E}}
\newcommand{\NB}{\mathrm{NB}}
\newcommand{\var}{\mathrm{var}}
\newcommand{\TV}{\mathrm{TV}}
\newcommand{\ours}{DeepNB-STAR}

\jmlrheading{}{2026}{1-\pageref{LastPage}}{11/26}{}{}{Hsin-Hsiung Huang}
\ShortHeadings{Deep Negative Binomial Autoregression}{Huang}
\firstpageno{1}

\begin{document}

\title{Deep Negative Binomial Autoregression for Sparse Spatio-Temporal Counts: Estimation and Predictive Distribution Bounds}

\author{\name Hsin-Hsiung Huang \email hsin.huang@ucf.edu \\
       \addr Dr.\ Yvette Kanouff School of Data, Mathematical, and Statistical Sciences\\
       University of Central Florida\\
       Orlando, FL 32816, USA}

\maketitle

\begin{abstract}
We study estimation and probabilistic forecasting for sparse, overdispersed count panels using DeepNB-STAR, a negative binomial autoregression whose log-mean and log-size are outputs of a deep ReLU network of temporal and neighboring summaries of the past, fitted by maximum likelihood. We prove an oracle inequality for the Hellinger risk along the observed trajectory without mixing assumptions, rates governed by compositional smoothness with a matching lower bound for a H\"older subclass, geometric absolute regularity, and bounds on the error of $h$-step predictive distributions of order $h^{1/2}$ times the root estimation risk plus a Monte Carlo term. All results hold for a size that depends on the covariates, and the estimation bounds hold under arbitrary dependence between sites, with the number of time points as sample size. On four Japanese surveillance panels, DeepNB-STAR has a weighted interval score 0.6\% to 1.0\% below that of a three-member ensemble of Temporal Fusion Transformers on every panel, distinguishable from zero only on the held-out panel. Fitted mean and size functions are reported with exploratory bootstrap intervals. Nominal 95\% prediction intervals for national totals cover only 81\% to 92\% of weeks: the joint predictive law, unlike the estimator, relies on conditional independence across sites.
\end{abstract}

\begin{keywords}
  Count Time Series, Deep Neural Networks, Maximum Likelihood, Nonparametric Estimation, Probabilistic Forecasting
\end{keywords}

\section{Introduction}\label{sec:intro}

Deep neural networks are now a standard tool for probabilistic forecasting of large panels of related time series. Architectures such as DeepAR \citep{salinas2020} and the Temporal Fusion Transformer \citep[TFT;][]{lim2021} are trained on all series jointly, output the parameters or the quantiles of a predictive distribution, and are applied in demand planning, energy and public health. Many of the panels encountered in these applications consist of counts that are sparse, bursty and overdispersed, and that are dependent across both time and space; examples include disease surveillance, traffic flows \citep{he2023traffic} and event records, for which deep temporal encoders have been paired with negative binomial and zero-inflated negative binomial likelihood heads in a Bayesian pipeline \citep{huang2026bayesdeep}. Weekly surveillance counts of a notifiable disease are a typical example: most site-weeks have no case, a few sites have bursts of several dozen cases, the variance is many times the mean, and neighboring sites move together. Despite the empirical success of deep forecasters on such data, their statistical behavior is largely unknown. It is not clear at what rate the estimated predictive distribution approaches the truth, whether that rate can be improved, or what the estimation error implies for a prediction interval issued several steps ahead.

A substantial body of work addresses the first two questions for nonparametric regression with independent observations. Deep networks with the rectified linear unit (ReLU) activation approximate smooth functions efficiently \citep{yarotsky2017}, and least squares estimators over such networks attain minimax optimal rates for smooth and for composition-structured regression functions \citep{bauer2019,schmidthieber2020,kohler2021,jiao2023}, as well as for non-smooth ones \citep{imaizumi2019}. The key insight of this line of work is that the rate depends on the intrinsic dimension of the components of the target function and not on the number of covariates. Extensions cover general loss functions \citep{farrell2021} and high-dimensional factor structure \citep{fan2024}, and network rates have been combined with structured response models, for instance for responses in metric spaces \citep{iao2025,han2026}. For dependent observations, generalization bounds are available under mixing conditions \citep{yu1994,mohri2008}, and rates over deep networks have been established for least squares with mixing sequences and nonlinear autoregressions \citep{ma2022,kurisu2025}, for Lipschitz losses under strong mixing \citep{kengne2025}, and for generalized Bayes estimators on Markov chains with a spectral gap \citep{alquier2025}. For a likelihood loss, \citet{yara2026} obtain Hellinger rates for nonparametric maximum likelihood over deep networks from independent replications of a covariate-driven Poisson process. These results do not cover deep count forecasting. The loss there is a likelihood for unbounded and overdispersed responses, with a second unknown function for the dispersion; the covariates are generated by the process itself, along a single trajectory that need be neither stationary nor mixing; and translating estimation error into count-valued predictive distributions several steps ahead requires an additional argument. Deriving mixing from model assumptions, as we do for the out-of-sample risk, is not unique to the present work \citep[Lemma~4.1]{kurisu2025}.

A second line of work, in statistics, models the conditional mean of a count given the past with a parametric form. Poisson and negative binomial autoregressions have a well-developed theory \citep{fokianos2009,neumann2011,davis2016}, nonparametric autoregression is classical for continuous responses \citep{fan2003}, and the endemic--epidemic framework of \citet{held2005} and \citet{meyer2014} adds spatial spillover between regions and is the standard tool of surveillance epidemiology. Recent parametric developments include parsimonious spatio-temporal count autoregressions with consistent and asymptotically normal quasi-likelihood estimators \citep{maletz2024} and negative binomial models whose dispersion follows its own observation-driven recursion \citep{silva2026}. Bayesian hierarchical models add zero inflation and spatial and temporal random effects with Gaussian process priors, with posterior computation by P\'olya--Gamma data augmentation \citep{polson2013,he2024zinb,huang2024mouse,scott2026landslide}, and shrinkage priors extend Bayesian count regression to sparse variable selection for mixed-type multivariate responses \citep{wang2025}. These models are interpretable and their properties are understood, but a mean that is log-linear in the covariates cannot express thresholds, saturation, or interactions between the level of a site and that of its neighbors. Nonparametric theory for count autoregression is scarce: \citet{wechsung2022} obtain a rate for a least squares estimator of the contractive link function of a univariate Poisson autoregression. Neural networks have entered count modeling in two other ways, neither with a rate theory. They serve as the response function of an integer-valued autoregression, a small network fitted as a parametric model \citep{jahn2024}; and they serve as amortized estimators: \citet{svensson2025} pre-trains a transformer on simulated data to return the three parameters of a two-group negative binomial regression with independent observations, and reports that closed-form moment estimators are the most efficient solution of that task. There the network replaces the optimizer, while the model stays parametric and the data independent. Here the network is the nonparametric component of the model, and the object of study is the conditional law of a dependent panel.

In this work, we study a model that connects these two lines and admits a statistical analysis under explicit restrictions. The deep negative binomial spatio-temporal autoregression, \ours, specifies that the count $Y_{s,t}$ at site $s$ in week $t$ is, given the past, negative binomial with mean $\exp\{f_0(X_{s,t})\}$ and size $\exp\{g_0(X_{s,t})\}$, where the covariate vector $X_{s,t}$ summarizes the recent counts of the site, those of its nearest neighbors, the overall level and the season, and $f_0$ and $g_0$ are unknown functions. Temporal dependence and spatial dependence enter through the arguments of both functions, and sparsity and overdispersion are handled by the likelihood. We estimate $f_0$ and $g_0$ jointly by maximum likelihood over deep ReLU networks with two outputs and obtain the predictive distribution at any horizon by simulating the fitted model forward. The network is not the contribution: it is the sieve that lets a likelihood model with two unknown functions of several dozen lagged and neighboring summaries adapt to low-dimensional structure. The contribution is statistical. It consists of a nonparametric count autoregression in which the overdispersion, like the mean, is an unknown function of the spatio-temporal past; of a theory for it that takes one dependent trajectory as it is, with a lower bound; and of guarantees for the object used in practice, the multi-step predictive distribution and its intervals. Because the model specifies the joint law of the panel, the questions raised above have precise answers.

Our main results are as follows. An oracle inequality bounds the Hellinger risk of the maximum likelihood estimator along the observed trajectory by the approximation error of the network class plus its metric entropy divided by the sample size. It requires only that the conditional model be correctly specified; no stationarity or mixing is assumed, so it applies to the autoregressive, spatially coupled and seasonal panel as it is. Conditional independence across sites, which the model also specifies, is not needed: if only the conditional law of each count is correct, the same inequality holds with the number of periods $T$ in place of $n=ST$ in its stochastic term, and this cannot be improved in general. When $f_0$ is a composition of H\"older functions, each depending on few variables, the inequality yields, up to logarithmic factors, the rate $\phi_n$ that \citet{schmidthieber2020} obtained for least squares regression with independent Gaussian errors, defined in \eqref{eq:phin}. For a H\"older subclass, we prove a matching lower bound, for any number of sites, in an experiment whose covariates include exogenous inputs; it does not cover a covariate map that depends on past counts alone, which we discuss. A model with bounded log-mean and finite memory, possibly driven by independent exogenous inputs, is shown to be geometrically absolutely regular \citep{bradley2005}, which yields the risk on an independent copy of the process. We then bound the total variation distance between the fitted and the true predictive distributions, one step and $h$ steps ahead, by the estimation risk plus the Monte Carlo error of the simulated quantiles; this gives explicit bounds on the difference between the fitted and the true coverage probabilities of prediction intervals at every horizon. Finally, all these results extend to a size that is unknown and may depend on the covariates, estimated jointly with the log-mean as a second output of the network. Averages of several fits inherit the rates, and so do functionals of the two functions that are Lipschitz in the empirical norm, such as averages over the observed site-weeks; this motivates the exploratory bootstrap summaries we report but does not validate their intervals, and functionals that evaluate the functions at unobserved covariate values, such as partial effects, are not covered.

The proofs combine known devices with new steps, and we separate the two. The conditional expectation of the square root of a likelihood ratio equals one minus the squared Hellinger distance, so an exponential supermartingale inequality bounds the Hellinger risk at the covariate values actually visited. This argument goes back to maximum likelihood for counting processes \citep{vandegeer1995} and is used for sequential density estimation \citep{zhang2006,agarwal2020}. For the horizon, the simulation lemma of reinforcement learning adds one-step total variation errors, so the bound grows linearly \citep{kearns2002,munos2008}, whereas Kullback--Leibler divergences are additive by the chain rule; trajectory-level analyses of the logarithmic loss exist for independent trajectories, in Hellinger distance \citep{foster2024} and in Kullback--Leibler divergence under bounded log-density ratios \citep{xu2026}. The error of the $h$-step predictive law is then of order $h^{1/2}$ times the root of the one-step risk. What is new is the analysis of a single trajectory of unbounded counts over a sieve: the unbounded likelihood ratio is controlled through conditional moment generating functions of the negative binomial law, which keeps the fast rate; the size is a second unknown function, separated from the mean in Hellinger distance through the first two factorial moments; a generalized H\"older inequality replaces conditional independence across sites in the supermartingale; and the rates are tied to the coverage of simulated prediction intervals, under the stated restrictions on the estimator and on the evaluation law (Section~\ref{sec:roadmap}).

We complement the theory with experiments. In simulations with a known log-mean, the risks of the practical fit and of a fit that satisfies the constraints of the theory both decrease at least at the guaranteed rate, trajectory and out-of-sample risks are close, and the error of predictive distributions decreases with the estimation error; when the size depends on the covariates, both functions are estimated at the guaranteed rate and moving-block bootstrap standard errors are close to the sampling standard deviations, although intervals built from them undercover where the estimator is biased. We then forecast weekly notifications of four diseases in the 47 prefectures of Japan from 1999 to 2024: syphilis, whose annual notifications rose about seventeen-fold from 2012 to 2024 \citep{takahashi2018}; enterohemorrhagic \emph{E.~coli} infection, a stable and strongly seasonal series; legionellosis, a sparse and rising panel; and amebiasis, a sparser and declining panel. The estimator was developed on syphilis and tested on new panels under protocols written before any model was fitted to them; amebiasis, chosen by a fixed rule, is the test panel of the final protocol. On all four panels \ours{} has a lower weighted interval score than an ensemble of three Temporal Fusion Transformers, by 0.6\% to 1.0\%; on amebiasis the difference, 0.9\%, is distinguishable from zero, on the other panels it is not, and in one period of one panel the transformer is better. \ours{} is clearly better than DeepAR and persistence, and better than or on a par with negative binomial regression and a Bayesian ZINB-GP model \citep{he2024zinb}. Because \ours{} is a likelihood model, its fit also describes the fitted mean and size functions, with exploratory pointwise bootstrap intervals: for instance, the fitted association with neighboring prefectures is strongest where the local level is low. On daily taxi-flow networks, a second domain in the supplement, the gain over a log-linear autoregression is 7.5\%.

Section~\ref{sec:model} sets up the model, the covariates and the function class; Section~\ref{sec:summary} summarizes the main results; Section~\ref{sec:est} gives the estimator, the forecasting algorithm and the inference procedure; Section~\ref{sec:theory} states the results and Section~\ref{sec:roadmap} outlines their proofs, which are in the appendix; Sections~\ref{sec:sim} and~\ref{sec:app} report simulations and the application, and Section~\ref{sec:disc} concludes.

\paragraph{Notations.} Sites are indexed by $s=1,\dots,S$ and weeks by $t=1,\dots,T$, and $n=ST$ is the number of site-weeks. We write $\NB(\mu,r)$ for the negative binomial distribution with mean $\mu>0$ and size $r>0$, whose probability mass function is
\[
p(y)=\frac{\Gamma(y+r)}{\Gamma(r)\,y!}\Big(\frac{r}{r+\mu}\Big)^{r}\Big(\frac{\mu}{r+\mu}\Big)^{y},\qquad y=0,1,2,\dots,
\]
with $\Gamma$ the gamma function; its variance is $\mu+\mu^2/r$. We write $p_f$ for this mass function when $\mu=e^f$, and call $f$ the log-mean. For probability mass functions $p$ and $q$ on the non-negative integers, $H^2(p,q)=1-\sum_y\{p(y)q(y)\}^{1/2}$ is the squared Hellinger distance, $\TV(p,q)=\tfrac12\sum_y|p(y)-q(y)|$ the total variation distance and $\mathrm{KL}(p\,\|\,q)=\sum_yp(y)\log\{p(y)/q(y)\}$ the Kullback--Leibler divergence; these satisfy $\TV\le\sqrt2\,H$ and Pinsker's inequality $\TV\le(\mathrm{KL}/2)^{1/2}$ \citep[Section~2.4]{tsybakov2009}. The same symbols are used for probability measures on general spaces. $\|g\|_\infty$ denotes the supremum norm of a function $g$ on its domain, $\|g\|_{L^2}$ its $L^2$ norm with respect to Lebesgue measure on $[0,1]^d$, and $|u|_\infty$ the maximum norm of a vector $u$. All logarithms are natural except $\log_2$. We write $a\wedge b=\min(a,b)$, $a\vee b=\max(a,b)$ and $u_+=\max(u,0)$; $\lfloor x\rfloor$ and $\lceil x\rceil$ are the integer part and the smallest integer not less than $x$, except that for a smoothness index $\beta>0$ we follow \citet{schmidthieber2020} and let $\lfloor\beta\rfloor$ denote the largest integer strictly smaller than $\beta$. For a multi-index $\bm\alpha=(\alpha_1,\dots,\alpha_k)$ of non-negative integers, $|\bm\alpha|=\sum_j\alpha_j$ and $\partial^{\bm\alpha}=\partial_1^{\alpha_1}\cdots\partial_k^{\alpha_k}$. $a\lesssim b$ indicates that $a\le cb$ for a constant $c>0$ that does not depend on the sample size, and $a\asymp b$ that $a\lesssim b$ and $b\lesssim a$. $\mathbf 1\{E\}$ is the indicator of the event $E$, $\E$ denotes expectation, and $\E_{f}$ expectation when the panel follows the model with log-mean function $f$. Constants denoted by $C$ and $c$ may change from line to line. For the function class and the networks we keep the notation of \citet{schmidthieber2020}, in which $\mathbf d$, $\mathbf t$, $\mathbf p$ and $\beta_i$ are unrelated to the covariate dimension $d$, the week $t$, the mass function $p_f$ and the mixing coefficients $\beta_X(\cdot)$.

\section{Problem Setup}\label{sec:model}

We first introduce the model, then the covariates that encode temporal and spatial dependence, and finally the class of target functions.

\subsection{Deep Negative Binomial Spatio-Temporal Autoregression}

Let $Y_{s,t}\in\{0,1,2,\dots\}$ be the count at site $s\in\{1,\dots,S\}$ in week $t\in\{1,\dots,T\}$, $\mathbf Y_t=(Y_{1,t},\dots,Y_{S,t})$, and let $\mathcal I_t$ be the $\sigma$-field generated by $\mathbf Y_u$, $u\le t$, together with any covariates known by the start of week $t+1$. The model has two ingredients. The first is a covariate map: for each site and week,
\begin{equation}\label{eq:cov}
X_{s,t}=\Phi_s(\mathbf Y_{t-1},\dots,\mathbf Y_{t-m};t)\in[0,1]^d
\end{equation}
is a vector of $d$ summaries of the last $m$ weeks, computed by a known measurable map $\Phi_s$; we call $m$ the memory of the model. Independent exogenous inputs can be added as further arguments of $\Phi_s$ (Assumption~\ref{ass:memory}). The second is a pair of functions, the log-mean $f_0:[0,1]^d\to[-F,F]$ and the log-size $g_0:[0,1]^d\to[\gamma_{\rm lo},\gamma_{\rm hi}]$. \ours{} specifies that, given $\mathcal I_{t-1}$,
\begin{equation}\label{eq:modelgen}
Y_{s,t}\mid\mathcal I_{t-1}\sim\NB\big(\exp\{f_0(X_{s,t})\},\,\exp\{g_0(X_{s,t})\}\big)\qquad\text{independently for }s=1,\dots,S,
\end{equation}
so that $\E(Y_{s,t}\mid\mathcal I_{t-1})=\mu_{s,t}=e^{f_0(X_{s,t})}$ and $\var(Y_{s,t}\mid\mathcal I_{t-1})=\mu_{s,t}+\mu_{s,t}^2e^{-g_0(X_{s,t})}$: the size $e^{g_0}$ controls the overdispersion, which can thus vary with the local level, the neighbors and the season. The special case of a constant size $r$,
\begin{equation}\label{eq:model}
Y_{s,t}\mid\mathcal I_{t-1}\sim\NB\big(\exp\{f_0(X_{s,t})\},\,r\big)\qquad\text{independently for }s=1,\dots,S,
\end{equation}
is the one analyzed in Sections~\ref{sec:oracle} to~\ref{sec:pred}; Section~\ref{sec:unknown} extends every result to \eqref{eq:modelgen} with unknown $g_0$. Other familiar models are special cases as well. When $\Phi_s$ uses only the history of site $s$ and the season, \eqref{eq:modelgen} is a deep negative binomial autoregression for each site with a function shared by all sites, which we call DeepNB-AR; when, in addition, $f_0$ is linear in $\log(1+Y_{s,t-1})$ and the season, it is a log-linear count autoregression of the kind studied by \citet{fokianos2009}, with negative binomial instead of Poisson errors; and a log-mean linear in transformed own, neighbor and seasonal summaries gives a log-linear negative binomial autoregression. This differs from the classical endemic--epidemic model of \citet{held2005}, whose conditional mean is an additive sum of endemic, own-history and neighboring contributions. Together with initial values $\mathbf Y_{0},\dots,\mathbf Y_{1-m}$, \eqref{eq:cov} and \eqref{eq:modelgen} define the joint law of the whole panel, so the model is generative: it can be simulated forward from any history.

Conditional independence across sites given the past is the standard assumption of endemic--epidemic models \citep{held2005}; spatial dependence between contemporaneous counts arises through the shared past. The function $f_0$ is common to all sites, which allows short and sparse series to inform one another, and site-specific behavior enters through the long-run level of the site, which is a coordinate of $X_{s,t}$, and, in the estimator of Section~\ref{sec:est}, through a learned encoding of the site. Sparsity needs no separate mechanism: under \eqref{eq:model} the probability of a zero is $\{r/(r+\mu)\}^r$, which is close to one for small means.

\subsection{Covariates: Temporal Memory and Nearest-Neighbor Encoding}\label{sec:covariates}

In the application of Section~\ref{sec:app}, $m=52$, and the covariate vector starts with 13 summaries. Let $\ell(y)=\min\{\log(1+y)/6,1\}$, which maps counts up to $e^6-1\approx402$, more than any weekly count in our data, into $[0,1]$. With $\bar Y_{s,t}^{(k)}$ the mean of $Y_{s,t-1},\dots,Y_{s,t-k}$, the coordinates of $X_{s,t}$ are as follows. Seven coordinates describe the history of the site itself, namely $\ell(Y_{s,t-1}),\dots,\ell(Y_{s,t-4})$, $\ell(\bar Y^{(8)}_{s,t})$, $\ell(\bar Y^{(26)}_{s,t})$ and $\ell(\bar Y^{(52)}_{s,t})$. Two coordinates describe its nearest neighbors, namely $\ell$ of the average of $\bar Y^{(4)}_{s',t}$ and of $\bar Y^{(26)}_{s',t}$ over the five sites $s'$ nearest to $s$ in great-circle distance. Two coordinates describe the overall level, namely $\ell$ of the average of $\bar Y^{(4)}_{s',t}$ and of $\bar Y^{(26)}_{s',t}$ over all sites. The last two coordinates are the seasonal harmonics $\{1+\sin(2\pi w_t/52.18)\}/2$ and $\{1+\cos(2\pi w_t/52.18)\}/2$, with $w_t\in\{1,\dots,53\}$ the week of the year and $52.18$ the average number of weeks in a year. The nearest-neighbor coordinates carry the spatial information that a nearest-neighbor Gaussian process \citep{datta2016} carries through its conditioning sets, and the lagged coordinates carry the temporal information that a recurrent or attention layer would extract. What the network must learn is how these summaries combine.

The estimator of Section~\ref{sec:est} adds two groups of inputs to these 13 summaries. The first is the raw recent history: $\ell(Y_{s,t-k})$, $\ell$ of the mean of $Y_{s',t-k}$ over the five nearest sites $s'$, and $\ell$ of the mean of $Y_{s',t-k}$ over all sites, for $k=1,\dots,26$, which lets the network learn its own temporal filters, as the recurrent encoder of a Temporal Fusion Transformer does. The second is the indicator $e_s\in\{0,1\}^S$ of the site, which enters the network only through a learned linear encoding of dimension three, the analogue of a static covariate encoder; it lets the log-mean and the log-size of sites with the same recent history differ. Both groups are functions of the past and of $s$, so they are coordinates of $X_{s,t}$ in \eqref{eq:cov}, now with $d=13+78+S$, and the covariate formulation of Section~\ref{sec:theory} accommodates them.

\subsection{Target Function with a Compositional Structure}\label{sec:structure}

The statistical rate we can achieve depends on the regularity of $f_0$. It is natural to suppose that $f_0$ combines a few low-dimensional components, for example
\begin{equation}\label{eq:comp}
f_0(x)=g_1\big(g_{01}(x_{\rm own}),\,g_{02}(x_{\rm nbr}),\,g_{03}(x_{\rm nat}),\,g_{04}(x_{\rm seas})\big),
\end{equation}
where $x_{\rm own}$, $x_{\rm nbr}$, $x_{\rm nat}$ and $x_{\rm seas}$ are the sub-vectors of $x$ formed by the seven, two, two and two coordinates of the four groups of Section~\ref{sec:covariates}, and in which the history of the site is first reduced to a local level and growth, the neighborhood and the overall level to an external pressure, the season to a phase effect, and these are then combined. An additive positive-component mean can be represented on the log scale by a log-sum-exp outer map applied to component log-means. Exact inclusion of an endemic--epidemic specification also requires suitable covariates and bounded ranges; it does not follow merely by making $f_0$ log-linear. We formalize \eqref{eq:comp} with the compositional H\"older class of \citet{schmidthieber2020}. For $\beta>0$ and $K>0$, let $\mathcal C^{\beta}(D,K)$ be the ball of radius $K$ of $\beta$-H\"older functions on $D\subset\mathbb R^{k}$, that is, the functions $g:D\to\mathbb R$ whose partial derivatives up to order $\lfloor\beta\rfloor$ exist and satisfy
\[
\sum_{|\bm\alpha|<\beta}\|\partial^{\bm\alpha}g\|_\infty+\sum_{|\bm\alpha|=\lfloor\beta\rfloor}\,\sup_{u\neq v}\frac{|\partial^{\bm\alpha}g(u)-\partial^{\bm\alpha}g(v)|}{|u-v|_\infty^{\beta-\lfloor\beta\rfloor}}\le K.
\]
For $q\ge0$, $\mathbf d=(d_0,\dots,d_{q+1})$ with $d_0=d$ and $d_{q+1}=1$, $\mathbf t=(t_0,\dots,t_q)$ with $t_i\le d_i$, and $\bm\beta=(\beta_0,\dots,\beta_q)$, define
\[
\begin{aligned}
\mathcal G(q,\mathbf d,\mathbf t,\bm\beta,K)=\big\{f=g_q\circ\cdots\circ g_0:\ &g_i=(g_{ij})_{j}:[a_i,b_i]^{d_i}\to[a_{i+1},b_{i+1}]^{d_{i+1}},\\
&g_{ij}\in\mathcal C^{\beta_i}_{t_i}([a_i,b_i]^{t_i},K),\ |a_i|,|b_i|\le K\big\},
\end{aligned}
\]
where $\mathcal C^{\beta_i}_{t_i}$ indicates that each $g_{ij}$ depends on at most $t_i$ of its $d_i$ arguments. Thus $q+1$ is the number of composed maps, $d_i$ the number of arguments of $g_i$, $t_i$ its intrinsic dimension and $\beta_i$ its smoothness. With the effective smoothness $\beta_i^*=\beta_i\prod_{l=i+1}^{q}(\beta_l\wedge1)$, the rate attached to this class is
\begin{equation}\label{eq:phin}
\phi_n=\max_{i=0,\dots,q}n^{-2\beta_i^*/(2\beta_i^*+t_i)}.
\end{equation}
In \eqref{eq:comp} with twice differentiable components, $q=1$, $\mathbf t=(7,4)$ and $\phi_n=n^{-4/11}$, whereas an unstructured twice differentiable function of 13 variables would give $n^{-4/17}$.

\section{Summary of Main Results}\label{sec:summary}

This section summarizes our main results and places them in the context of prior work on nonparametric estimation with deep neural networks. Let $n=ST$ be the number of site-weeks. The key result is that, under the conditional model \eqref{eq:model} and the compositional assumption on $f_0$, the maximum likelihood estimator over sparse deep ReLU networks satisfies
\[
\frac1n\sum_{s,t}\{\hat f(X_{s,t})-f_0(X_{s,t})\}^2\lesssim\phi_n\log^3n
\]
with high probability, under the stated conditional distribution and within-week independence assumptions (Theorems~\ref{thm:oracle} and~\ref{thm:rate}). The same rate, with one more logarithmic factor, holds for the risk on an independent copy of the process (Theorem~\ref{thm:oos}), because the model is geometrically absolutely regular (Proposition~\ref{prop:mixing}). In the regression experiment defined in Section~\ref{sec:lower}, Theorem~\ref{thm:lower} shows that no estimator can improve on $n^{-2\beta/(2\beta+d^*)}$ uniformly over $\beta$-H\"older log-means that depend on $d^*$ variables, which is $\phi_n$ for that class. Theorem~\ref{thm:pred} turns the risk into prediction error bounds: the difference between true and fitted-model coverage of a one-step prediction interval is at most a constant multiple of $|\hat f-f_0|$ at the covariate value in question, and the independent-copy average coverage shortfall of $h$-step intervals computed from $B$ simulated paths, under the additional conditions stated after Theorem~\ref{thm:pred}, is bounded in order by $(hS\phi_n)^{1/2}\log^2n+B^{-1/2}$. Theorem~\ref{thm:unknown} shows that the same rates hold when the size is unknown and may depend on the covariates, so that $f_0$ and the log-size $g_0$ are estimated jointly, and that functionals satisfying the empirical Lipschitz condition have absolute error of the square-root order of the squared-function-risk bound. This is consistency, not a distributional limit or a bootstrap validity theorem. Theorem~\ref{thm:dep} shows that the estimation bounds need only the conditional law of each count: under arbitrary dependence between the sites they hold with $T=n/S$ in place of $n$.

Supplementary Table~\ref{S-tab:s-related} compares these results with the closest works. \citet{schmidthieber2020} establishes the rate $\phi_n\log^3n$ with a lower bound of order $\phi_n$ for least squares with Gaussian noise and independent observations, and \citet{yara2026} the analogous Hellinger rate for the likelihood of a Poisson process observed in independent replications. \citet{ma2022} and \citet{kurisu2025} obtain rates for least squares under mixing, \citet{kurisu2025} also verifying geometric mixing for their autoregressive class, and \citet{alquier2025} for least squares and logistic losses on Markov chains with a spectral gap. Our results differ in four respects: the loss is the likelihood of an unbounded, overdispersed count, handled through the Hellinger distance and its equivalence with the squared difference of log-means (Lemma~\ref{lem:hell}); the dispersion is a second unknown function, estimated at the same rate (Theorem~\ref{thm:unknown}); the main bound holds along one observed trajectory without stationarity or mixing, through a martingale argument; and the estimation risk is connected to multi-step predictive distributions and the coverage of their intervals.

The rate theorems concern sparse networks with weights bounded by one and an approximate maximizer of the likelihood, and the oracle inequalities hold for any bounded class with finite covering numbers (Section~\ref{sec:unknown}); the networks of the case study are unconstrained and trained by a stochastic gradient method, whose optimization error our results do not certify. Simulation~A therefore reports both a fit that respects the constraints of the theory and the practical fit.

\section{Estimation by Maximum Likelihood and Forecasting by Simulation}\label{sec:est}

We next describe the estimator and the procedure that turns it into predictive distributions.

\subsection{Maximum Likelihood over ReLU Networks}

The log-likelihood of a candidate $f$ is
\begin{equation}\label{eq:loglik}
\ell_n(f)=\sum_{t=1}^{T}\sum_{s=1}^{S}\log p_{f(X_{s,t})}(Y_{s,t}).
\end{equation}
Let $\sigma(u)=\max(u,0)$ be the ReLU activation and, for a shift vector $v=(v_1,\dots,v_k)$, let $\sigma_v(u_1,\dots,u_k)=(\sigma(u_1-v_1),\dots,\sigma(u_k-v_k))$. A ReLU network with $L$ hidden layers and width vector $\mathbf p=(p_0,\dots,p_{L+1})$, $p_0=d$ and $p_{L+1}=1$, is a function
\[
f(x)=W_L\,\sigma_{v_L}W_{L-1}\,\sigma_{v_{L-1}}\cdots W_1\,\sigma_{v_1}W_0\,x,\qquad x\in[0,1]^d,
\]
with weight matrices $W_l\in\mathbb R^{p_{l+1}\times p_l}$ and shift vectors $v_l\in\mathbb R^{p_l}$. Following \citet{schmidthieber2020}, let $\mathcal F(L,\mathbf p,s_n,F)$ be the set of such functions with all entries of the $W_l$ and $v_l$ in $[-1,1]$, at most $s_n$ of them non-zero, and $\|f\|_\infty\le F$. The estimator is any measurable $\hat f\in\mathcal F_n=\mathcal F(L,\mathbf p,s_n,F)$ with
\begin{equation}\label{eq:mle}
\ell_n(\hat f)\ge\sup_{f\in\mathcal F_n}\ell_n(f)-n\Delta_n,
\end{equation}
where $\Delta_n\ge0$ is the optimization error per observation, which measures how far the optimizer is from the maximum. In the experiments we use a fully connected network with three hidden layers of width 64 and output truncated to $[-F,F]$ with $F=7$, trained by Adam \citep{kingma2015} with weight decay $10^{-5}$ and batches of 1{,}024. The estimator uses the inputs of Section~\ref{sec:covariates} including the raw history and the site encoding, gives the network a second output, the log-size $g(X_{s,t})$, truncated to $[\log0.5,\log200]$, and maximizes the likelihood \eqref{eq:mle2} of the model with covariate-dependent size jointly over both outputs (Section~\ref{sec:unknown}). Within every fit and yearly refit, the learning rate decreases linearly from $10^{-3}$ to $5\times10^{-5}$; this reduces the optimization error at the end of each fit and, in our experiments, most of the variation between fits from different initializations. The first fit, to the data up to the first forecast origin, uses 40 epochs; each yearly refit continues from the previous estimate for 8 epochs on all data to date. With 13 summaries, a constant size and a constant learning rate, the same procedure gives the special case that we evaluated first; its results and the development that led from it to the present estimator are reported in the supplementary material.

The likelihood is not concave in the network weights, so the fit depends on the random initialization. In the application we therefore use the average $\bar f=n_{\rm fit}^{-1}\sum_{k=1}^{n_{\rm fit}}\hat f^{(k)}$ of fits from different initializations, together with the average $\bar g$ of the log-sizes (for a constant size, the geometric mean of the estimated sizes); $\bar f$ and $\bar g$ each average $n_{\rm fit}=3$ fits. Averaging preserves the error bounds, by convexity of the squared norms and a union bound over the fits (Theorem~\ref{thm:unknown}(d)), and Theorems~\ref{thm:pred} and~\ref{thm:unknown}(c) apply to the averaged model as to any bounded function. All of these choices were made on the syphilis panel, the development panel of Section~\ref{sec:app}, and fixed in advance for amebiasis, the test panel.

\subsection{Predictive Distributions}

One step ahead, the predictive distribution at site $s$ is $\NB(\exp\{\hat f(X_{s,t+1})\},\hat r)$, available in closed form; here $\hat r$ is the estimate of the size, $r$ itself when it is known, or $\exp\{\hat g(X_{s,t+1})\}$ when the size depends on the covariates. For horizons $h\ge2$ the predictive distribution of $\mathbf Y_{t+h}$ given $\mathcal I_t$ is a mixture over the unobserved intermediate weeks, and we compute it by simulating \eqref{eq:cov} and \eqref{eq:modelgen} forward with $(\hat f,\hat g)$ in place of $(f_0,g_0)$, as described by Algorithm~\ref{alg:forecast}. The simulated paths are joint over sites and horizons, so they also give predictive distributions of any functional, such as the total over all sites or the cumulative count over four weeks. For the marginal predictive distribution of a single count we Rao--Blackwellize: given the path up to week $t+h-1$, the count of week $t+h$ is negative binomial with known parameters, so the $h$-step predictive law is estimated by the mixture $B^{-1}\sum_b\NB(\exp\{\hat f(X^{(b)}_{s,t+h})\},\exp\{\hat g(X^{(b)}_{s,t+h})\})$ over the paths rather than by the empirical law of the simulated counts. The distribution functions of both have expectation equal to the predictive distribution function of the fitted model, and that of the mixture has the smaller variance, by the Rao--Blackwell theorem.

\begin{algorithm}[t]
\caption{\ours: estimation and forecasting}\label{alg:forecast}
\begin{algorithmic}[1]
\State \textbf{input:} counts $\mathbf Y_1,\dots,\mathbf Y_t$, covariate map $\Phi=(\Phi_1,\dots,\Phi_S)$, largest horizon $h_{\max}$, number of paths $B$
\State Compute $X_{s,u}$ by \eqref{eq:cov} for $u\le t$ and maximize the likelihood over $(f,r)$, or over $(f,g)$ for a covariate-dependent size, to obtain the fitted model, or the average of $n_{\rm fit}$ such fits
\For{$b=1,\dots,B$}
  \State Set $\mathbf Y^{(b)}_u=\mathbf Y_u$ for $u\le t$
  \For{$k=1,\dots,h_{\max}$}
     \State $X^{(b)}_{s,t+k}\gets\Phi_s(\mathbf Y^{(b)}_{t+k-1},\dots,\mathbf Y^{(b)}_{t+k-m};t+k)$ for all $s$
     \State Draw $Y^{(b)}_{s,t+k}\sim\NB(\exp\{\hat f(X^{(b)}_{s,t+k})\},\hat r)$ independently for all $s$, with $\hat r=\exp\{\hat g(X^{(b)}_{s,t+k})\}$ for a covariate-dependent size
  \EndFor
\EndFor
\State \textbf{return} for each $(s,k)$ the quantiles of the mixture $B^{-1}\sum_b\NB(\exp\{\hat f(X^{(b)}_{s,t+k})\},\hat r^{(b)}_{s,t+k})$ (exact negative binomial quantiles for $k=1$), and the paths for any joint functional
\end{algorithmic}
\end{algorithm}

\subsection{Statistical Inference}\label{sec:inference}

\ours{} is a likelihood model, so it supports inference about the data-generating process and not only prediction. Three kinds of inference are reported in Section~\ref{sec:app}. The first is predictive: prediction intervals and predictive probabilities, whose coverage error is bounded by Theorems~\ref{thm:pred} and~\ref{thm:unknown}(c) and checked against the model-implied coverage. The second concerns the conditional mean and size functions themselves, through interpretable functionals: the fitted expected count of the whole panel in every week, the predicted mean and size at stated histories, the seasonal profile, and average partial effects. For a group $G$ of covariates (the history of the site, of its neighbors, or of the whole panel), let $x^{(G)}$ be the covariate vector obtained from $x$ by replacing each selected count summary $a$ by $2a$, recomputing its transformed coordinate, and holding all other coordinates fixed; the average partial effect of $G$ is
\begin{equation}\label{eq:ape}
\mathrm{APE}_G(f)=\exp\Big[\frac1n\sum_{s,t}\big\{f\big(X^{(G)}_{s,t}\big)-f(X_{s,t})\big\}\Big],
\end{equation}
the geometric mean, over the observed site-weeks, of the fitted mean ratio. This is a finite feature contrast, not a derivative or a causal intervention. Since own, neighboring and national summaries overlap, holding the other groups fixed can produce combinations that no single panel history realizes; these contrasts describe the fitted response surface. Functionals that average $f$ and $g$ over observed site-weeks, such as the fitted expected count of the panel, are estimated at the rate of Theorem~\ref{thm:unknown}(e); partial effects and curves at stated histories evaluate the functions at covariate values that need not have been observed, and the theorem does not cover them. Approximate pointwise intervals come from a moving-block bootstrap of the conditional likelihood: the weeks are resampled in blocks of consecutive weeks, keeping all sites of a week and their covariates together, the estimator is refitted to each resample, and the standard deviation of a functional over the refits, on the logarithmic scale for a positive functional, is its standard error $\widehat{\rm se}$. For a positive functional $\Psi$ the 95\% interval is $\Psi(\bar f,\bar g)\exp(\pm1.96\,\widehat{\rm se})$. This resampling preserves the observed response--history pairs and within-block dependence, but does not generate a new autoregressive trajectory; under correct specification the score of the conditional likelihood is a martingale difference sequence, which makes it reasonable for parametric autoregressions. We do not prove that the bootstrap is valid for deep network estimators: geometric $\beta$-mixing alone does not justify the stationary moving-block bootstrap \citep{kunsch1989} for these seasonal, changing panels, and regularization bias and the small number of refits can affect the intervals. Simulation~C checks the standard errors and the coverage. The third kind is comparative: whether a group of covariates or a feature of the model improves prediction is assessed by paired weekly differences of proper scores between the model with and without it, with block-bootstrap intervals, on forecasts that are out of sample.

\section{Main Results}\label{sec:theory}

We present the assumptions in Section~\ref{sec:assumptions} and the main results in Sections~\ref{sec:oracle} to~\ref{sec:unknown}.

\subsection{Assumptions and Risks}\label{sec:assumptions}

\begin{assumption}[Conditional model]\label{ass:model}
For every $t$, $X_{s,t}$ is $\mathcal I_{t-1}$-measurable with values in $[0,1]^d$ and, given $\mathcal I_{t-1}$, the counts $Y_{1,t},\dots,Y_{S,t}$ are independent with $Y_{s,t}\sim\NB(\exp\{f_0(X_{s,t})\},r)$, where $r>0$ is known.
\end{assumption}
\begin{assumption}[Boundedness]\label{ass:bounded}
$\|f_0\|_\infty\le F$, and every $f\in\mathcal F_n$ satisfies $\|f\|_\infty\le F$.
\end{assumption}
\begin{assumption}[Compositional smoothness]\label{ass:structure}
$f_0\in\mathcal G(q,\mathbf d,\mathbf t,\bm\beta,K)$ with $K\le F$.
\end{assumption}
\begin{assumption}[Finite memory]\label{ass:memory}
$X_{s,t}=\Phi_s(\mathbf Y_{t-1},\dots,\mathbf Y_{t-m},U_t;t)$ for measurable maps $\Phi_s$, a fixed integer $m\ge1$ and exogenous innovations $U_1,U_2,\dots$ that are independent and identically distributed on a Polish space $\mathcal U$, with $U_t$ independent of $(\mathbf Y_u,U_u)_{u<t}$ and $\mathcal I_{t-1}$-measurable.
\end{assumption}
\begin{assumption}[Absolute regularity]\label{ass:mixing}
The $\beta$-mixing coefficients \eqref{eq:betamix} of the covariate process $\mathbf X_t=(X_{1,t},\dots,X_{S,t})$ satisfy $\beta_X(a)\le C_\beta e^{-c_\beta a}$ for all integers $a\ge1$ and some constants $C_\beta\ge1$ and $c_\beta>0$.
\end{assumption}

In Assumption~\ref{ass:mixing}, the coefficient of absolute regularity, or $\beta$-mixing coefficient, between two random elements $U$ and $V$ is the total variation distance $\beta(U,V)=\TV(P_{(U,V)},P_U\otimes P_V)$ between their joint law and the product of their marginal laws \citep{bradley2005,yu1994}, and
\begin{equation}\label{eq:betamix}
\beta_X(a)=\sup_t\beta\big\{(\mathbf X_u)_{u\le t},(\mathbf X_u)_{u\ge t+a}\big\}.
\end{equation}
Assumption~\ref{ass:model} is the model \eqref{eq:model}; it allows any dependence of the covariates on the past and does not require stationarity. Assumption~\ref{ass:bounded} bounds the weekly mean between $e^{-F}$ and $e^{F}$. Assumption~\ref{ass:memory} is used for model-derived mixing and multi-step forecasting, and Assumption~\ref{ass:mixing} for the independent-copy risk bound; Proposition~\ref{prop:mixing} shows that the former implies the latter. The covariate map of Section~\ref{sec:covariates} has no innovation ($\mathcal U$ a single point); the innovations cover exogenous inputs such as the uniform covariate of Simulation~A.

Two risks are used. Along the observed trajectory,
\[
R_n^H(f)=\frac1n\sum_{s,t}H^2\big(p_{f(X_{s,t})},p_{f_0(X_{s,t})}\big),\qquad\|f-f_0\|_n^2=\frac1n\sum_{s,t}\{f(X_{s,t})-f_0(X_{s,t})\}^2 .
\]
Out of sample, let $(\mathbf Y'_t)$ be an independent copy of the panel, with covariates $\mathbf X'_t=(X'_{1,t},\dots,X'_{S,t})$ and states $Z'_t$ defined as $Z_t$ in Section~\ref{sec:pred}, and let
\[
R(f)=\frac1n\sum_{s,t}\E\{f(X'_{s,t})-f_0(X'_{s,t})\}^2 ,
\]
the average squared error of $f$ at covariate values that were not used for estimation; $R(\hat f)$ denotes $R(f)$ evaluated at $f=\hat f$. The two scales are equivalent: under Assumption~\ref{ass:bounded}, $c_1(f-f_0)^2\le H^2(p_f,p_{f_0})\le c_2(f-f_0)^2$ pointwise, with explicit constants $c_1,c_2$ depending only on $(F,r)$ (Lemma~\ref{lem:hell}).

\subsection{An Oracle Inequality along the Trajectory}\label{sec:oracle}

For $\varepsilon>0$ let $N_n(\varepsilon)$, the covering number of the class, be the smallest number of sup-norm balls of radius $\varepsilon$, centered in $\mathcal F_n$, that cover $\mathcal F_n$.

\begin{theorem}[Oracle inequality]\label{thm:oracle}
Let Assumptions~\ref{ass:model} and \ref{ass:bounded} hold, let $\hat f$ satisfy \eqref{eq:mle}, and let $f^*\in\mathcal F_n$ be any fixed function. For every $\varepsilon\in(0,1]$ and $\delta\in(0,1/3)$, with probability at least $1-3\delta$,
\begin{equation}\label{eq:oracle}
\begin{aligned}
R_n^H(\hat f)\le{}&\bar c\,\|f^*-f_0\|_\infty^2+\Delta_n+\frac{2\log N_n(\varepsilon)+(2+\lambda_0^{-1}+\varepsilon\lambda_1^{-1})\log(1/\delta)}{n}\\
&+\varepsilon(\bar\mu+r)+2c_2\varepsilon^2,
\end{aligned}
\end{equation}
and $\|\hat f-f_0\|_n^2\le R_n^H(\hat f)/c_1$. The constants $\bar c,\bar\mu,\lambda_0,\lambda_1,c_1,c_2$ are given in \eqref{eq:constants} and depend only on $(F,r)$.
\end{theorem}

The bound has the fast form, entropy over $n$ and not its square root, and its three parts are the approximation error of the class, the optimization error, and the complexity of the class. No assumption is made on the dependence of the data beyond the conditional specification.

The maximizer itself is not needed, and the optimization error can be measured against the truth. The proof shows that, with probability at least $1-2\delta$, every estimator $\hat f$ with values in $\mathcal F_n$, however it is computed, satisfies
\begin{equation}\label{eq:cert}
R_n^H(\hat f)\le\frac{\ell_n(f_0)-\ell_n(\hat f)}{n}+\frac{2\log N_n(\varepsilon)+(2+\varepsilon\lambda_1^{-1})\log(1/\delta)}{n}+\varepsilon(\bar\mu+r)+2c_2\varepsilon^2 .
\end{equation}
A fit whose likelihood on the trajectory is at least that of the truth thus attains the complexity term alone, whatever optimizer produced it.

\subsection{Rate of Convergence}

\begin{theorem}[Rate along the trajectory]\label{thm:rate}
Let Assumptions~\ref{ass:model}--\ref{ass:structure} hold and let $\mathcal F_n=\mathcal F(L,\mathbf p,s_n,F)$ with (i) $F\ge\max(K,1)$, (ii) $\sum_{i=0}^{q}\frac{\beta_i+t_i}{2\beta_i^*+t_i}\log_2(4t_i\vee4\beta_i)\log_2n\le L\lesssim n\phi_n$, (iii) $n\phi_n\lesssim\min_{1\le l\le L}p_l$ and (iv) $s_n\asymp n\phi_n\log n$. Then there is a constant $C$, depending only on $q,\mathbf d,\mathbf t,\bm\beta,F,r$ and the implicit constants in (ii)--(iv), such that for every $\delta\in(0,1/3)$, with probability at least $1-3\delta$,
\begin{equation}\label{eq:rate}
\|\hat f-f_0\|_n^2\le C\Big\{\phi_nL\log^2n+\Delta_n+\frac{\log(1/\delta)}{n}\Big\}.
\end{equation}
In particular, if $L\asymp\log n$ and $\Delta_n\lesssim\phi_n\log^3n$, then $\|\hat f-f_0\|_n^2\lesssim\phi_n\log^3n$ with probability at least $1-3/n$.
\end{theorem}

The rate is that of \citet{schmidthieber2020} for Gaussian regression with independent observations; here it holds for a negative binomial likelihood and for the dependent panel, in the empirical norm at the visited covariates. Its exponent depends on the component dimensions $t_i$; constants and architecture requirements can still depend on the ambient dimension $d$.

The depth condition uses the correction of \citet{schmidthieber2024correction}; the asymptotic rate is unchanged. The approximation statements are understood for sufficiently large $n$.

\subsection{Out-of-Sample Risk}\label{sec:oos}

To pass from the trajectory to new data we need the covariate process to forget its past. For \ours{} this is a consequence of the model.

\begin{proposition}[The model is geometrically absolutely regular]\label{prop:mixing}
Under Assumptions~\ref{ass:model}, \ref{ass:bounded} and \ref{ass:memory}, $\beta_X(a)\le(1-\epsilon_0)^{\lfloor a/m\rfloor}$ for all integers $a\ge1$, where $\epsilon_0=\{r/(r+e^F)\}^{rSm}$. Hence Assumption~\ref{ass:mixing} holds with $C_\beta=(1-\epsilon_0)^{-1}$ and $c_\beta=-m^{-1}\log(1-\epsilon_0)$.
\end{proposition}

The constant $\epsilon_0$ is very small for realistic $S$ and $m$, so the proposition should be read as a qualitative statement that no separate mixing assumption is needed. Seasonal covariates and non-stationarity are allowed because the bound is uniform in time.

\begin{theorem}[Out-of-sample rate]\label{thm:oos}
Let Assumptions~\ref{ass:model}--\ref{ass:structure} and \ref{ass:mixing} hold, with $\mathcal F_n$ as in Theorem~\ref{thm:rate}, $L\asymp\log n$ and $S$ fixed. There are constants $C$ and $T_0$, depending only on the quantities in Theorem~\ref{thm:rate} and on $(S,C_\beta,c_\beta)$, such that for $T\ge T_0$ and every $\delta\in(0,1/4)$, with probability at least $1-4\delta-C_\beta/T$,
\begin{equation}\label{eq:oos}
R(\hat f)\le C\Big\{\phi_n\log^4n+\Delta_n+\frac{\log n\,\log(1/\delta)}{n}\Big\},
\end{equation}
and $\E R(\hat f)\le C(\phi_n\log^4n+\Delta_n)$.
\end{theorem}

\subsection{Minimax Lower Bound}\label{sec:lower}

Let $\mathcal G_{\beta,d^*}(K)$ be the functions $f(x)=g(x_J)$ with $J\subset\{1,\dots,d\}$, $|J|=d^*$ and $g\in\mathcal C^{\beta}([0,1]^{d^*},K)$. It is the class $\mathcal G(q,\mathbf d,\mathbf t,\bm\beta,K)$ with $q=0$ and $t_0=d^*$, for which $\phi_n=n^{-2\beta/(2\beta+d^*)}$. For a class $\mathcal G$ of log-mean functions and fixed constants $(F,r,C_\beta,c_\beta)$, let $\mathcal P_n(\mathcal G)$ be the set of laws $P$ of $(X_{s,t},Y_{s,t})_{s\le S,t\le T}$, for which Assumption~\ref{ass:model} holds with some $f_0\in\mathcal G$, $\|f_0\|_\infty\le F$, and Assumption~\ref{ass:mixing} holds, for a fixed number of sites $S\ge1$ and $T=n/S$ weeks. The risk $R$ depends on $P$ through $f_0$ and through the law of the independent copy, and $\E_P$ denotes expectation under $P$.

\begin{theorem}[Lower bound]\label{thm:lower}
Let $0<K\le F$ and $1\le d^*\le d$. There is a constant $c>0$, depending only on $(\beta,d^*,K,F,r)$ and not on $S$, such that for all $n$,
\[
\inf_{\hat f}\ \sup_{P\in\mathcal P_n(\mathcal G_{\beta,d^*}(K))}\E_PR(\hat f)\ \ge\ c\,n^{-2\beta/(2\beta+d^*)},
\]
where the infimum is over all estimators, that is, all measurable maps from the data to functions on $[0,1]^d$.
\end{theorem}

Theorem~\ref{thm:oos} applies to every law in $\mathcal P_n(\mathcal G_{\beta,d^*}(K))$, so for every fixed $S$ the two theorems match up to logarithmic factors when the optimization error has the required order. The least favorable laws in the proof have independent uniform covariates; they satisfy all of Assumptions~\ref{ass:model}--\ref{ass:mixing}, with the covariates as exogenous innovations in Assumption~\ref{ass:memory}, and Simulation~A uses a design of this type. The theorem does not establish minimax optimality for a covariate map that depends on past counts alone, such as the map of Section~\ref{sec:covariates}. For such a map the covariates take countably many values, the law of the design is determined by $f_0$ itself, and the risk $R$ weights the error of $f$ only on the support of that law, so the packing argument in $L^2$ does not apply; the minimax rate for a fixed map of past counts is an open question. We also do not state a lower bound for the full compositional class, which requires a separate packing argument.

\subsection{Prediction Error Bounds}\label{sec:pred}

Forecast users care about predictive distributions more than about $f_0$. Let $P_f(\cdot\mid x)=\NB(e^{f(x)},r)$ be the one-step predictive law and, under Assumption~\ref{ass:memory}, let $\mathbb P^{(h)}_f(\cdot\mid z,t)$ be the joint law of $(\mathbf Y_{t+1},\dots,\mathbf Y_{t+h})$ when the last $m$ weeks are $z$ and the panel evolves according to \eqref{eq:cov}--\eqref{eq:model} with $f$ in place of $f_0$. Write $\TV_h(z,t)=\TV\{\mathbb P^{(h)}_{f_0}(\cdot\mid z,t),\mathbb P^{(h)}_{f}(\cdot\mid z,t)\}$ and $Z_t=(\mathbf Y_t,\dots,\mathbf Y_{t-m+1},U_{t+1})$ for the state of the panel at the end of week $t$ (without $U_{t+1}$ when there are no innovations); future innovations are drawn from their common law under both models, and $\E_{f_0}[\,\cdot\mid Z_t=z]$ is the expectation under the true model started from $z$.

\begin{theorem}[Prediction error]\label{thm:pred}
Let Assumptions~\ref{ass:model} and \ref{ass:bounded} hold and let $f$ be any function with $\|f\|_\infty\le F$.
\begin{enumerate}
\item[(a)] One step. For every $x$ and every set $C$ of counts,
\begin{equation}\label{eq:cov1}
|P_f(C\mid x)-P_{f_0}(C\mid x)|\le\TV\{P_f(\cdot\mid x),P_{f_0}(\cdot\mid x)\}\le(2c_2)^{1/2}|f(x)-f_0(x)|.
\end{equation}
Consequently, for $\alpha\in(0,1)$, a prediction set $\hat C_\alpha(x)$ with $P_f(\hat C_\alpha(x)\mid x)\ge1-\alpha$ has true coverage at least $1-\alpha-(2c_2)^{1/2}|f(x)-f_0(x)|$, and its coverage averaged over an independent copy of the process is at least $1-\alpha-\{2c_2R(f)\}^{1/2}$. Moreover the excess mean squared prediction error of the point forecast $e^{f(x)}$ satisfies $\E[(Y-e^{f(x)})^2-(Y-e^{f_0(x)})^2\mid X=x]\le e^{2F}\{f(x)-f_0(x)\}^2$.
\item[(b)] $h$ steps. Under Assumption~\ref{ass:memory} in addition, for every state $z$, week $t$ and $h\ge1$,
\begin{equation}\label{eq:simlemma}
\TV_h(z,t)\le\Big\{\frac{V_{\max}}{4}\,\E_{f_0}\Big[\sum_{k=1}^{h}\sum_{s=1}^{S}\big\{f(X_{s,t+k})-f_0(X_{s,t+k})\big\}^2\,\Big|\,Z_t=z\Big]\Big\}^{1/2},
\end{equation}
with $V_{\max}=e^F+e^{2F}/r$, and, averaging over forecast origins of an independent copy,
\begin{equation}\label{eq:simavg}
\frac{1}{T-h}\sum_{t=1}^{T-h}\E\,\TV_h(Z'_t,t)\le\Big\{\frac{V_{\max}\,h\,S\,T}{4(T-h)}R(f)\Big\}^{1/2}.
\end{equation}
Any prediction region for a function of $(\mathbf Y_{t+1},\dots,\mathbf Y_{t+h})$ that has probability at least $1-\alpha$ under the fitted model therefore has true coverage at least $1-\alpha-\TV_h(Z_t,t)$.
\item[(c)] Monte Carlo quantiles. Let $G$ be the fitted-model distribution function of a real-valued function of $(\mathbf Y_{t+1},\dots,\mathbf Y_{t+h})$ given $Z_t$, let $G_B$ be the empirical distribution function of $B$ independent draws from it, and let $\hat q_-\le\hat q_+$ satisfy $G_B(\hat q_+)\ge1-\alpha/2-\varkappa_B$ and $G_B(\hat q_--)\le\alpha/2+\varkappa_B$ for some $\varkappa_B\ge0$, where $G_B(q-)$ is the left limit. For every $\eta\in(0,1)$, with probability at least $1-\eta$ over the draws, the interval $[\hat q_-,\hat q_+]$ has fitted-model probability at least $1-\alpha-2\varkappa_B-2\{\log(2/\eta)/(2B)\}^{1/2}$. The empirical quantiles $\hat q_\pm=G_B^{-1}(\cdot)$ satisfy the condition with $\varkappa_B=0$, and quantiles computed by linear interpolation of the order statistics and rounded to an integer satisfy it with $\varkappa_B=1/B$.
\item[(d)] Rao--Blackwellized marginal quantiles. Conditional on the fitted model and a forecast origin, let $V_1,\ldots,V_B$ be independent intermediate histories for a fixed site and horizon, and let $K_b(q)$ be the conditional distribution function of its terminal count given $V_b$. Put $\widetilde G_B(q)=B^{-1}\sum_bK_b(q)$, so $G(q)=\E K_b(q)$. With probability at least $1-\eta$,
\[
\sup_q|\widetilde G_B(q)-G(q)|\le e_B(\eta)
:=\sqrt{\frac{\pi}{2B}}+\sqrt{\frac{\log(1/\eta)}{2B}}.
\]
If $\widetilde q_\pm$ satisfy the two quantile inequalities in (c) with $\widetilde G_B$ in place of $G_B$, their interval has fitted-model probability at least $1-\alpha-2\varkappa_B-2e_B(\eta)$. Exact mixture quantiles have $\varkappa_B=0$, and a numerical inversion that is accurate to a tolerance $\tau$ in probability has $\varkappa_B=\tau$. One step ahead, the conditional negative binomial distribution is known exactly and this Monte Carlo term is zero.
\end{enumerate}
\end{theorem}

Combining (b) with (c) or (d) with Theorem~\ref{thm:oos}, for $h<T$ with $T/(T-h)$ bounded and optimization error $\Delta_n\lesssim\phi_n\log^4n$, the coverage of the $h$-step intervals of Algorithm~\ref{alg:forecast}, averaged over the origins of an independent copy, falls short of their fitted-model probability by at most a quantity of order $(hS\phi_n)^{1/2}\log^2n+B^{-1/2}$, and the excess one-step mean squared prediction error is of order $\phi_n\log^4n$. Three qualifications apply. The bounds compare true and fitted-model probabilities; discrete intervals can cover more than their nominal level even under the true model. The average is over the time-indexed marginal laws of the observation period, so it does not bound conditional coverage at a particular origin or under a shift of the covariate law in the future. And the dependence on $h$ is an upper bound.

The bound also speaks to a common objection to recursive forecasting, that feeding forecasts back into the model accumulates errors over the horizon, which direct multi-horizon forecasters such as the Temporal Fusion Transformer avoid. For a likelihood model, \eqref{eq:simavg} quantifies the accumulation: the error of the $h$-step predictive law grows at most like the square root of the horizon, and its rate in $n$ is the one-step rate. Moreover, the predictive distributions of all sites and horizons are marginals of one joint law, so quantiles never cross, and the predictive distribution of any functional of the path, such as a national total or a four-week cumulative count, is available with the same guarantee. In our development experiments, a direct multi-horizon version of \ours{} was worse than recursive simulation at every horizon (supplementary Section~\ref{S-sec:s-dev}), and in the application \ours{} has a lower score than the three-seed transformer ensemble at every horizon on every panel, significantly so one week ahead on three panels and at all horizons on amebiasis (Table~\ref{tab:vstft}).

\subsection{Estimated and Covariate-Dependent Dispersion}\label{sec:unknown}

In practice the size is unknown, and it need not be constant: how overdispersed a count is can depend on the local level, on the neighbors and on the season. We therefore replace the known size of Assumption~\ref{ass:model} by the general model \eqref{eq:modelgen}, with an unknown log-size function $g_0$ taking values in $[\gamma_{\rm lo},\gamma_{\rm hi}]$, where $r_{\rm lo}=e^{\gamma_{\rm lo}}\le r_{\rm hi}=e^{\gamma_{\rm hi}}$ are known; a constant size $r_0$ is the case $g_0\equiv\log r_0$. Write $p_{f,r}$ for the $\NB(e^f,r)$ mass function. Let $\mathcal P_n$ be a class of pairs $(f,g)$ of measurable functions with $\|f\|_\infty\le F$ and $\gamma_{\rm lo}\le g\le\gamma_{\rm hi}$, and let $N_n(\varepsilon)$ be its covering number for the norm $\max(\|f-f'\|_\infty,\|g-g'\|_\infty)$. Two examples matter. For a constant size, $\mathcal P_n=\mathcal F_n\times\{\text{constants}\}$, and $N_n(\varepsilon)\le N^{\mathcal F}_n(\varepsilon)\{\lceil(\gamma_{\rm hi}-\gamma_{\rm lo})/(2\varepsilon)\rceil+1\}$, with $N^{\mathcal F}_n$ the covering number of $\mathcal F_n$. For a covariate-dependent size, $\mathcal P_n=\mathcal F^{(2)}(L,\mathbf p,s_n,F)$ is the class of sparse ReLU networks with two outputs, the first truncated to $[-F,F]$ and the second to $[\gamma_{\rm lo},\gamma_{\rm hi}]$; this is a constrained theoretical analogue of the estimator used in Section~\ref{sec:app}. The estimator is any measurable $(\hat f,\hat g)\in\mathcal P_n$ with
\begin{equation}\label{eq:mle2}
\ell_n(\hat f,\hat g)\ge\sup_{(f,g)\in\mathcal P_n}\ell_n(f,g)-n\Delta_n,\qquad \ell_n(f,g)=\sum_{s,t}\log p_{f(X_{s,t}),\exp g(X_{s,t})}(Y_{s,t}),
\end{equation}
and its trajectory risk is the Hellinger risk of the fitted one-step predictive laws,
$R_n^H(f,g)=n^{-1}\sum_{s,t}H^2(p_{f(X_{s,t}),e^{g(X_{s,t})}},p_{f_0(X_{s,t}),e^{g_0(X_{s,t})}})$; $R^H(f,g)$ is the same average of expectations over an independent copy, and $R_g(g)=n^{-1}\sum_{s,t}\E\{g(X'_{s,t})-g_0(X'_{s,t})\}^2$. The constants $\bar c_\gamma,\lambda_\gamma,\rho,s_0,\bar\mu_\gamma,c_0,\kappa_\gamma,c_f,c_\gamma$ of the next theorem are given in \eqref{eq:constants2a} and \eqref{eq:constants2} and depend only on $(F,\gamma_{\rm lo},\gamma_{\rm hi})$.

\begin{theorem}[Estimated and covariate-dependent dispersion]\label{thm:unknown}
Let \eqref{eq:modelgen} and Assumption~\ref{ass:bounded} hold, and let $(\hat f,\hat g)$ satisfy \eqref{eq:mle2}.
\begin{enumerate}
\item[(a)] For every fixed $(f^*,g^*)\in\mathcal P_n$ with $a=\max(\|f^*-f_0\|_\infty,\|g^*-g_0\|_\infty)$, every $\varepsilon\in(0,1]$ and $\delta\in(0,1/3)$, with probability at least $1-3\delta$,
\[
\begin{aligned}
R_n^H(\hat f,\hat g)\le{}&\bar c_\gamma a^2+\Delta_n+\frac{2\log N_n(\varepsilon)+(2+\lambda_\gamma^{-1}+2\rho\varepsilon s_0^{-1})\log(1/\delta)}{n}\\
&+2\rho\varepsilon(\bar\mu_\gamma+c_0)+\kappa_\gamma\varepsilon^2,
\end{aligned}
\]
and $c_f\|\hat f-f_0\|_n^2+c_\gamma\|\hat g-g_0\|_n^2\le2R_n^H(\hat f,\hat g)$.
\item[(b)] Let $f_0$ and $g_0$ belong to classes $\mathcal G(q,\mathbf d,\mathbf t,\bm\beta,K)$, possibly different, let $\phi_n$ be the larger of their rates \eqref{eq:phin}, and let $\mathcal P_n=\mathcal F^{(2)}(L,\mathbf p,s_n,F)$ satisfy conditions (i)--(iv) of Theorem~\ref{thm:rate} for both. Then $\|\hat f-f_0\|_n^2+\|\hat g-g_0\|_n^2\le C\{\phi_nL\log^2n+\Delta_n+\log(1/\delta)/n\}$ with probability at least $1-3\delta$, and under the conditions of Theorem~\ref{thm:oos}, $R(\hat f)+R_g(\hat g)\le C\{\phi_n\log^4n+\Delta_n+\log n\log(1/\delta)/n\}$ with probability at least $1-4\delta-C_\beta/T$. The constants $C$ depend in addition on $(\gamma_{\rm lo},\gamma_{\rm hi})$.
\item[(c)] Proposition~\ref{prop:mixing} holds with $\epsilon_0=\{(r_{\rm hi}/(r_{\rm hi}+e^F))^{r_{\rm hi}}\}^{Sm}$, and Theorem~\ref{thm:pred} holds for fitted laws $p_{f,e^g}$ with $\gamma_{\rm lo}\le g\le\gamma_{\rm hi}$, with $(2c_2)^{1/2}|f(x)-f_0(x)|$ in \eqref{eq:cov1} replaced by $\tfrac12\kappa_\gamma^{1/2}[\{f(x)-f_0(x)\}^2+\{g(x)-g_0(x)\}^2]^{1/2}$, $V_{\max}$ in \eqref{eq:simlemma} and \eqref{eq:simavg} replaced by $\kappa_\gamma$, the squared error of $f$ in \eqref{eq:simlemma} by the sum of the squared errors of $f$ and $g$, and $R(f)$ in \eqref{eq:simavg} by $R(f)+R_g(g)$.
\item[(d)] If $(\bar f,\bar g)=n_{\rm fit}^{-1}\sum_{k=1}^{n_{\rm fit}}(\hat f_k,\hat g_k)$ is the average of $n_{\rm fit}$ pairs with values in the same ranges, then $R_n^H(\bar f,\bar g)\le\{\kappa_\gamma/(2c_{\min})\}n_{\rm fit}^{-1}\sum_kR_n^H(\hat f_k,\hat g_k)$ with $c_{\min}=\min(c_f,c_\gamma)$, and the same holds for $R^H$.
\item[(e)] Let $\Psi$ be a real functional of $(f,g)$ such that $|\Psi(f,g)-\Psi(f_0,g_0)|\le L_\Psi(\|f-f_0\|_n+\|g-g_0\|_n)$ for all pairs in $\mathcal P_n$. Under the conditions of (b), $|\Psi(\hat f,\hat g)-\Psi(f_0,g_0)|\le CL_\Psi\{\phi_nL\log^2n+\Delta_n+\log(1/\delta)/n\}^{1/2}$ with probability at least $1-3\delta$. This applies, with $L_\Psi\le e^Fw^{-1/2}$, to the average of $e^{f}$ over any set of site-weeks that contains a fraction $w$ of all site-weeks, for example the expected count of the whole panel averaged over a period; for averages of $e^{g}$ the constant is $e^{\gamma_{\rm hi}}w^{-1/2}$, and averages of one-step predictive probabilities are covered by Lemma~\ref{lem:gamma}(b).
\end{enumerate}
\end{theorem}

The size enters as a second function. With a constant size it adds $\log\lceil(\gamma_{\rm hi}-\gamma_{\rm lo})/(2\varepsilon)\rceil\lesssim\log n$ to the entropy, and with a covariate-dependent size it doubles the output layer of the network; in both cases the rates are those of Theorems~\ref{thm:rate} and~\ref{thm:oos}, and the log-size is estimated at the same rate as the log-mean. Part~(d) covers the estimator of Section~\ref{sec:est}, which averages the log-means and log-sizes of several fits: any rate that holds for each fit with probability at least $1-\eta$ holds for the average with probability at least $1-n_{\rm fit}\eta$, up to the constant $\kappa_\gamma/(2c_{\min})$. Unlike the constant-size model, the model \eqref{eq:modelgen} can represent a variance that grows faster or slower than the square of the mean across the range of the covariates, which matters when sites of very different size share one network.

Theorems~\ref{thm:oracle} and~\ref{thm:unknown}(a) hold for any class of bounded functions, or pairs of functions, with finite covering numbers. In particular they apply to fully connected networks whose weights and shifts are bounded by one, the class $\mathcal F(L,\mathbf p,s_n,F)$ with $s_n$ equal to the number of parameters, for which $\log N_n(\varepsilon)\le(s_n+1)\log\{4\varepsilon^{-1}(L+1)V^2\}$ with $V=\prod_{l=0}^{L+1}(p_l+1)$, by Lemma~\ref{lem:sh}. Simulation~A fits such a network by projected stochastic gradient steps; the networks of the case study, with unconstrained weights and weight decay, lie outside this class.

\subsection{Dependence Between Sites}\label{sec:dep}

Conditional independence across sites is the part of the model that data contradict most visibly: national totals vary more than a sum of independent counts allows (Section~\ref{sec:calib}). The estimation theory does not need it. Suppose only that every count has the conditional law of the model.

\begin{assumption}[Marginal specification]\label{ass:marginal}
For every $s$ and $t$, $X_{s,t}$ is $\mathcal I_{t-1}$-measurable with values in $[0,1]^d$ and the conditional law of $Y_{s,t}$ given $\mathcal I_{t-1}$ is $\NB(\exp\{f_0(X_{s,t})\},\exp\{g_0(X_{s,t})\})$, with $\|f_0\|_\infty\le F$ and $\gamma_{\rm lo}\le g_0\le\gamma_{\rm hi}$. The joint conditional law of $(Y_{1,t},\dots,Y_{S,t})$ is arbitrary.
\end{assumption}

Under Assumption~\ref{ass:marginal} the criterion $\ell_n$ of \eqref{eq:mle2} is no longer the log-likelihood of the panel; it is a composite likelihood \citep{varin2011}, the sum of the log-likelihoods of the sites.

\begin{theorem}[Dependent sites]\label{thm:dep}
Let Assumption~\ref{ass:marginal} hold and let $(\hat f,\hat g)\in\mathcal P_n$ satisfy \eqref{eq:mle2}. Write $T=n/S$.
\begin{enumerate}
\item[(a)] For every fixed $(f^*,g^*)\in\mathcal P_n$ with $a=\max(\|f^*-f_0\|_\infty,\|g^*-g_0\|_\infty)$, every $\varepsilon\in(0,1]$ and $\delta\in(0,1/3)$, with probability at least $1-3\delta$,
\[
R_n^H(\hat f,\hat g)\le\bar c_\gamma a^2+\Delta_n+\frac{2\log N_n(\varepsilon)+(2+\lambda_\gamma^{-1}+2\rho\varepsilon s_0^{-1})\log(1/\delta)}{T}+2\rho\varepsilon(\bar\mu_\gamma+c_0)+\kappa_\gamma\varepsilon^2,
\]
and $c_f\|\hat f-f_0\|_n^2+c_\gamma\|\hat g-g_0\|_n^2\le2R_n^H(\hat f,\hat g)$. This is the bound of Theorem~\ref{thm:unknown}(a) with $T$ in place of $n$ in the stochastic term.
\item[(b)] Let $f_0$ and $g_0$ be as in Theorem~\ref{thm:unknown}(b) and let $\mathcal P_n=\mathcal F^{(2)}(L,\mathbf p,s,F)$ satisfy conditions (i)--(iv) of Theorem~\ref{thm:rate} with $T$ in place of $n$. Then $\|\hat f-f_0\|_n^2+\|\hat g-g_0\|_n^2\le C\{\phi_TL\log^2T+\Delta_n+\log(1/\delta)/T\}$ with probability at least $1-3\delta$, and Theorem~\ref{thm:unknown}(d) and (e) hold with this bound.
\item[(c)] The one-step bound \eqref{eq:cov1}, in the form of Theorem~\ref{thm:unknown}(c), holds for every site.
\item[(d)] The replacement of $n$ by $T$ cannot be avoided. For a known constant size, let $(X_{1,t},Y_{1,t})_{t\le T}$ follow a law of the class $\mathcal P_T(\mathcal G_{\beta,d^*}(K))$ of Theorem~\ref{thm:lower} for one site, and let $X_{s,t}=X_{1,t}$ and $Y_{s,t}=Y_{1,t}$ for $s=2,\dots,S$. These laws satisfy Assumption~\ref{ass:marginal}, and over them $\inf_{\hat f}\sup\E R(\hat f)\ge c\,T^{-2\beta/(2\beta+d^*)}$ with the constant $c$ of Theorem~\ref{thm:lower}.
\end{enumerate}
\end{theorem}

The proof replaces one step of the martingale argument: by the generalized H\"older inequality, the conditional expectation of $\prod_s\{p_{f,g}/p_{f_0,g_0}\}^{1/(2S)}(Y_{s,t})$ is at most $\prod_s(1-H_{s,t}^2)^{1/S}$ whatever the dependence between the sites, so the likelihood ratio raised to the power $1/(2S)$ still yields a supermartingale, at the price of a factor $S$ in the deviation terms. The two functions are thus estimated consistently, with the number of periods as sample size in the worst case and with the rate $\phi_n$ of the independent case, up to a constant, when $S$ is fixed; every statement about one site and one step remains valid. What is lost is the joint law. The bound of Theorem~\ref{thm:pred}(b) on the distance between the fitted and the true $h$-step laws, and with it the coverage of intervals for several steps ahead or for sums over sites, requires the dependence to be specified; beyond one step even the law of a single site depends on it, through the covariates. Parts~(c) and~(d) of that theorem remain valid, because they bound the Monte Carlo error relative to the fitted law, but they no longer translate into true coverage. The surveillance panels (Section~\ref{sec:calib}), Simulation~D with a Gaussian copula between sites (supplementary Section~\ref{S-sec:s-simD}) and daily taxi-flow networks (Section~\ref{S-sec:s-taxi}) all show this pattern: intervals for single counts are calibrated and those for totals over sites too narrow, and in Simulation~D the risk decreases at the same rate whatever the copula correlation.

\section{Roadmap of Proofs}\label{sec:roadmap}

The complete proofs are in the appendix; here we outline the arguments and where they depart from existing analyses of deep nonparametric estimation.

\paragraph{A martingale argument for the likelihood ratio.} Existing rates for deep networks compare an empirical risk with its population counterpart, which requires independent or mixing observations and a bounded or sub-Gaussian loss. We instead order the $n$ observations in time and, within a week, by site. For a fixed $f$, the likelihood ratio $\Lambda_i=p_{f(X_i)}(Y_i)/p_{f_0(X_i)}(Y_i)$ satisfies $\E[\Lambda_i^{1/2}\mid\mathcal H_{i-1}]=1-H^2(p_{f(X_i)},p_{f_0(X_i)})$ whatever the dependence between $X_i$ and the past, so an exponential supermartingale inequality (Lemma~\ref{lem:mart}) gives $\sum_iH^2(p_{f(X_i)},p_{f_0(X_i)})\le\tfrac12\{\ell_n(f_0)-\ell_n(f)\}+\log(1/\delta)$ with probability at least $1-\delta$, and a union bound over a sup-norm net makes this uniform at the price of the metric entropy. The device is that of \citet{vandegeer1995}, \citet{zhang2006} and \citet{agarwal2020}; what is new is its combination with sparse-network approximation for dependent negative binomial data, where the likelihood ratio is unbounded. The passage from the net to the estimator costs $\varepsilon\sum_i(Y_i+r)$, because the log-likelihood is Lipschitz in $f$ with constant $y+r$ (Lemma~\ref{lem:lip}), and the sum of the counts is bounded by a second supermartingale; the approximation term $\ell_n(f_0)-\ell_n(f^*)$ is controlled through its conditional moment generating function, rather than its second moment, which keeps the fast rate. Lemma~\ref{lem:hell} then gives Theorem~\ref{thm:oracle}, and Theorem~\ref{thm:rate} follows with the approximation and entropy bounds of \citet{schmidthieber2020}.

\paragraph{New data, the lower bound and the horizon.} Proposition~\ref{prop:mixing} is a Doeblin minorization \citep{meyn2009}: whatever the history, the whole panel is zero for $m$ consecutive weeks with probability at least $\epsilon_0$, after which the past is forgotten. Theorem~\ref{thm:oos} follows by blocking \citep{yu1994} and Bernstein's inequality on the net; mixing is needed only for this comparison. Theorem~\ref{thm:lower} applies Fano's method \citep[Theorem~2.5]{tsybakov2009} to a Varshamov--Gilbert packing in negative binomial regression with independent uniform covariates. For the horizon, the simulation lemma of reinforcement learning \citep{kearns2002,munos2008} adds one-step total variation distances, which grows linearly in $h$; Theorem~\ref{thm:pred}(b) uses the chain rule for the Kullback--Leibler divergence, as does \citet{xu2026} for independent trajectories, and applies Pinsker's inequality once, so that the sum over $S$ sites and $h$ steps sits under one square root. Parts~(c) and~(d) add the Monte Carlo errors by the Dvoretzky--Kiefer--Wolfowitz inequality and by conditional Jensen with bounded differences.

\paragraph{Unknown and covariate-dependent size.} The martingale steps use only that the true conditional law belongs to the model. The two-parameter family (Lemma~\ref{lem:two}) has a log-likelihood that is Lipschitz with constants linear in the count, a Kullback--Leibler divergence bounded by a quadratic form, and a Hellinger distance bounded below by the squared differences of both parameters, through the first two factorial moments, so that the Hellinger risk controls the size as well as the log-mean.

\section{Simulations}\label{sec:sim}

\subsection{Rates of Convergence}\label{sec:simA}

The first experiment illustrates the rate scales of Theorems~\ref{thm:rate}, \ref{thm:oos} and~\ref{thm:pred} in a model where $f_0$ is known. A single series ($S=1$) follows
\[
Y_t\mid\mathcal I_{t-1}\sim\NB\big(\exp\{f_0(X_t)\},3\big),\qquad X_t=\big(\ell_0(Y_{t-1}),\,\ell_0(Y_{t-2}),\,U_t\big),
\]
with memory $m=2$, $\ell_0(y)=\min\{\log(1+y)/\log61,1\}$, exogenous covariates $U_t$ that are independent and uniform on $[0,1]$, and
\[
f_0(x)=0.4+1.8x_1-0.9(x_1-0.5)_++0.8\,|x_2-0.4|+2\min(x_3,1-x_3).
\]
The log-mean is a sum of three Lipschitz functions of one variable, so $f_0\in\mathcal G(q,\mathbf d,\mathbf t,\bm\beta,K)$ with $q=1$, $\mathbf t=(1,3)$, $\beta_0=1$ and $\beta_1$ arbitrarily large, and $\phi_n=n^{-2/3}$. The process satisfies Assumptions~\ref{ass:model}--\ref{ass:mixing}, with $U_t$ an exogenous innovation in Assumption~\ref{ass:memory}, and it uses exogenous inputs as allowed in Theorem~\ref{thm:lower}, although its lagged covariates are dependent and it is not the independent-design subexperiment of that proof.

For $n$ between 250 and 16{,}000 we fit two estimators to the same data. The \emph{practical} fit is a configuration similar to that of the case study: a fully connected ReLU network with three hidden layers of width 32 and $F=4$, trained by Adam with $r=3$ known on the first $0.8n$ observations, with the number of epochs (passes through the training data) chosen on the remaining $0.2n$. The \emph{theory-aligned} fit is the estimator of Theorem~\ref{thm:unknown} over a fully connected class: the same architecture, with all weights and shifts projected onto $[-1,1]$ after every step and the output truncated to $[-F,F]$, so that it lies in $\mathcal F(L,\mathbf p,s_n,F)$ with $s_n$ the number of parameters; the size estimated jointly in $[r_{\rm lo},r_{\rm hi}]=[0.5,50]$; all $n$ observations; and a fixed budget of 3{,}000 projected Adam steps on batches of 512, with no data-dependent stopping. Neither fit certifies the optimization error $\Delta_n$. For each $n$ we record, over 30 replications, the trajectory risk $\|\hat f-f_0\|_n^2$, the out-of-sample risk $R(\hat f)$, estimated on an independent trajectory of length 5{,}000, and the total variation distance between the fitted and true one-step predictive laws (with the estimated size for the aligned fit), averaged over 1{,}000 new covariate values. In 5 replications we also record the coverage error of the nominal 90\% four-step prediction interval of Algorithm~\ref{alg:forecast} at 50 origins drawn from the true process. From each origin, three independent sets of 10{,}000 paths give the interval (fitted model), the probability that the fitted model assigns to it (fitted model) and its true coverage (true model); we report the mean over origins of the absolute difference between the last two, and the same procedure with $f_0$ in place of the fit gives the Monte Carlo floor of this statistic. A negative binomial regression that is log-linear in $X_t$ serves as a parametric benchmark.

\begin{table}[t]
\caption{Simulation A. Means over 30 replications of the trajectory risk $\|\hat f-f_0\|_n^2$ (Traj.), the out-of-sample risk $R(\hat f)$ (Out) and the total variation distance between fitted and true one-step predictive laws (TV); and, over 5 replications, of the absolute coverage error of 90\% four-step prediction intervals (Cov.), whose Monte Carlo floor is 0.0025. For the theory-aligned fit, $\hat r$ is the median estimated size (true value 3). ``Slope'' is the least squares slope of the logarithm of each column on $\log n$ for $n\ge500$; ``Reference'' is the exponent of the upper bounds in Theorems~\ref{thm:rate}, \ref{thm:oos} and~\ref{thm:pred}, without logarithmic factors.}\label{tab:simA}
\centering\footnotesize
\setlength{\tabcolsep}{4pt}
\begin{tabular}{rrrrrrrrrrr}
\toprule
 & \multicolumn{4}{c}{Practical fit ($r$ known)} & \multicolumn{5}{c}{Theory-aligned fit ($r$ estimated)} & NB reg. \\
\cmidrule(lr){2-5}\cmidrule(lr){6-10}\cmidrule(lr){11-11}
$n$ & Traj. & Out & TV & Cov. & Traj. & Out & $\hat r$ & TV & Cov. & Out \\
\midrule
250 & 0.1655 & 0.1701 & 0.1733 & 0.0368 & 0.0974 & 0.1168 & 4.62 & 0.1499 & 0.0191 & 0.1040 \\
500 & 0.0397 & 0.0408 & 0.0779 & 0.0204 & 0.0449 & 0.0494 & 3.54 & 0.0945 & 0.0187 & 0.1007 \\
1,000 & 0.0235 & 0.0242 & 0.0615 & 0.0204 & 0.0216 & 0.0223 & 3.21 & 0.0644 & 0.0088 & 0.0978 \\
2,000 & 0.0092 & 0.0093 & 0.0391 & 0.0049 & 0.0122 & 0.0124 & 3.07 & 0.0472 & 0.0112 & 0.0949 \\
4,000 & 0.0059 & 0.0059 & 0.0315 & 0.0062 & 0.0063 & 0.0063 & 3.04 & 0.0335 & 0.0098 & 0.0947 \\
8,000 & 0.0034 & 0.0034 & 0.0241 & 0.0058 & 0.0040 & 0.0040 & 2.96 & 0.0271 & 0.0060 & 0.0946 \\
16,000 & 0.0023 & 0.0023 & 0.0198 & 0.0036 & 0.0034 & 0.0034 & 2.98 & 0.0259 & 0.0052 & 0.0944 \\
\midrule
Slope & $-0.85$ & $-0.86$ & $-0.41$ & $-0.50$ & $-0.77$ & $-0.79$ & & $-0.39$ & $-0.32$ & $-0.02$ \\
Reference & $-0.67$ & $-0.67$ & $-0.33$ & $-0.33$ & $-0.67$ & $-0.67$ & & $-0.33$ & $-0.33$ & \\
\bottomrule
\end{tabular}

\end{table}

\begin{figure}[t]
\centering
\includegraphics[width=0.9\textwidth]{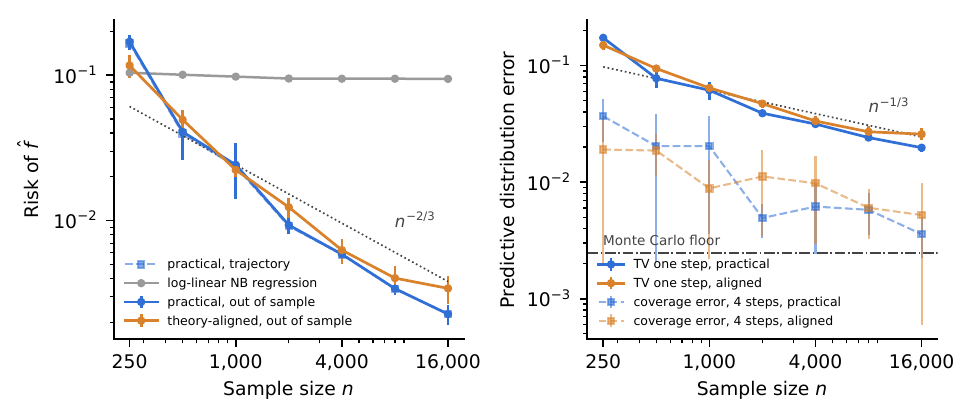}
\caption{Simulation A. Left: out-of-sample risk of the practical fit (blue, with two standard errors) and of the theory-aligned fit (orange), trajectory risk of the practical fit (dashed; it nearly coincides with the out-of-sample risk), and risk of a log-linear negative binomial regression (gray), against the sample size. Right: total variation error of the one-step predictive law and coverage error of 90\% four-step prediction intervals for both fits; the dash-dotted line is the Monte Carlo floor of the coverage error. Dotted lines show the reference slopes.}\label{fig:rates}
\end{figure}

Table~\ref{tab:simA} and Figure~\ref{fig:rates} give the results. First, the risks of both fits decrease with $n$, from 0.041 (practical) and 0.049 (aligned) at $n=500$ to 0.0023 and 0.0034 at $n=16{,}000$, with log-log slopes of $-0.86$ and $-0.79$. These are steeper than the reference $-2/3$; the guarantee is a worst case over the class, and this $f_0$ is piecewise linear, hence exactly representable by a ReLU network. The parametric benchmark stays at its misspecification error of 0.094. At $n=250$ both networks are worse than the parametric model, which is the price of flexibility in small samples. Second, the theory-aligned fit, which respects the weight bound, uses no validation data and estimates $r$, behaves much like the practical fit: it is better at $n=250$ and $n=1{,}000$, worse at $n=500$, and worse by 7\% to 48\% for $n\ge2{,}000$, where its fixed budget of gradient steps, rather than the sample size, appears to limit it. Its estimate of the size has median 4.62 at $n=250$ and between 2.96 and 3.07 for $n\ge2{,}000$, and its mean squared error falls from 0.15 at $n=1{,}000$ to 0.012 at $n=16{,}000$, in line with Theorem~\ref{thm:unknown}(b). Third, the trajectory risk and the out-of-sample risk agree to within 3\% for the practical fit at every $n$, and for the aligned fit from $n=2{,}000$ on; the theorems bound their orders but do not assert equality. Fourth, the total variation error of the one-step predictive law has slopes $-0.41$ and $-0.39$, about half those of the risks, and for the practical fit its ratio to the square root of the risk lies between 0.39 and 0.42 at every $n$, consistent with \eqref{eq:cov1}. Fifth, the coverage error of four-step intervals, computed from independent path sets, falls from 0.037 (practical) and 0.019 (aligned) at $n=250$ to 0.004 and 0.005 at $n=16{,}000$, against a Monte Carlo floor of 0.0025. The signed differences show that, where the error is distinguishable from the floor, the true coverage is below the probability that the fitted model assigns to its interval, which is the direction controlled by Theorem~\ref{thm:pred}(b) and~(c); the bound itself, which holds for every event simultaneously, is far from tight for a single interval.

\subsection{Forecasting a Spatial Panel}\label{sec:simB}

The second experiment, reported in full in supplementary Section~\ref{S-sec:s-simB}, simulates panels of 47 sites at the locations of the Japanese prefectures, with negative binomial counts whose log-mean is either log-linear in lagged own, neighbor and seasonal summaries, so that the NB regression benchmark of Section~\ref{sec:protocol} is correctly specified one week ahead, or nonlinear, with saturating self-excitation and a neighborhood effect that switches on above a threshold only where the local level is low. Over 10 replications and 104 test weeks, the constant-size special case of \ours{} with the 13 summaries loses 2.8\% in WIS to the correctly specified regression under the log-linear process, the cost of a nonparametric log-mean, and is better than the regression by 3.4\% under the nonlinear process, in each replication; both improve on persistence by more than 20\%.

\subsection{Covariate-Dependent Dispersion and Bootstrap Confidence Intervals}\label{sec:simC}

The third experiment is the variant of Simulation~A in which the dispersion depends on the covariates, the setting of Theorem~\ref{thm:unknown}, and it checks the confidence intervals of Section~\ref{sec:inference}. The process is that of Simulation~A with the size $r=3$ replaced by $\exp\{g_0(X_t)\}$,
\[
g_0(x)=\log3+1.2\,(x_3-0.5)+0.8\,(x_1-0.5),
\]
so that the size ranges from about 1.1 to 8.2 and the overdispersion of a count depends on the previous count and on the exogenous input. The estimator is the two-output network of the application, with three hidden layers of width 32, the log-mean truncated to $[-4,4]$ and the log-size to $[\log0.5,\log50]$, fitted jointly by maximum likelihood with Adam and a learning rate decreasing linearly from $2\times10^{-3}$ to $10^{-4}$ over $\max(2{,}000,n)$ steps on batches of 256, and averaged over two fits. For $n$ between 500 and 16{,}000 we record, over 20 replications, the out-of-sample risks $R(\hat f)$ and $R_g(\hat g)=\E\{\hat g(X)-g_0(X)\}^2$, estimated on an independent trajectory of length 5{,}000, and the total variation distance between the fitted and true one-step laws. For $n=2{,}000$ and $8{,}000$ we estimate, in each of 100 replications, the standard errors of $\hat f(x)$ and $\hat g(x)$ at three covariate values fixed in advance, $x^{(1)}=(0.3,0.3,0.25)$, $x^{(2)}=(0.6,0.5,0.5)$ and $x^{(3)}=(0.8,0.2,0.75)$, and of two averaged functionals like those of the application: the mean expected count $n^{-1}\sum_t\exp\{f(X_t)\}$ and the average partial effect $\exp[n^{-1}\sum_t\{f(X_t')-f(X_t)\}]$ of doubling one plus the previous count (the transformation $y\mapsto2y+1$ before clipping), where $X_t'$ is $X_t$ with its first coordinate increased by $\log2/\log61$ (and truncated at one). The standard errors come from a moving-block bootstrap of the pairs $(X_t,Y_t)$ with blocks of 20 observations and 20 refits, each of which, like the estimate, averages two fits; the intervals are normal, on the logarithmic scale for the two functionals, and we record how often they contain the true values.

\begin{table}[t]
\caption{Simulation C (covariate-dependent dispersion). Top: means over 20 replications of the out-of-sample risks of the log-mean and the log-size and of the total variation distance between fitted and true one-step laws; ``Slope'' is the least squares slope of the logarithm of each column on $\log n$ for $n\ge1{,}000$, and ``Reference'' the exponent of the bounds of Theorem~\ref{thm:unknown}, without logarithmic factors. Bottom: coverage of 95\% normal intervals with moving-block bootstrap standard errors for $f_0$ and $g_0$ at three covariate values and for the mean expected count (Mean) and the average feature contrast for doubling one plus the previous count (APE), with the number of replications in parentheses (100, so that the binomial standard error is about 0.02 near 95\%).}\label{tab:simC}
\centering\footnotesize\setlength{\tabcolsep}{4pt}
\begin{tabular}{rrrr}
\toprule
$n$ & $R(\hat f)$ & $R_g(\hat g)$ & TV \\
\midrule
500 & 0.0210 & 0.0969 & 0.0717 \\
1,000 & 0.0114 & 0.0448 & 0.0527 \\
2,000 & 0.0060 & 0.0202 & 0.0381 \\
4,000 & 0.0047 & 0.0129 & 0.0321 \\
8,000 & 0.0025 & 0.0085 & 0.0243 \\
16,000 & 0.0017 & 0.0057 & 0.0192 \\
Slope & $-0.67$ & $-0.72$ & $-0.36$ \\
Reference & $-0.67$ & $-0.67$ & $-0.33$ \\
\bottomrule
\end{tabular}

\vspace{6pt}
\begin{tabular}{lrrrrrrrr}
\toprule
 & \multicolumn{3}{c}{$f_0(x)$} & \multicolumn{3}{c}{$g_0(x)$} & \multicolumn{2}{c}{Averaged functionals} \\
\cmidrule(lr){2-4}\cmidrule(lr){5-7}\cmidrule(lr){8-9}
$n$ (replications) & $x^{(1)}$ & $x^{(2)}$ & $x^{(3)}$ & $x^{(1)}$ & $x^{(2)}$ & $x^{(3)}$ & Mean & APE \\
\midrule
2,000 (100) & 0.94 & 0.82 & 0.82 & 0.97 & 0.93 & 0.91 & 0.94 & 0.87 \\
8,000 (100) & 0.97 & 0.79 & 0.93 & 1.00 & 0.93 & 0.91 & 0.91 & 0.92 \\
\bottomrule
\end{tabular}

\end{table}

Table~\ref{tab:simC} gives the results. The observed slopes are consistent with the reference exponents: the out-of-sample risk of the log-mean falls from 0.021 at $n=500$ to 0.0017 at $n=16{,}000$, with log-log slope $-0.67$, and that of the log-size from 0.097 to 0.0057, with slope $-0.72$; the size is harder to estimate than the mean, and its risk is between 2.7 and 4.6 times larger. The total variation error of the one-step law has slope $-0.36$, close to half that of the risks, consistent with the upper-bound scale in Theorem~\ref{thm:unknown}(c), which does not assert equality of empirical slopes. The bootstrap standard errors are close to the sampling standard deviations: averaged over replications, their ratio to the Monte Carlo standard deviation of the estimate is between 0.96 and 1.22 for $\hat f(x)$ and $\hat g(x)$ at the three points and between 0.99 and 1.05 for the two averaged functionals. With 100 replications the binomial standard error of an estimated coverage near 95\% is about 0.02. The coverage of the 95\% intervals is between 0.87 and 0.94 for the averaged functionals, between 0.91 and 1.00 for $\hat g$ at all three points, and between 0.93 and 0.97 for $\hat f$ at $x^{(1)}$ and, at $n=8{,}000$, at $x^{(3)}$. It is 0.82 and 0.79 for $\hat f$ at $x^{(2)}$, where $f_0$ has a kink in $x_3$: there the network estimator rounds the kink, its bias ($-0.056$ and $-0.049$) is comparable to or larger than its standard deviation (0.056 and 0.035), and an interval that is centered at the estimate cannot correct it. At $n=2{,}000$ the coverage is also low at $x^{(3)}$, a point in a sparse region of the covariate distribution (0.82), where the bias falls from 0.088 to 0.037 at $n=8{,}000$ and the coverage recovers to 0.93. The intervals of the application are therefore exploratory: they reflect the sampling variability of the estimator well, but not its bias where the fitted functions bend sharply or the data are sparse.

\section{Application to Infectious Disease Surveillance in Japan}\label{sec:app}

We forecast four panels of weekly notifications in the 47 prefectures of Japan and summarize the fitted model. Section~\ref{sec:data} describes the data, Section~\ref{sec:protocol} the forecasting protocol and the benchmarks, Section~\ref{sec:accuracy} the accuracy of the forecasts, Section~\ref{sec:inf-app} exploratory inference for the fitted model, and Section~\ref{sec:calib} the calibration of its predictive distributions.

\subsection{Data}\label{sec:data}

Syphilis, enterohemorrhagic \emph{Escherichia coli} (EHEC) infection, legionellosis and amebiasis are notifiable diseases in Japan, and every diagnosed case is reported to the National Epidemiological Surveillance of Infectious Diseases \citep{takahashi2018}. The Japan Institute for Health Security publishes the weekly number of notifications in each of the 47 prefectures as open data, which we obtained through the R package \texttt{jpinfect} \citep{hoshi2025}. Each series runs from the week ending 11 April 1999, when the current surveillance law took effect, to the week ending 29 December 2024: 1{,}342 weeks and 63{,}074 prefecture-weeks. Provisional counts for 2025 and later are subject to revision and are not used. EHEC infection was renamed in 2006, and the two series are joined.

\begin{figure}[p]
\centering
\includegraphics[width=\textwidth]{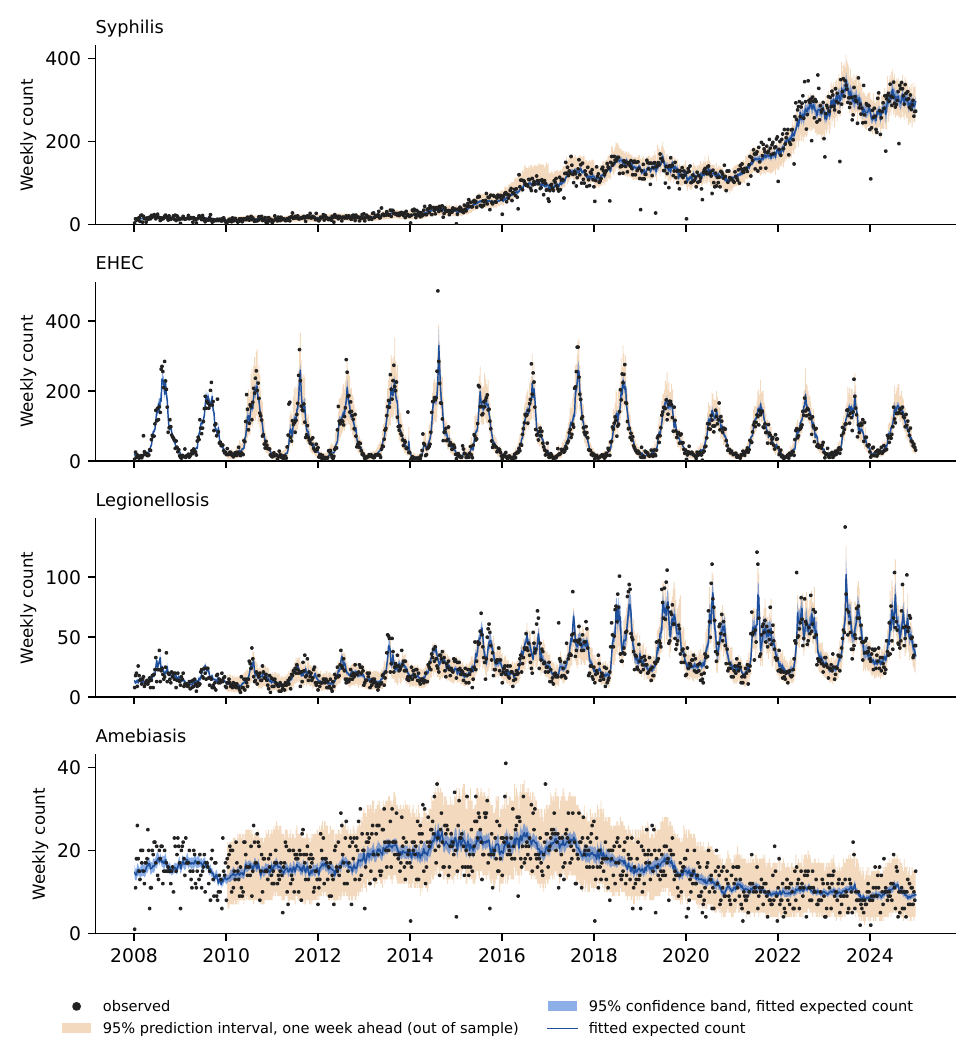}
\caption{Weekly national notifications (black) of the four panels. Dark band: the fitted expected national count $\sum_s\exp\{\hat f(X_{s,t})\}$ of \ours{} fitted to all weeks (line), with an exploratory 95\% moving-block bootstrap band. Light band: the out-of-sample one-week-ahead 95\% prediction interval for the national count, 2010--2024, from the yearly refits. The dark exploratory bands are normal intervals on the logarithmic scale with standard errors from 40 moving-block bootstrap refits (syphilis) or 30 (the other panels).}\label{fig:data}
\end{figure}

The panels differ in the ways that matter for a forecaster (Table~\ref{tab:data} and Figure~\ref{fig:data}). Syphilis notifications rose about seventeen-fold, from 875 in 2012 to 14{,}829 in 2024, after a decade between 500 and 900 a year; in 2022 to 2024 Tokyo accounts for 26\% of them, while 13 prefectures still record no case in more than a quarter of the weeks. EHEC infection has a stable level and a strong season, with a national weekly mean of 203 at its peak in week~34 against 10 at its trough, and outbreak clusters. Legionellosis rose from a few hundred notifications a year to about 2{,}400 and is sparse; amebiasis is sparser still, with 78\% of the prefecture-weeks in the evaluation period equal to zero, a mean of 0.34 notifications a week and a level that has declined since 2016. Syphilis is the panel on which the estimator was developed; EHEC and legionellosis were the test panels of the earlier protocols; amebiasis is the test panel of the final protocol (Section~\ref{sec:protocol}). It was selected by a rule fixed before it was examined: among the diseases of the same source that are reported in every week from 1999 to 2024, the most notified other than syphilis and legionellosis. EHEC infection does not qualify, because each of its two names is reported in only part of the period. Before the protocol was written we inspected only its total and its fraction of zero prefecture-weeks.

\begin{table}[t]
\caption{The four panels: total notifications in 1999--2024, notifications in three years, and, for the prefecture-weeks of the evaluation period 2012--2024, the fraction of zeros, the mean, the ratio of variance to mean pooled over prefectures and the maximum. ``Season'' is the ratio of the largest to the smallest national mean by week of the year over 1999--2024.}\label{tab:data}
\centering\small\setlength{\tabcolsep}{4pt}
\begin{tabular}{lrrrrrrrrr}
\toprule
 & & \multicolumn{3}{c}{Notifications per year} & \multicolumn{4}{c}{Prefecture-weeks, 2012--2024} & \\
\cmidrule(lr){3-5}\cmidrule(lr){6-9}
Panel & Total & 2000 & 2012 & 2024 & Zero & Mean & Var./mean & Max. & Season \\
\midrule
Syphilis & 95,997 & 759 & 875 & 14,829 & 0.45 & 2.75 & 7.9 & 99 & 2.4 \\
EHEC & 97,458 & 3,642 & 3,768 & 3,748 & 0.55 & 1.51 & 7.7 & 216 & 20.8 \\
Legionellosis & 29,101 & 154 & 899 & 2,428 & 0.59 & 0.74 & 1.6 & 40 & 3.4 \\
Amebiasis & 19,092 & 378 & 932 & 523 & 0.78 & 0.34 & 1.2 & 13 & 2.2 \\
\bottomrule
\end{tabular}

\end{table}

\subsection{Forecasting Protocol and Benchmarks}\label{sec:protocol}

Forecasts are issued every week for horizons of one to four weeks and are evaluated on the target weeks from January 2012 to 29 December 2024 for which every method has forecasts at every horizon. Every method is refitted once a year: the refit at the final observed week of year $y-1$ is used for the origins of year $y$, and training targets never exceed the forecast origin. These are rolling historical forecasts of finalized counts, not a prospective real-time evaluation.

\ours{} is the estimator of Section~\ref{sec:est}: the 13 summaries of Section~\ref{sec:covariates} together with the raw history of 26 weeks and a three-dimensional site encoding, a network with three hidden layers of width 64 and two outputs, the log-mean truncated to $[-7,7]$ and the log-size to $[\log0.5,\log200]$, the average of three fits with seeds 1, 2 and 3, and $B=1{,}000$ simulated paths with Rao--Blackwellized quantiles. It was developed on the syphilis panel, starting from the constant-size special case; the development, including direct multi-horizon heads that did not help, is reported in supplementary Sections~\ref{S-sec:s-dev}--\ref{S-sec:s-versions}. Each version was fixed in a protocol, written with SHA-256 digests of the code and data before any model was fitted to a new test panel; the final protocol fixed the estimator described here and selected amebiasis as its test panel. On syphilis the comparison is therefore post hoc, on EHEC and legionellosis the estimator had been tested in an earlier version, and amebiasis is the one held-out-panel test. A regeneration from scratch of all \ours{} forecasts, and of every benchmark of the primary comparison, reproduced every stored quantile exactly (supplementary Section~\ref{S-sec:s-final}).

We compare \ours{} with five benchmarks. The first is \emph{persistence}: the last observed count plus the empirical quantiles of the symmetrized $h$-week increments of the same prefecture over the previous three years, truncated at zero. The second is a negative binomial regression (\emph{NB regression}) with lagged spatial summaries, distinct from the additive endemic--epidemic model of \citet{held2005}, with prefecture-specific intercepts and a log-mean that is linear in the logarithms of one plus the last count, the 4-week and 26-week means, the 4-week mean of the five nearest prefectures and the national 4-week mean, with annual harmonics, fitted by maximum likelihood separately for each horizon to the most recent three years. The third, on syphilis only because of its computational cost, is a Bayesian zero-inflated negative binomial regression with Gaussian process random effects \citep[\emph{ZINB-GP};][]{he2024zinb}, fitted by the P\'olya--Gamma Gibbs sampler of the R package \texttt{ZINB.GP} \citep{scott2026zinbgp}, which we reimplemented in Python and compared with the package (Section~\ref{S-sec:s-zinb}); some parameters of both samplers mix slowly, so this benchmark carries Monte Carlo uncertainty that we do not quantify. The fourth is a compact implementation of the Temporal Fusion Transformer \citep[\emph{TFT};][]{lim2021} with variable selection, a recurrent encoder over the last 26 weeks, self-attention and a four-week decoder (105{,}137 parameters), trained with the quantile loss \citep{koenker1978} on $\log(1+y)$ for 20 epochs and 5 per refit; it forecasts all four horizons directly. The fifth is DeepAR \citep{salinas2020} with a negative binomial likelihood: a two-layer recurrent network with 40 cells, dropout 0.1 and a prefecture embedding, shared by all prefectures, conditioned on the last 52 weeks of the prefecture and trained by maximum likelihood with the scaling of the original paper (22{,}058 parameters, 20 epochs and 5 per refit); it does not see the neighbors. The neural benchmarks are run with seeds 1, 2 and 3, and their ensembles average the count-scale quantiles of the three seeds, so that every neural method combines three fits. The TFT and DeepAR implementations are compact custom benchmarks, run with default settings, and their performance should not be generalized to every implementation of these architectures. Because counts are integers, all quantile forecasts are rounded to the nearest integer.

Forecasts are scored by the weighted interval score \citep{bracher2021} in its quantile form,
\[
\mathrm{WIS}(q,y)=\frac{1}{n_\tau}\sum_{k=1}^{n_\tau}2\{\mathbf 1(y\le q_k)-\tau_k\}(q_k-y),
\]
where $y$ is the observed count and $q=(q_1,\dots,q_{n_\tau})$ are the predictive quantiles at the $n_\tau=9$ levels $\tau_k\in\{0.025,0.05,0.1,0.25,0.5,0.75,0.9,0.95,0.975\}$. The score is the average of the quantile losses, is negatively oriented, is a proper scoring rule and approximates the continuous ranked probability score \citep{gneiting2007}. We also report the mean absolute error (MAE) of the predictive median and the coverage of the central prediction intervals $[q_{\alpha/2},q_{1-\alpha/2}]$ of nominal level $1-\alpha$. Differences in WIS between two methods are assessed by a moving-block bootstrap \citep{kunsch1989} of the weekly mean differences, with 4{,}000 resamples and blocks of 8 weeks; the intervals for blocks of 4, 13 and 26 weeks are similar (supplementary Section~\ref{S-sec:s-final}). The primary comparison of the final protocol is the paired difference between the TFT ensemble and \ours{} on amebiasis.

\subsection{Forecast Accuracy}\label{sec:accuracy}

\begin{table}[t]
\caption{\ours{} and the benchmarks on the four panels, 677 common target weeks of 2012 to 2024 in each: WIS over horizons 1--4 of \ours{} and, for each competitor, its WIS and the paired weekly difference, competitor minus \ours, with a moving-block bootstrap 95\% interval (blocks of 8 weeks, 4{,}000 resamples). Positive differences favor \ours. Amebiasis is the test panel of the final protocol; the TFT ensemble of six seeds was run on legionellosis only and ZINB-GP on syphilis only.}\label{tab:summary}
\centering\footnotesize\setlength{\tabcolsep}{3pt}
\begin{tabular}{@{}lrrrr@{}}
\toprule
 & Syphilis & EHEC & Legionellosis & Amebiasis \\
\midrule
\ours{} (WIS) & 0.5032 & 0.5936 & 0.2858 & 0.1476 \\
\midrule
TFT (3 seeds) & 0.5073 & 0.5996 & 0.2877 & 0.1490 \\
\quad difference & $+0.0040$ & $+0.0061$ & $+0.0018$ & $+0.0014$ \\
\quad 95\% interval & $(-0.0032, +0.0106)$ & $(-0.0008, +0.0126)$ & $(-0.0007, +0.0043)$ & $(+0.0006, +0.0021)$ \\[3pt]
TFT (6 seeds) & --- & --- & 0.2863 & --- \\
\quad difference &  &  & $+0.0005$ &  \\
\quad 95\% interval &  &  & $(-0.0020, +0.0028)$ &  \\[3pt]
NB regression & 0.5291 & 0.6034 & 0.2879 & 0.1491 \\
\quad difference & $+0.0258$ & $+0.0099$ & $+0.0020$ & $+0.0015$ \\
\quad 95\% interval & $(+0.0168, +0.0353)$ & $(+0.0040, +0.0166)$ & $(-0.0008, +0.0047)$ & $(+0.0004, +0.0028)$ \\[3pt]
ZINB-GP & 0.5246 & --- & --- & --- \\
\quad difference & $+0.0213$ &  &  &  \\
\quad 95\% interval & $(+0.0125, +0.0311)$ &  &  &  \\[3pt]
DeepAR (3 seeds) & 0.5565 & 0.6237 & 0.3150 & 0.1615 \\
\quad difference & $+0.0533$ & $+0.0301$ & $+0.0291$ & $+0.0139$ \\
\quad 95\% interval & $(+0.0420, +0.0646)$ & $(+0.0233, +0.0374)$ & $(+0.0240, +0.0347)$ & $(+0.0124, +0.0156)$ \\[3pt]
Persistence & 0.6900 & 0.8885 & 0.3975 & 0.2128 \\
\quad difference & $+0.1868$ & $+0.2950$ & $+0.1117$ & $+0.0652$ \\
\quad 95\% interval & $(+0.1671, +0.2081)$ & $(+0.2440, +0.3557)$ & $(+0.1018, +0.1223)$ & $(+0.0614, +0.0690)$ \\[3pt]
\bottomrule
\end{tabular}

\end{table}

\begin{table}[t]
\caption{Where the TFT ensemble and \ours{} differ: paired weekly WIS difference, TFT ensemble minus \ours, by horizon, by third of the prefectures ordered by their mean weekly count in 2012--2024, and by period. Positive values favor \ours; an asterisk marks a moving-block bootstrap 95\% interval (blocks of 8 weeks) that excludes zero.}\label{tab:vstft}
\centering\small
\begin{tabular}{@{}lrrrr@{}}
\toprule
 & Syphilis & EHEC & Legionellosis & Amebiasis \\
\midrule
Horizon 1 & $+0.0054$ & $+0.0081^{*}$ & $+0.0033^{*}$ & $+0.0016^{*}$ \\
Horizon 2 & $+0.0040$ & $+0.0068$ & $+0.0012$ & $+0.0014^{*}$ \\
Horizon 3 & $+0.0033$ & $+0.0058$ & $+0.0011$ & $+0.0013^{*}$ \\
Horizon 4 & $+0.0035$ & $+0.0035$ & $+0.0018$ & $+0.0011^{*}$ \\
\midrule
Small prefectures & $+0.0048^{*}$ & $-0.0008$ & $+0.0011$ & $+0.0009^{*}$ \\
Medium prefectures & $+0.0056^{*}$ & $+0.0075^{*}$ & $-0.0005$ & $+0.0007$ \\
Large prefectures & $+0.0018$ & $+0.0116$ & $+0.0047$ & $+0.0025^{*}$ \\
\midrule
2012--2015 & $+0.0054^{*}$ & $-0.0106^{*}$ & $+0.0028^{*}$ & $+0.0023^{*}$ \\
2016--2019 & $+0.0007$ & $+0.0062$ & $+0.0013$ & $+0.0015^{*}$ \\
2020--2021 & $+0.0028$ & $+0.0256^{*}$ & $+0.0015$ & $+0.0000$ \\
2022--2024 & $+0.0075$ & $+0.0151^{*}$ & $+0.0015$ & $+0.0009$ \\
\bottomrule
\end{tabular}

\end{table}

Table~\ref{tab:summary} gives the main comparison; scores by horizon, absolute errors and coverages for each panel are in supplementary Tables~\ref{S-tab:s-fin-scores-syphilis}--\ref{S-tab:s-fin-scores-amebiasis}. In the primary comparison of the final protocol, on amebiasis, \ours{} has a WIS of 0.1476 and the TFT ensemble 0.1490; the difference of 0.0014, or 0.9\% of the score of the TFT ensemble, favors \ours, and its interval, $(0.0006,0.0021)$, excludes zero, as it does for blocks of 4, 13 and 26 weeks. On amebiasis \ours{} also has the lowest score at every horizon, and it is better than the NB regression by 1.0\% (interval $(0.0004,0.0028)$), than DeepAR by 8.6\% and than persistence by 31\%. On the other three panels its WIS is also below that of the TFT ensemble: by 0.8\% on syphilis (0.5032 against 0.5073), by 1.0\% on EHEC (0.5936 against 0.5996), and by 0.6\% on legionellosis (0.2858 against 0.2877), but on these panels the intervals for blocks of 8 weeks include zero. On legionellosis a TFT ensemble of six seeds is as good as \ours{} (0.2863), and the constant-size special case of \ours{} that we evaluated first scores slightly better, 0.2841, a difference that is not distinguishable from zero either (supplementary Section~\ref{S-sec:s-versions}). \ours{} is clearly better than DeepAR, by 4.8\% to 9.6\%, and than persistence on every panel, better than the NB regression on syphilis, EHEC and amebiasis, and better than ZINB-GP on syphilis; on legionellosis the NB regression is within 0.7\% of it. Its median has the lowest absolute error on three panels and is within 0.1\% of the lowest on syphilis, and over all horizons its central intervals cover at least their nominal level on every panel.

Table~\ref{tab:vstft} shows where the two best methods differ. The advantage of \ours{} is largest one week ahead, where its predictive law is exact, and it is significant one week ahead on three panels and at every horizon on amebiasis; it persists at four weeks, where the Rao--Blackwellized simulation of Algorithm~\ref{alg:forecast} is used. The transformer, which forecasts each horizon directly and is fitted to the quantiles, is therefore not more accurate at long horizons than recursive simulation from the likelihood model. By size of prefecture and by period the picture is the same, except that the transformer ensemble is ahead of \ours{} in three of the 44 cells of the table: in the small prefectures of EHEC and the medium prefectures of legionellosis, by differences below 0.001 that are not distinguishable from zero, and on EHEC in 2012--2015, by 0.011 with an interval that excludes zero. In that period, before the disruptions of the COVID-19 pandemic, the transformer forecasts the seasonal EHEC series better; in 2020--2021 and 2022--2024 \ours{} is better, by 0.026 and 0.015. \ours{} is ahead with an interval that excludes zero in 17 cells. These subgroup comparisons are exploratory and unadjusted for multiplicity.

\subsection{Exploratory Inference for the Fitted Model}\label{sec:inf-app}

The fitted likelihood describes associations in the conditional mean and size functions, but the rate theorems do not establish the validity of bootstrap confidence intervals; we therefore present the intervals as exploratory uncertainty summaries, alongside the predictive evaluation. We fit the estimator to all weeks of each panel (40 epochs, three fits) and compute the functionals of Section~\ref{sec:inference}. Their standard errors are the standard deviations, on the logarithmic scale, of the functionals over 40 moving-block bootstrap refits with blocks of 26 weeks for syphilis and 30 refits for the other panels, and the 95\% intervals are the estimates multiplied and divided by $\exp(1.96\,\mathrm{se})$. Simulation~C (Section~\ref{sec:simC}) provides a limited check in a smaller model and shows undercoverage where the estimator is biased. The intervals below are exploratory pointwise uncertainty summaries, not simultaneous confidence bands or established frequentist coverage guarantees. We report the syphilis panel in detail; Table~\ref{tab:ape} gives the average partial effects for all four panels.

\begin{table}[t]
\caption{Average partial effects of the fitted \ours, with 95\% intervals from moving-block bootstrap standard errors (40 refits for syphilis and 30 for the other panels, blocks of 26 weeks). The first three rows are the factor by which the mean count changes, averaged over all prefecture-weeks on the log scale, when the past counts of the prefecture, of its five nearest neighbors, or of the whole country are doubled; the last row is the ratio of the average fitted size in the tenth of prefecture-weeks with the largest fitted mean to that in the tenth with the smallest.}\label{tab:ape}
\centering\footnotesize\setlength{\tabcolsep}{3pt}
\begin{tabular}{lllll}
\toprule
Effect & Syphilis & EHEC & Legionellosis & Amebiasis \\
\midrule
Own history doubled & 1.359 (1.335, 1.384) & 1.298 (1.265, 1.333) & 1.207 (1.176, 1.239) & 1.070 (1.043, 1.098) \\
Neighbors doubled & 1.110 (1.080, 1.140) & 1.149 (1.103, 1.197) & 1.152 (1.085, 1.223) & 1.156 (1.095, 1.220) \\
National level doubled & 1.156 (1.120, 1.194) & 0.999 (0.888, 1.124) & 1.394 (1.284, 1.513) & 1.979 (1.810, 2.164) \\
Size ratio, top to bottom tenth & 19.74 (9.22, 42.26) & 3.25 (1.32, 8.02) & 10.26 (1.68, 62.61) & 5.80 (2.20, 15.30) \\
\bottomrule
\end{tabular}

\end{table}

\begin{figure}[t]
\centering
\includegraphics[width=0.9\textwidth]{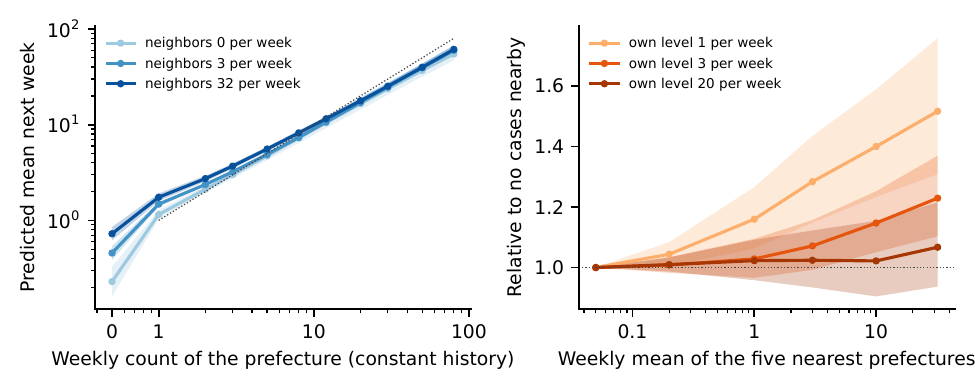}
\caption{Syphilis: the fitted conditional mean of \ours{} with exploratory 95\% bootstrap bands. Left: predicted mean for next week against the weekly count of the prefecture, for three levels of its five nearest neighbors. Right: predicted mean relative to its value when the neighbors have no cases, against the level of the neighbors, for three levels of the prefecture itself. The history of the prefecture is constant at the stated weekly count, the national level is held at its 2022--2024 mean, and the predicted mean is averaged over the weeks of the year and the site encodings.}\label{fig:surface}
\end{figure}

\begin{figure}[t]
\centering
\includegraphics[width=0.9\textwidth]{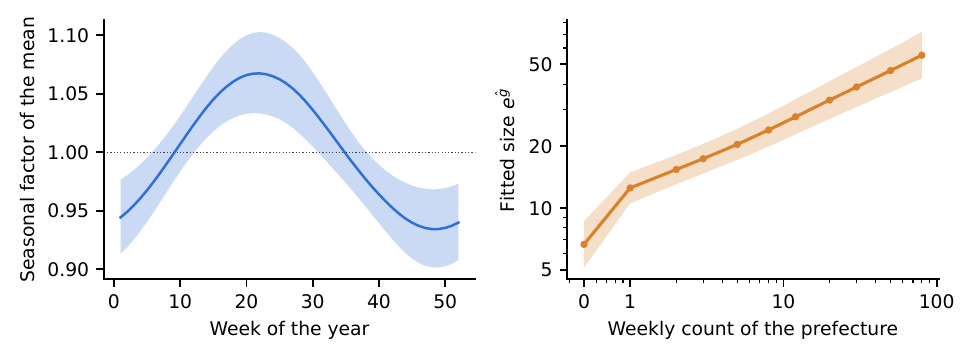}
\caption{Syphilis: seasonal factor of the mean (left), at a constant history of 3 cases a week in the prefecture and its neighbors, and fitted size $e^{\hat g}$ against the weekly count of the prefecture, with neighbors at 3 cases a week (right), with exploratory 95\% bootstrap bands.}\label{fig:effects}
\end{figure}

The fitted expected national count follows the observed series closely (Figure~\ref{fig:data}); its exploratory bootstrap band has a median half-width of 5.3\% of the fitted value on syphilis. Table~\ref{tab:ape} quantifies the three sources of information in the log-mean. On syphilis, doubling the past counts of a prefecture raises its expected count by 36\% (95\% interval 33\% to 38\%), a strongly sublinear response, doubling the counts of the five nearest prefectures raises it by 11\% (8\% to 14\%), and doubling the national level by 16\% (12\% to 19\%); none of the three intervals includes one. The neighbor effect is not constant (Figure~\ref{fig:surface}): as the neighbors go from no cases to 32 a week, the predicted mean of a prefecture with one case a week rises by a factor of 1.52 (1.31 to 1.76), of a prefecture with three cases by 1.23 (1.10 to 1.37), and of a prefecture with 20 cases by 1.07 (0.94 to 1.22), an interval that includes one. The fitted neighbor association is strongest at low local levels; this interaction cannot be expressed by a log-linear model with a single neighbor coefficient and does not identify transmission between prefectures. The seasonal factor of the mean varies only between 0.93 and 1.07 (Figure~\ref{fig:effects}, left). The size is far from constant: at a constant history the fitted size increases from 6.6 (5.1 to 8.6) for a prefecture without cases to 55 (43 to 72) at 80 cases a week (Figure~\ref{fig:effects}), and the ratio of the average fitted size in the tenth of prefecture-weeks with the largest mean to that in the tenth with the smallest is 19.7 (9.2 to 42.3). A larger fitted size means a smaller coefficient $1/r$ of the quadratic variance term, not necessarily a smaller variance-to-mean ratio $1+\mu/r$. Across the panels (Table~\ref{tab:ape}), the response to a prefecture's own history weakens as the counts become sparser, from a factor of 1.36 for doubling on syphilis to 1.30 on EHEC, 1.21 on legionellosis and 1.07 on amebiasis, whereas the response to the national level strengthens, to 1.39 on legionellosis and 1.98 (1.81 to 2.16) on amebiasis: where a prefecture has few cases, the country carries most of the information about its next week. The neighbor effect is between 1.11 and 1.16 on all four panels, and on every panel the fitted size is larger where the mean is larger, with intervals for the ratio that exclude one.

These intervals are conditional on the model, pointwise, unadjusted for the many contrasts, and not causal; they include the variability between optimizer runs through the refits.

\subsection{Calibration}\label{sec:calib}

\begin{figure}[t]
\centering
\includegraphics[width=0.9\textwidth]{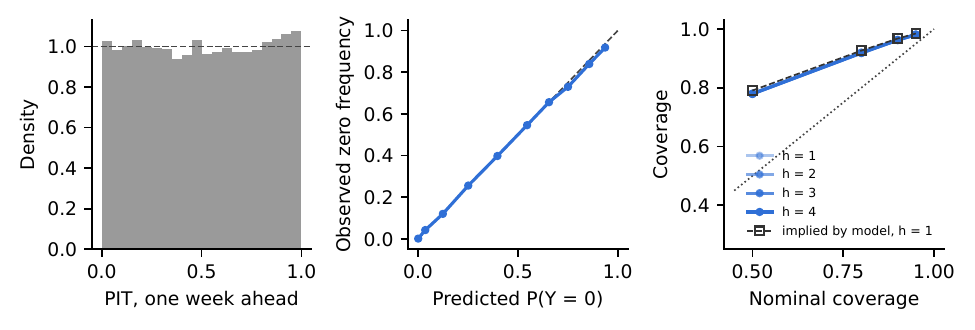}
\caption{Syphilis: calibration of \ours, target weeks of 2012 to 2024. Left: randomized probability integral transform of the one-week-ahead predictive distributions. Center: observed frequency of zero counts against the predicted probability of zero, in tenths of the predicted probability. Right: empirical coverage of central prediction intervals at horizons of one to four weeks, and the coverage implied by the fitted model one week ahead, against the nominal level.}\label{fig:calibration}
\end{figure}

Figure~\ref{fig:calibration} examines the predictive distributions on syphilis. The randomized probability integral transform \citep{czado2009} is uniformly distributed when the predictive distribution is correct. One week ahead it is close to uniform: its empirical distribution function differs from the uniform one by at most 0.011, a descriptive statistic, since the forecasts are dependent. The predicted probability of a zero count averages 0.456 against an observed frequency of 0.451, and the agreement holds across the range of predicted probabilities. Central prediction intervals cover more than their nominal level at every horizon (Figure~\ref{fig:calibration}, right), largely because an interval of integers that contains at least the nominal probability usually contains more. The comparison addressed by Theorem~\ref{thm:pred} is between the probability that the fitted model assigns to its own interval and the realized coverage. One week ahead these are 0.791 and 0.781 for the nominal 50\% interval, 0.927 and 0.920 for 80\%, 0.968 and 0.964 for 90\%, and 0.985 and 0.981 for 95\%; the moving-block bootstrap intervals for the differences (blocks of 8 weeks) exclude zero. Theorem~\ref{thm:unknown}(c) bounds the absolute gap under correct specification; it neither predicts its sign nor attributes these gaps to estimation error, and no numerical bound is evaluated here. On the other panels the realized coverage is at most 1.4 points below and 2.7 points above the model-implied one (supplementary Table~\ref{S-tab:s-fin-calib}). These pooled one-step comparisons do not establish conditional calibration at each prefecture and time.

Calibration of the prefecture counts does not imply calibration of their sum, and here the model falls short. Out of sample, the one-week-ahead 95\% prediction interval for the national count (Figure~\ref{fig:data}), computed exactly from the joint predictive law under conditional independence across prefectures, covers only 83\% of the target weeks of 2012--2024 on syphilis, 86\% on EHEC, 81\% on legionellosis and 92\% on amebiasis. Two sources can be separated (supplementary Table~\ref{S-tab:s-national}). The first is reporting holidays: on syphilis the interval covers 41\% of the 63 weeks that contain New Year, Golden Week or Obon days, when reporting falls in all prefectures at once, and 87\% of the other weeks. The second is contemporaneous dependence between prefectures. The standardized residuals of different prefectures have an average pairwise correlation of only 0.01 to 0.03, but over 47 sites this raises the variance of the standardized national residual to between 1.3 and 3.2 outside holiday weeks, where the model implies one. Shocks common to all sites are outside the conditional model \eqref{eq:modelgen}. By Theorem~\ref{thm:dep} they leave the estimation bounds and the one-step comparison of fitted and true count probabilities intact, but not the guarantees for joint or multi-step predictive laws, so the national intervals should not be used as they stand. Inflating their variance by the factor estimated from earlier forecast weeks gives a coverage of 92\% to 96\%. Forecast paths for four prefectures are in supplementary Figure~\ref{S-fig:s-forecasts}.

\FloatBarrier

\section{Conclusion}\label{sec:disc}

\ours{} models each site's count given the past by a negative binomial law whose log-mean and log-size are functions of temporal and nearest-neighbor summaries, estimated by maximum likelihood over deep ReLU networks. The oracle inequality holds along the observed trajectory without mixing; mixing implied by the bounded model permits an independent-copy risk bound. A lower bound applies to an exogenous-covariate subexperiment. The Kullback--Leibler chain rule controls multi-step prediction error with a square-root dependence on the horizon, and the results extend to an unknown covariate-dependent size. Empirically, \ours{} had a lower weighted interval score than a three-member Temporal Fusion Transformer ensemble on four panels; the primary difference was distinguishable from zero only on the held-out amebiasis panel. The fitted functions describe associations with local, neighboring and national history, changes in dispersion and seasonality.

Several limitations should be kept in mind. First, the rate theorems concern sparse networks with weights bounded by one and an approximate maximizer of the likelihood, whereas the case study trains fully connected networks with unconstrained weights by a stochastic gradient method; in Simulation~A a fit that respects the constraints behaves much like the practical fit, but the optimization error $\Delta_n$ is not certified, and rates for fully connected networks without a sparsity constraint \citep{kohler2021} motivate a separate analysis of the implemented class. Second, the lower bound is attained by laws with exogenous covariates; for a covariate map of past counts alone the design law is determined by $f_0$, and the minimax rate is open. Third, the constant $\epsilon_0$ of Proposition~\ref{prop:mixing} is extremely small for 47 sites, so the proposition is qualitative and Theorem~\ref{thm:oracle}, which needs no mixing, carries the quantitative content; likewise the factor $S^{1/2}$ in Theorem~\ref{thm:pred}(b) is conservative for a single site. Fourth, the theory assumes a correctly specified negative binomial law; under misspecification the likelihood targets a distribution-dependent approximation, and the calibration diagnostics are descriptive. Fifth, the bootstrap intervals are not justified by a theorem; in Simulation~C, intervals centered at the estimate covered as little as 0.79 where the estimator is biased. Sixth, the model treats sites as independent given the past, so it cannot represent shocks common to all sites, such as reporting holidays or outbreaks that span several prefectures. Theorem~\ref{thm:dep} shows that the estimator survives such dependence, but the joint predictive law does not: the national one-week-ahead 95\% intervals covered only 81\% to 92\% of the weeks, because of holiday weeks and of a weak but pervasive correlation between prefectures (Section~\ref{sec:calib}); holiday indicators or a latent common factor are natural extensions. Finally, the empirical comparison is less conclusive than the theory. \ours{} was developed on syphilis after its first version had been behind the transformer ensemble, earlier versions had been tested on EHEC and legionellosis, and only on the held-out amebiasis panel is its advantage, about 1\%, distinguishable from zero; in some strata the transformer is ahead (Table~\ref{tab:vstft}). The scores are competitive with a transformer benchmark, but neither general superiority nor noninferiority has been established, the evidence comes from four panels of one surveillance system and, with fewer benchmarks, one transport panel, and the neural benchmarks are compact implementations with default hyperparameters.

Natural extensions are zero inflation or a hurdle in the likelihood \citep{he2024zinb}, a factor shared by the sites within a period, further covariates such as testing volume or mobility, and, for operational use with provisional counts, a nowcasting step.

\long\def\acks#1{\vskip 0.3in\noindent{\large\bf Acknowledgments and Disclosure of Funding}\nopagebreak\vskip 0.2in\nopagebreak
\noindent #1}
\acks{H.-H. Huang was supported by National Science Foundation grants DMS-1924792 and DMS-2318925 and by the National Institutes of Health, National Institute of Allergy and Infectious Diseases, award R01AI206603. The weekly notifications of the four diseases are taken from the R package \texttt{jpinfect} \citep{hoshi2025}, which redistributes the open data of the Japan Institute for Health Security. \ifx\repourl\empty A replication archive with the code, data, cached forecasts and evaluation protocols accompanies the paper.\else A demonstration notebook and the replication archive, with the code, data, cached forecasts and evaluation protocols, are available at \url{\repourl}.\fi{} The author has no competing interests to declare.}

\appendix
\section{Preliminaries on the Negative Binomial Family}\label{app:prelim}

Throughout, $r>0$ is fixed, $\mu=e^{f}$, and the $\mathrm{NB}(\mu,r)$ probability mass function is written in exponential-family form
\begin{equation}\label{eq:expfam}
\begin{gathered}
p_f(y)=c_y\exp\{y\,\theta(f)-A(\theta(f))\},\qquad c_y=\frac{\Gamma(y+r)}{\Gamma(r)\,y!},\\
\theta(f)=f-\log(e^{f}+r),\qquad A(\theta)=-r\log(1-e^{\theta}),
\end{gathered}
\end{equation}
for $y=0,1,2,\dots$, where $\theta(f)<0$ is the natural parameter and $A$ the cumulant function; this is the mass function of Section~\ref{sec:intro} with $\mu=e^f$. Direct differentiation gives
\begin{equation}\label{eq:Aderiv}
A'(\theta)=\frac{re^{\theta}}{1-e^{\theta}}=\mu,\quad A''(\theta)=\frac{re^{\theta}}{(1-e^{\theta})^{2}}=\mu+\frac{\mu^{2}}{r},\quad \theta'(f)=\frac{r}{e^{f}+r}\in(0,1],
\end{equation}
and both $A'$ and $A''$ are increasing on $(-\infty,0)$. For $u<-\theta$ the moment generating function is
\begin{equation}\label{eq:mgf}
\E_\theta e^{uY}=\sum_y c_y e^{y(\theta+u)-A(\theta)}=\exp\{A(\theta+u)-A(\theta)\}.
\end{equation}
For $|f|\le F$ write $\theta_{\rm lo}=\theta(-F)$, $\theta_{\rm hi}=\theta(F)<0$, $V_{\min}=A''(\theta_{\rm lo})=e^{-F}+e^{-2F}/r$ and $V_{\max}=A''(\theta_{\rm hi})=e^{F}+e^{2F}/r$. By \eqref{eq:Aderiv} and the mean value theorem, for $|f_1|,|f_2|\le F$,
\begin{equation}\label{eq:thetaLip}
\frac{r}{e^{F}+r}\,|f_1-f_2|\le|\theta(f_1)-\theta(f_2)|\le|f_1-f_2|.
\end{equation}
Here $\E_\theta$ is the expectation when $Y$ has the mass function \eqref{eq:expfam} with natural parameter $\theta$. The squared Hellinger distance of Section~\ref{sec:intro} is also $H^2(p,q)=\tfrac12\sum_y(\sqrt{p(y)}-\sqrt{q(y)})^2$, and $H$ is a metric. For probability measures $P$ and $Q$ on a general space, with densities $p$ and $q$ with respect to a dominating measure $\lambda$, $H^2(P,Q)=1-\int\sqrt{pq}\,d\lambda$ and $\mathrm{TV}(P,Q)=\sup_B|P(B)-Q(B)|=\tfrac12\int|p-q|\,d\lambda$, and $P\otimes Q$ denotes a product measure. $\mathbb N_0=\{0,1,2,\dots\}$.

The constants used below depend only on $(F,r)$:
\begin{equation}\label{eq:constants}
\begin{gathered}
c_2=\frac{V_{\max}}{8},\qquad c_1=\frac{V_{\min}}{8\,(1+V_{\max}F^{2}/2)}\Big(\frac{r}{e^{F}+r}\Big)^{2},\qquad
\omega=\Big(\frac{e^{F}}{e^{F}+r}\Big)^{1/2}=e^{\theta_{\rm hi}/2},\\
\bar c=\frac{r\omega}{(1-\omega)^{2}}=A''(\theta_{\rm hi}/2),\qquad \bar\mu=\frac{r\omega}{1-\omega}=A'(\theta_{\rm hi}/2),\\
\lambda_0=\min\Big\{1,\frac{|\theta_{\rm hi}|}{2(\theta_{\rm hi}-\theta_{\rm lo})}\Big\},\qquad \lambda_1=\frac{|\theta_{\rm hi}|}{2}.
\end{gathered}
\end{equation}

\begin{lemma}[Hellinger and Kullback--Leibler versus the log-mean]\label{lem:hell}
Let $|f_1|,|f_2|\le F$. Then
\[
c_1(f_1-f_2)^2\;\le\;H^2(p_{f_1},p_{f_2})\;\le\;c_2(f_1-f_2)^2,\qquad
\mathrm{KL}(p_{f_1}\,\|\,p_{f_2})\le\frac{V_{\max}}{2}(f_1-f_2)^2 .
\]
\end{lemma}
\begin{proof}
Write $\theta_i=\theta(f_i)$ and $\bar\theta=(\theta_1+\theta_2)/2<0$. By \eqref{eq:expfam}, $\sum_y\sqrt{p_{f_1}(y)p_{f_2}(y)}=\sum_yc_y\exp\{y\bar\theta-(A(\theta_1)+A(\theta_2))/2\}=e^{-J}$ with $J=\{A(\theta_1)+A(\theta_2)\}/2-A(\bar\theta)$, because $\sum_yc_ye^{y\bar\theta}=e^{A(\bar\theta)}$. Put $\vartheta=(\theta_2-\theta_1)/2$. Taylor's theorem at $\bar\theta$ gives $A(\bar\theta\pm\vartheta)=A(\bar\theta)\pm\vartheta A'(\bar\theta)+\tfrac12\vartheta^2A''(\tilde\theta_\pm)$ for some $\tilde\theta_\pm$ between $\theta_1$ and $\theta_2$, so $J=\tfrac14\vartheta^2\{A''(\tilde\theta_+)+A''(\tilde\theta_-)\}$ and, since $A''$ is increasing and $\theta_{\rm lo}\le\theta_i\le\theta_{\rm hi}$,
\[
\frac{V_{\min}}{8}(\theta_1-\theta_2)^2\le J\le\frac{V_{\max}}{8}(\theta_1-\theta_2)^2 .
\]
Now $H^2=1-e^{-J}\le J\le c_2(f_1-f_2)^2$ by \eqref{eq:thetaLip}. For the lower bound, $e^{J}\ge1+J$ gives $1-e^{-J}\ge J/(1+J)$, and $J\le V_{\max}(2F)^2/8=V_{\max}F^2/2$ because $|\theta_1-\theta_2|\le|f_1-f_2|\le2F$; hence $H^2\ge J/(1+V_{\max}F^2/2)\ge c_1(f_1-f_2)^2$ by \eqref{eq:thetaLip}. Finally, for an exponential family $\mathrm{KL}(p_{f_1}\|p_{f_2})=A(\theta_2)-A(\theta_1)-(\theta_2-\theta_1)A'(\theta_1)=\tfrac12(\theta_2-\theta_1)^2A''(\tilde\theta)\le\tfrac12V_{\max}(f_1-f_2)^2$ for some $\tilde\theta$ between $\theta_1$ and $\theta_2$, by Taylor's theorem, the monotonicity of $A''$ and \eqref{eq:thetaLip}.
\end{proof}

\begin{lemma}[Lipschitz log-likelihood]\label{lem:lip}
For all $f_1,f_2\in\mathbb R$ and $y\in\{0,1,\dots\}$, $|\log p_{f_1}(y)-\log p_{f_2}(y)|\le(y+r)\,|f_1-f_2|$.
\end{lemma}
\begin{proof}
By \eqref{eq:expfam}--\eqref{eq:Aderiv}, $\partial_f\log p_f(y)=\{y-A'(\theta(f))\}\theta'(f)=(y-\mu)r/(\mu+r)$, which lies between $-\mu r/(\mu+r)\ge-r$ and $yr/(\mu+r)\le y$. Hence $|\partial_f\log p_f(y)|\le y+r$, and the claim follows from the mean value theorem.
\end{proof}

\begin{lemma}[Exponential supermartingale bound]\label{lem:mart}
Let $(\xi_i)_{i=1}^n$ be adapted to a filtration $(\mathcal H_i)$ with $\E[e^{\xi_i}\mid\mathcal H_{i-1}]\le1$ almost surely for every $i$. Then $P(\sum_{i=1}^n\xi_i\ge\log(1/\delta))\le\delta$ for every $\delta\in(0,1)$.
\end{lemma}
\begin{proof}
$M_k=\exp(\sum_{i\le k}\xi_i)$ satisfies $\E[M_k\mid\mathcal H_{k-1}]=M_{k-1}\E[e^{\xi_k}\mid\mathcal H_{k-1}]\le M_{k-1}$, so $\E M_n\le1$ and Markov's inequality gives $P(M_n\ge1/\delta)\le\delta$.
\end{proof}

\section{Proof of Theorem~\ref{thm:oracle}}\label{app:oracle}

Order the $n=ST$ observations lexicographically, $i=(t-1)S+s$. Set $\mathcal H_0=\mathcal I_0$ and
\[
\mathcal H_i=\sigma(\mathcal I_{t-1},Y_{1,t},\dots,Y_{s,t})\quad(s<S),
\qquad \mathcal H_{tS}=\mathcal I_t.
\] Thus any newly revealed predictable covariates are included before the next response is observed. Write $(X_i,Y_i)$ for the $i$th pair. By Assumption~\ref{ass:model}, $X_i$ is $\mathcal H_{i-1}$-measurable and, because the sites are conditionally independent given $\mathcal I_{t-1}$, the conditional law of $Y_i$ given $\mathcal H_{i-1}$ is $p_{f_0(X_i)}$. Put $\ell_n(f)=\sum_i\log p_{f(X_i)}(Y_i)$ and $H_i^2(f)=H^2(p_{f(X_i)},p_{f_0(X_i)})$.

\smallskip\noindent\textit{Step 1 (one fixed function).} Fix a measurable $f$ with $\|f\|_\infty\le F$, write $\Lambda_i=p_{f(X_i)}(Y_i)/p_{f_0(X_i)}(Y_i)$ and let $\xi_i=\tfrac12\log\Lambda_i-\log\{1-H_i^2(f)\}$. Since
\[
\E\big[\Lambda_i^{1/2}\mid\mathcal H_{i-1}\big]=\sum_y\{p_{f(X_i)}(y)\,p_{f_0(X_i)}(y)\}^{1/2}=1-H_i^2(f),
\]
we have $\E[e^{\xi_i}\mid\mathcal H_{i-1}]=1$, and Lemma~\ref{lem:mart} with $-\log(1-x)\ge x$ yields, with probability at least $1-\delta$,
\begin{equation}\label{eq:step1}
\sum_{i=1}^nH_i^2(f)\le\tfrac12\{\ell_n(f_0)-\ell_n(f)\}+\log(1/\delta).
\end{equation}

\smallskip\noindent\textit{Step 2 (a net).} Let $f_1,\dots,f_N\in\mathcal F_n$, $N=N_n(\varepsilon)$, be such that every $f\in\mathcal F_n$ has some $j$ with $\|f-f_j\|_\infty\le\varepsilon$. By \eqref{eq:step1} and a union bound there is an event $\Omega_1$ of probability at least $1-\delta$ on which
\begin{equation}\label{eq:step2}
\sum_{i}H_i^2(f_j)\le\tfrac12\{\ell_n(f_0)-\ell_n(f_j)\}+\log(N/\delta)\qquad\text{for all }j=1,\dots,N.
\end{equation}

\smallskip\noindent\textit{Step 3 (approximation term).} Let $f^*\in\mathcal F_n$ be non-random with $a=\|f^*-f_0\|_\infty$, and let
\[
D_i=\log\frac{p_{f_0(X_i)}(Y_i)}{p_{f^*(X_i)}(Y_i)}=Y_i\zeta_i-\{A(\theta_{0i})-A(\theta^*_i)\},
\]
where $\theta_{0i}=\theta(f_0(X_i))$, $\theta^*_i=\theta(f^*(X_i))$ and $\zeta_i=\theta_{0i}-\theta^*_i$, so that $|\zeta_i|\le\min\{a,\theta_{\rm hi}-\theta_{\rm lo}\}$ by \eqref{eq:thetaLip}. For $\lambda=\lambda_0$ in \eqref{eq:constants}, $\theta_{0i}+\lambda\zeta_i\le\theta_{\rm hi}+|\theta_{\rm hi}|/2=\theta_{\rm hi}/2<0$, so \eqref{eq:mgf} applies and
\[
\log\E[e^{\lambda D_i}\mid\mathcal H_{i-1}]=-\lambda\{A(\theta_{0i})-A(\theta_{0i}-\zeta_i)\}+A(\theta_{0i}+\lambda\zeta_i)-A(\theta_{0i}).
\]
By Taylor's theorem, $A(\theta_{0i})-A(\theta_{0i}-\zeta_i)=\zeta_iA'(\theta_{0i})-\tfrac12\zeta_i^2A''(\tilde\theta_{1i})$ and $A(\theta_{0i}+\lambda\zeta_i)-A(\theta_{0i})=\lambda\zeta_iA'(\theta_{0i})+\tfrac12\lambda^2\zeta_i^2A''(\tilde\theta_{2i})$ for intermediate points $\tilde\theta_{1i},\tilde\theta_{2i}\le\theta_{\rm hi}/2$. The first-order terms cancel and, as $A''$ is increasing and $\lambda\le1$,
\[
\log\E[e^{\lambda D_i}\mid\mathcal H_{i-1}]=\tfrac12\zeta_i^2\{\lambda A''(\tilde\theta_{1i})+\lambda^2A''(\tilde\theta_{2i})\}\le\lambda\,\bar c\,a^2 .
\]
Lemma~\ref{lem:mart} applied to $\xi_i=\lambda_0(D_i-\bar ca^2)$ gives an event $\Omega_2$ of probability at least $1-\delta$ on which
\begin{equation}\label{eq:step3}
\ell_n(f_0)-\ell_n(f^*)=\sum_iD_i\le n\bar c\,a^2+\lambda_0^{-1}\log(1/\delta).
\end{equation}

\smallskip\noindent\textit{Step 4 (sum of the counts).} Since $\theta_{0i}+\lambda_1\le\theta_{\rm hi}/2$, \eqref{eq:mgf} and the monotonicity of $A'$ give $\log\E[e^{\lambda_1Y_i}\mid\mathcal H_{i-1}]=A(\theta_{0i}+\lambda_1)-A(\theta_{0i})\le\lambda_1A'(\theta_{\rm hi}/2)=\lambda_1\bar\mu$. Lemma~\ref{lem:mart} with $\xi_i=\lambda_1(Y_i-\bar\mu)$ gives an event $\Omega_3$ of probability at least $1-\delta$ on which
\begin{equation}\label{eq:step4}
\sum_iY_i\le n\bar\mu+\lambda_1^{-1}\log(1/\delta).
\end{equation}

\smallskip\noindent\textit{Step 5 (assembly).} Work on $\Omega_1\cap\Omega_2\cap\Omega_3$, which has probability at least $1-3\delta$, and let $j$ be an index with $\|\hat f-f_j\|_\infty\le\varepsilon$. Because $H$ is a metric, the inequality $(x+x')^2\le2x^2+2x'^2$ and Lemma~\ref{lem:hell} give $H_i^2(\hat f)\le2H_i^2(f_j)+2c_2\varepsilon^2$. By Lemma~\ref{lem:lip}, $\ell_n(\hat f)-\ell_n(f_j)\le\varepsilon\sum_i(Y_i+r)$, and by the definition of $\hat f$, $\ell_n(f^*)-\ell_n(\hat f)\le n\Delta_n$. Therefore, by \eqref{eq:step2}--\eqref{eq:step4},
\begin{align*}
\sum_iH_i^2(\hat f)&\le\{\ell_n(f_0)-\ell_n(f_j)\}+2\log(N/\delta)+2nc_2\varepsilon^2\\
&=\{\ell_n(f_0)-\ell_n(f^*)\}+\{\ell_n(f^*)-\ell_n(\hat f)\}+\{\ell_n(\hat f)-\ell_n(f_j)\}\\
&\qquad+2\log(N/\delta)+2nc_2\varepsilon^2\\
&\le n\bar ca^2+n\Delta_n+2\log N+\Big(2+\frac1{\lambda_0}+\frac{\varepsilon}{\lambda_1}\Big)\log\frac1\delta+n\varepsilon(\bar\mu+r)+2nc_2\varepsilon^2 .
\end{align*}
Dividing by $n$ proves \eqref{eq:oracle}. The bound on $\|\hat f-f_0\|_n^2$ follows from $H_i^2(\hat f)\ge c_1\{\hat f(X_i)-f_0(X_i)\}^2$ (Lemma~\ref{lem:hell}). For \eqref{eq:cert}, stop after the first line of the display on $\Omega_1\cap\Omega_3$, which does not involve $f^*$ or the definition of $\hat f$, and write $\ell_n(f_0)-\ell_n(f_j)=\{\ell_n(f_0)-\ell_n(\hat f)\}+\{\ell_n(\hat f)-\ell_n(f_j)\}$. \qed

\section{Proof of Theorem~\ref{thm:rate}}\label{app:rate}

We use the approximation and entropy results of \citet{schmidthieber2020}, with the corrected depth condition and parameter discretization in \citet{schmidthieber2024correction}.

\begin{lemma}[\citealp{schmidthieber2020}, Lemma~5 and corrected proof of Theorem~1]\label{lem:sh}
For a class $\mathcal F$ of functions on $[0,1]^d$ and $\varepsilon>0$, let $\mathcal N(\varepsilon,\mathcal F,\|\cdot\|_\infty)$ be the smallest number of sup-norm balls of radius $\varepsilon$, with arbitrary centers, that cover $\mathcal F$. Let $f_0\in\mathcal G(q,\mathbf d,\mathbf t,\bm\beta,K)$ and let $\mathcal F_n=\mathcal F(L,\mathbf p,s_n,F)$ satisfy conditions (i)--(iv) of Theorem~\ref{thm:rate}. For sufficiently large $n$, there is a constant $C_{\rm SH}$, depending only on $q$, $\mathbf d$, $\mathbf t$, $\bm\beta$, $F$ and the implicit constants in (ii)--(iv), such that
\[
\begin{gathered}
\inf_{f\in\mathcal F_n}\|f-f_0\|_\infty^2\le C_{\rm SH}\,\phi_n,\\
\log\mathcal N\big(\varepsilon,\mathcal F(L,\mathbf p,s_n,\infty),\|\cdot\|_\infty\big)\le(s_n+1)\log\big(2\varepsilon^{-1}(L+1)V^2\big),\quad V=\prod_{l=0}^{L+1}(p_l+1),
\end{gathered}
\]
for every $\varepsilon>0$, where $\mathcal F(L,\mathbf p,s_n,\infty)$ is the class without the bound on $\|f\|_\infty$, and, after removing inactive nodes so that $p_l\le s_n$,
\[
\log\mathcal N\big(1/(2n),\mathcal F_n,\|\cdot\|_\infty\big)\le C_{\rm SH}\,n\phi_nL\log^2n.
\]
\end{lemma}

An internal $\varepsilon$-net (centers in $\mathcal F_n$) is obtained from an external $\varepsilon/2$-net by replacing each center whose ball meets $\mathcal F_n$ with a point of $\mathcal F_n$ in that ball, so $N_n(\varepsilon)\le\mathcal N(\varepsilon/2,\mathcal F_n,\|\cdot\|_\infty)$. Apply Theorem~\ref{thm:oracle} with $\varepsilon=1/n$ and with a fixed $f^*\in\mathcal F_n$ such that $\|f^*-f_0\|_\infty^2\le2C_{\rm SH}\phi_n$, which exists by Lemma~\ref{lem:sh}. Then $\bar c\|f^*-f_0\|_\infty^2\le2\bar cC_{\rm SH}\phi_n$ and $2\log N_n(1/n)/n\le2C_{\rm SH}\phi_nL\log^2n$, while $\varepsilon(\bar\mu+r)+2c_2\varepsilon^2\le(\bar\mu+r+2c_2)/n$ and $(2+1/\lambda_0+\varepsilon/\lambda_1)\le2+1/\lambda_0+1/\lambda_1$. Since $\phi_n\ge n^{-1}$, all terms are bounded by a constant multiple of $\phi_nL\log^2n+\Delta_n+\log(1/\delta)/n$, and dividing by $c_1$ gives \eqref{eq:rate}. \qed

\section{Proofs for Section~\ref{sec:oos}: Mixing and Out-of-Sample Risk}\label{app:oos}

Recall from Section~\ref{sec:assumptions} that $\beta(U,V)=\mathrm{TV}(P_{(U,V)},P_U\otimes P_V)$ for random elements $U,V$ defined on a common probability space, where $P_U$ denotes the law of $U$; for Polish spaces this is the absolute regularity coefficient between the $\sigma$-fields generated by $U$ and $V$ \citep{bradley2005}. The coefficients $\beta_X(a)$ of the covariate process are defined in \eqref{eq:betamix}. A Markov kernel $K$ from a measurable space $\mathcal Z$ to a measurable space $\mathcal W$ assigns to each $z\in\mathcal Z$ a probability measure $K(z,\cdot)$ on $\mathcal W$, measurably in $z$. For a probability measure $\pi$ on $\mathcal Z$, $\pi K=\int\pi(dz)K(z,\cdot)$ is a probability measure on $\mathcal W$ and $(\pi\otimes K)(dz,dw)=\pi(dz)K(z,dw)$ one on $\mathcal Z\times\mathcal W$; $K_1K_2$ denotes the composition of two kernels and $\delta_z$ the point mass at $z$.

\begin{lemma}[Markov kernels contract total variation]\label{lem:contract}
Let $K$ be a Markov kernel and $\pi,\pi'$ probability measures. (a) $\mathrm{TV}(\pi K,\pi'K)\le\mathrm{TV}(\pi,\pi')$. (b) If $K(z,\cdot)\ge\epsilon_0\,\nu_0(\cdot)$ for all $z$, some $\epsilon_0\in(0,1]$ and a probability measure $\nu_0$, then $\mathrm{TV}(\pi K,\pi'K)\le(1-\epsilon_0)\mathrm{TV}(\pi,\pi')$. (c) For Markov kernels $K_1,K_2$ between the same spaces, $\mathrm{TV}(\pi\otimes K_1,\pi\otimes K_2)=\int\mathrm{TV}(K_1(z,\cdot),K_2(z,\cdot))\,\pi(dz)$.
\end{lemma}
\begin{proof}
(a) For measurable $B$, $(\pi-\pi')K(B)=\int K(z,B)(\pi-\pi')(dz)$ with $0\le K(z,B)\le1$, so $|(\pi K-\pi'K)(B)|\le\mathrm{TV}(\pi,\pi')$. (b) If $\epsilon_0=1$ then $K(z,\cdot)=\nu_0$ and the left side is zero. Otherwise $\tilde K=(K-\epsilon_0\nu_0)/(1-\epsilon_0)$ is a Markov kernel and $(\pi-\pi')K=(1-\epsilon_0)(\pi-\pi')\tilde K$ because $\pi-\pi'$ has total mass zero; apply (a) to $\tilde K$. (c) With $\lambda$ a measure dominating $K_1(z,\cdot)$ and $K_2(z,\cdot)$ for $\pi$-almost every $z$ (counting measure in our application) and densities $k_1,k_2$, both sides equal $\tfrac12\iint|k_1(z,w)-k_2(z,w)|\,\lambda(dw)\,\pi(dz)$.
\end{proof}

\begin{proof}\textbf{of Proposition~\ref{prop:mixing}.}
Let $Z_u=(\mathbf Y_u,\dots,\mathbf Y_{u-m+1},U_{u+1})\in\mathcal Z=(\mathbb N_0^S)^m\times\mathcal U$, the state of Section~\ref{sec:pred}. Under Assumptions~\ref{ass:model} and \ref{ass:memory}, $(Z_u)$ is a (time-inhomogeneous) Markov chain: $\mathbf X_{u+1}$ is a function of $(Z_u,u)$, the law of $\mathbf Y_{u+1}$ given the past is the product measure $\bigotimes_{s=1}^Sp_{f_0(X_{s,u+1})}$, and $U_{u+2}$ is independent of the past and of $\mathbf Y_{u+1}$ with law $P_U$. (Without innovations, $\mathcal U$ is a single point and $\mathcal Z$ is countable.) Write $K_{u,v}$ for the transition kernel from time $u$ to time $v\ge u$, $\mathbf 0$ for the element of $(\mathbb N_0^S)^m$ in which all counts are zero, and $\nu_0=\delta_{\mathbf 0}\otimes P_U$. Since $P(\mathrm{NB}(\mu,r)=0)=\{r/(r+\mu)\}^r$ is decreasing in $\mu\le e^F$, the event $A_u=\{\mathbf Y_{u+1}=\dots=\mathbf Y_{u+m}=0\}$ satisfies, for every $z$ and $u$,
\[
\begin{aligned}
P(A_u\mid Z_u=z)&=\prod_{k=1}^{m}P(\mathbf Y_{u+k}=0\mid\mathbf Y_{u+k-1}=0,\dots,\mathbf Y_{u+1}=0,Z_u=z)\\
&\ge\Big\{\Big(\frac{r}{r+e^F}\Big)^{r}\Big\}^{Sm}=\epsilon_0,
\end{aligned}
\]
because each conditional probability is an average over the intermediate innovations of a product of $S$ such zero probabilities. On $A_u$, $Z_{u+m}=(\mathbf 0,U_{u+m+1})$, and $U_{u+m+1}$ is independent of $Z_u$ and of $A_u$ with law $P_U$. Hence $K_{u,u+m}(z,B)\ge P(A_u\cap\{Z_{u+m}\in B\}\mid Z_u=z)\ge\epsilon_0\,\nu_0(B)$ for every measurable $B\subset\mathcal Z$, that is, $K_{u,u+m}(z,\cdot)\ge\epsilon_0\nu_0$. Write $a=km+j$ with integers $k\ge0$ and $0\le j<m$, and factor $K_{u,u+a}=K_{u,u+m}K_{u+m,u+2m}\cdots K_{u+(k-1)m,u+km}K_{u+km,u+a}$. Lemma~\ref{lem:contract}(a,b), applied to the point masses $\delta_z$ and $\delta_{z'}$, then gives
\[
\sup_{z,z'}\mathrm{TV}\big(K_{u,u+a}(z,\cdot),K_{u,u+a}(z',\cdot)\big)\le(1-\epsilon_0)^{k}.
\]

Fix $t$ and an integer $a\ge1$ and let $U=(Z_u)_{u\le t}$, $V=(Z_u)_{u\ge t+a}$. Let $\bar u$ and $\bar v$ denote generic values of $U$ and $V$, with coordinates $z_t$ and $z_{t+a}$ at times $t$ and $t+a$. By the Markov property,
\[
\begin{aligned}
P_{(U,V)}(d\bar u,d\bar v)&=P_U(d\bar u)\,K_{t,t+a}(z_t,dz_{t+a})\,Q(z_{t+a},d\bar v),\\
(P_U\otimes P_V)(d\bar u,d\bar v)&=P_U(d\bar u)\,\nu(dz_{t+a})\,Q(z_{t+a},d\bar v),
\end{aligned}
\]
where $Q$ is the conditional law of the future $(Z_u)_{u\ge t+a}$ given $Z_{t+a}$ and
\[
\nu=\int P_{Z_t}(dz')\,K_{t,t+a}(z',\cdot)
\]
is the law of $Z_{t+a}$. By Lemma~\ref{lem:contract}(c) and (a),
\[
\begin{aligned}
\beta(U,V)&=\int\mathrm{TV}\big(K_{t,t+a}(z_t,\cdot)Q,\nu Q\big)P_U(d\bar u)\le\int\mathrm{TV}\big(K_{t,t+a}(z_t,\cdot),\nu\big)P_U(d\bar u)\\
&\le\sup_{z,z'}\mathrm{TV}\big(K_{t,t+a}(z,\cdot),K_{t,t+a}(z',\cdot)\big),
\end{aligned}
\]
where the last step uses that $\nu$ is a mixture of the measures $K_{t,t+a}(z',\cdot)$ and that total variation is convex. Hence $\beta\{(Z_u)_{u\le t},(Z_u)_{u\ge t+a}\}\le(1-\epsilon_0)^{\lfloor a/m\rfloor}$. Finally $(\mathbf X_u)_{u\le t}$ is a function of $(Z_u)_{u\le t-1}$ and $(\mathbf X_u)_{u\ge t+a}$ a function of $(Z_u)_{u\ge t+a-1}$, and $\beta$ does not increase under measurable maps, so $\beta_X(a)\le(1-\epsilon_0)^{\lfloor a/m\rfloor}$. When $\mathcal U$ is uncountable, Lemma~\ref{lem:contract}(c) and the disintegrations above are used on the Polish space $\mathcal Z$, where regular conditional laws exist.
\end{proof}

\begin{lemma}[Independent blocks]\label{lem:blocks}
Let $B_1,\dots,B_{N_{\rm b}}$ be random elements such that, for $k=1,\dots,N_{\rm b}-1$, the coefficient $\beta\{(B_1,\dots,B_k),B_{k+1}\}$ is at most $b$. Then \[\mathrm{TV}\Big(P_{(B_1,\dots,B_{N_{\rm b}})},\bigotimes_{k=1}^{N_{\rm b}}P_{B_k}\Big)\le(N_{\rm b}-1)b.\]
\end{lemma}
\begin{proof}
By induction on $N_{\rm b}$; the case $N_{\rm b}=1$ is trivial. Let $N_{\rm b}\ge2$. Write $B_{1:k}=(B_1,\dots,B_k)$. By the triangle inequality,
\[
\begin{aligned}
\mathrm{TV}\Big(P_{B_{1:N_{\rm b}}},\bigotimes_{k\le N_{\rm b}}P_{B_k}\Big)\le{}&\mathrm{TV}\big(P_{B_{1:N_{\rm b}}},P_{B_{1:N_{\rm b}-1}}\otimes P_{B_{N_{\rm b}}}\big)\\
&+\mathrm{TV}\Big(P_{B_{1:N_{\rm b}-1}}\otimes P_{B_{N_{\rm b}}},\bigotimes_{k\le N_{\rm b}}P_{B_k}\Big).
\end{aligned}
\]
The first term is $\beta(B_{1:N_{\rm b}-1},B_{N_{\rm b}})\le b$. The second term equals $\mathrm{TV}(P_{B_{1:N_{\rm b}-1}},\bigotimes_{k\le N_{\rm b}-1}P_{B_k})$, which is at most $(N_{\rm b}-2)b$ by the induction hypothesis.
\end{proof}

\begin{lemma}[From the trajectory to an independent copy]\label{lem:transfer}
Let $(\mathbf X_t)_{t=1}^T$ have mixing coefficients $\beta_X(\cdot)$, let $\mathcal G$ be a finite class of measurable functions with values in $[0,1]$, and let $1\le a\le T/2$ be an integer and $N_{\rm b}=\lfloor T/(2a)\rfloor$. Then, for every $\delta\in(0,1)$, with probability at least $1-\delta-2(N_{\rm b}-1)\beta_X(a)$, for all $g\in\mathcal G$,
\[
\sum_{t=1}^{T}\E g(\mathbf X_t)\le2\sum_{t=1}^{T}g(\mathbf X_t)+\frac{16}{3}\,a\log\frac{2|\mathcal G|}{\delta}+2a .
\]
\end{lemma}
\begin{proof}
For $k=1,\dots,N_{\rm b}$ let $O_k=\{2(k-1)a+1,\dots,(2k-1)a\}$ and $E_k=\{(2k-1)a+1,\dots,2ka\}$, write $\mathbf X_{O_k}=(\mathbf X_t)_{t\in O_k}$, and put $U_k=\sum_{t\in O_k}g(\mathbf X_t)\in[0,a]$ for a fixed $g\in\mathcal G$. Consecutive odd blocks are separated by $a$ time points, so $\beta\{(\mathbf X_{O_1},\dots,\mathbf X_{O_k}),\mathbf X_{O_{k+1}}\}\le\beta_X(a)$ and, by Lemma~\ref{lem:blocks}, the probability of any event defined through $(\mathbf X_{O_k})_{k\le N_{\rm b}}$ differs by at most $(N_{\rm b}-1)\beta_X(a)$ from its probability $\tilde P$ under independent blocks with the same marginal laws. Under $\tilde P$, the variables $\E U_k-U_k\le a$ are independent and centered with $\sum_k\mathrm{Var}(U_k)\le\sum_k\E U_k^2\le a\sum_k\E U_k$, so Bernstein's inequality \citep[Theorem~2.10]{boucheron2013} gives, with $\tilde P$-probability at least $1-e^{-x}$,
\[
\sum_k(\E U_k-U_k)\le\Big(2a\,x\sum_k\E U_k\Big)^{1/2}+\frac{ax}{3}\le\frac12\sum_k\E U_k+ax+\frac{ax}{3},
\]
using $\sqrt{2uv}\le u/2+v$. Rearranging, $\sum_k\E U_k\le2\sum_kU_k+\tfrac83ax$. The same holds for the even blocks. Taking $x=\log(2|\mathcal G|/\delta)$ and a union bound over $\mathcal G$ and over the two families of blocks, with probability at least $1-\delta-2(N_{\rm b}-1)\beta_X(a)$ and for all $g\in\mathcal G$, $\sum_{t\le2N_{\rm b}a}\E g(\mathbf X_t)\le2\sum_{t\le2N_{\rm b}a}g(\mathbf X_t)+\tfrac{16}{3}ax$. The at most $2a$ remaining indices contribute at most $2a$ to the left side.
\end{proof}

\begin{proof}\textbf{of Theorem~\ref{thm:oos}.}
For a function $f$ with $\|f\|_\infty\le F$ let
\[
\begin{gathered}
g_f(\mathbf x)=\frac1S\sum_{s=1}^{S}H^2(p_{f(x_s)},p_{f_0(x_s)})\in[0,1],\\
R_n^H(f)=\frac1T\sum_{t=1}^{T}g_f(\mathbf X_t),\qquad R^H(f)=\frac1T\sum_{t=1}^{T}\E g_f(\mathbf X'_t).
\end{gathered}
\] Let $f_1,\dots,f_N$ be the net of the proof of Theorem~\ref{thm:oracle} with $\varepsilon=1/n$; these are non-random functions. Apply Lemma~\ref{lem:transfer} to $\mathcal G=\{g_{f_j}\}$ with $a=a_T=\lceil2c_\beta^{-1}\log T\rceil$, for which $2(N_{\rm b}-1)\beta_X(a)\le(T/a)C_\beta T^{-2}\le C_\beta/T$; this requires $a_T\le T/2$, which holds for $T$ larger than a constant depending on $c_\beta$. On the intersection of this event with the event of Theorem~\ref{thm:oracle}, which has probability at least $1-4\delta-C_\beta/T$, let $j$ satisfy $\|\hat f-f_j\|_\infty\le\varepsilon$. As in Step~5 of that proof, $g_{\hat f}\le2g_{f_j}+2c_2\varepsilon^2$ and $g_{f_j}\le2g_{\hat f}+2c_2\varepsilon^2$ pointwise, so
\begin{align*}
R^H(\hat f)&\le2R^H(f_j)+2c_2\varepsilon^2\le4R_n^H(f_j)+\frac{2}{T}\Big\{\frac{16}{3}a_T\log\frac{2N}{\delta}+2a_T\Big\}+2c_2\varepsilon^2\\
&\le8R_n^H(\hat f)+10c_2\varepsilon^2+\frac{Sa_T}{n}\Big\{\frac{32}{3}\log\frac{2N}{\delta}+4\Big\}.
\end{align*}
Here $R^H(\hat f)$ denotes $R^H(f)$ evaluated at $f=\hat f$, the copy $(\mathbf X'_t)$ being independent of the data. By Theorem~\ref{thm:oracle} and Lemma~\ref{lem:sh} (as in the proof of Theorem~\ref{thm:rate}), $R_n^H(\hat f)\le C\{\phi_nL\log^2n+\Delta_n+\log(1/\delta)/n\}$ and $\log N\le C_{\rm SH}n\phi_nL\log^2n$. Since $a_T\le C\log n$, $L\asymp\log n$ and $S$ is fixed, $R^H(\hat f)\le C\{\phi_n\log^4n+\Delta_n+\log n\log(1/\delta)/n\}$, and $R(\hat f)\le R^H(\hat f)/c_1$ by Lemma~\ref{lem:hell}. For the bound in expectation take $\delta=1/n$ and use $R(\hat f)\le4F^2$ on the exceptional event, whose probability is at most $4/n+C_\beta S/n$.
\end{proof}

\section{Proof of Theorem~\ref{thm:lower}}\label{app:lower}

Fix $S\ge1$. It suffices to consider the sub-model of $\mathcal P_n(\mathcal G_{\beta,d^*}(K))$ in which the covariates $X_{s,t}$, $s\le S$, $t\le T$, are independent and uniform on $[0,1]^d$, the covariates of week $t$ are revealed at the start of that week (take $\mathcal I_{t-1}=\sigma(\mathbf X_u,u\le t;\ \mathbf Y_u,u\le t-1)$), and $Y_{s,t}\mid\mathcal I_{t-1}\sim\mathrm{NB}(e^{f_0(X_{s,t})},r)$ independently. It satisfies Assumptions~\ref{ass:model}, \ref{ass:bounded} and~\ref{ass:mixing} with $\beta_X\equiv0$, and also Assumption~\ref{ass:memory} with innovations $U_t=\mathbf X_t$ on $\mathcal U=[0,1]^{dS}$; for it $R(f)=\|f-f_0\|_{L^2}^2$, and the $n=ST$ pairs $(X_{s,t},Y_{s,t})$ are independent and identically distributed. Index them by $i=1,\dots,n$ as in Appendix~\ref{app:oracle}.

\smallskip\noindent\textit{Hypotheses.} Let $\psi:\mathbb R^{d^*}\to\mathbb R$ be infinitely differentiable, supported in $[0,1]^{d^*}$ and not identically zero, and let $C_\psi$ be its $\beta$-H\"older norm, that is, the left side of the inequality that defines $\mathcal C^{\beta}(D,K)$ in Section~\ref{sec:structure} with $g=\psi$. For an integer $M\ge1$ and $\mathbf j\in\{0,\dots,M-1\}^{d^*}$ let $\psi_{\mathbf j}(u)=\kappa M^{-\beta}\psi(Mu-\mathbf j)$ with $\kappa=K/(2C_\psi)$, and for $w\in\{0,1\}^{M^{d^*}}$ let
\[
f_w(x)=\sum_{\mathbf j}w_{\mathbf j}\,\psi_{\mathbf j}(x_1,\dots,x_{d^*}).
\]
The functions $\psi_{\mathbf j}$ have supports with disjoint interiors, and $\partial^{\bm\alpha}\psi_{\mathbf j}(u)=\kappa M^{|\bm\alpha|-\beta}(\partial^{\bm\alpha}\psi)(Mu-\mathbf j)$, so each $\psi_{\mathbf j}$ has $\beta$-H\"older norm at most $\kappa C_\psi$. For $u,v$ in different cubes, let $\bar u$ and $\bar v$ be the points where the segment from $u$ to $v$ leaves the cube of $u$ and enters the cube of $v$; all derivatives of $f_w$ vanish there, so for $|\bm\alpha|=\lfloor\beta\rfloor$,
\[
|\partial^{\bm\alpha}f_w(u)-\partial^{\bm\alpha}f_w(v)|\le|\partial^{\bm\alpha}f_w(u)-\partial^{\bm\alpha}f_w(\bar u)|+|\partial^{\bm\alpha}f_w(\bar v)-\partial^{\bm\alpha}f_w(v)|\le2\kappa C_\psi|u-v|_\infty^{\beta-\lfloor\beta\rfloor},
\]
because $|u-\bar u|_\infty$ and $|\bar v-v|_\infty$ are at most $|u-v|_\infty$. Hence $f_w$ has $\beta$-H\"older norm at most $2\kappa C_\psi=K$, that is, $f_w\in\mathcal G_{\beta,d^*}(K)$, and $\|f_w\|_\infty\le K\le F$. Moreover
\begin{equation}\label{eq:sep}
\|f_w-f_{w'}\|_{L^2}^2=\kappa^2\|\psi\|_{L^2}^2M^{-2\beta-d^*}\rho_H(w,w'),
\end{equation}
where $\rho_H(w,w')=\sum_{\mathbf j}\mathbf 1\{w_{\mathbf j}\ne w'_{\mathbf j}\}$ is the Hamming distance and $\|\cdot\|_{L^2}$ the norm of $L^2[0,1]^d$ or $L^2[0,1]^{d^*}$, as appropriate.

\smallskip\noindent\textit{Packing and Kullback--Leibler bound.} By the Varshamov--Gilbert bound \citep[Lemma~2.9]{tsybakov2009}, if $M^{d^*}\ge8$ there are $w^{(0)}=\mathbf 0,w^{(1)},\dots,w^{(N)}$ with $N\ge2^{M^{d^*}/8}$ and $\rho_H(w^{(j)},w^{(k)})\ge M^{d^*}/8$ for $j\ne k$. By \eqref{eq:sep}, $\|f_{w^{(j)}}-f_{w^{(k)}}\|_{L^2}\ge2s_M$ with $s_M=\kappa\|\psi\|_{L^2}M^{-\beta}/(4\sqrt2)$. Let $P_j$ be the law of $(X_i,Y_i)_{i\le n}$ under $f_{w^{(j)}}$. By the chain rule for the Kullback--Leibler divergence, Lemma~\ref{lem:hell} and \eqref{eq:sep},
\[
\mathrm{KL}(P_j\,\|\,P_0)=n\,\E\,\mathrm{KL}\big(p_{f_{w^{(j)}}(X_1)}\,\|\,p_{f_{w^{(0)}}(X_1)}\big)\le\frac{nV_{\max}}{2}\|f_{w^{(j)}}\|_{L^2}^2\le\frac{nV_{\max}}{2}\kappa^2\|\psi\|_{L^2}^2M^{-2\beta}.
\]
Fix $\alpha_0=1/16$ and choose $M=\lceil(c_*n)^{1/(2\beta+d^*)}\rceil$ with $c_*=4V_{\max}\kappa^2\|\psi\|_{L^2}^2/(\alpha_0\log2)$. Then $M^{2\beta+d^*}\ge c_*n$ gives $\mathrm{KL}(P_j\|P_0)\le\alpha_0(\log2)M^{d^*}/8\le\alpha_0\log N$ for every $j$.

\smallskip\noindent\textit{Conclusion.} By Theorem~2.5 of \citet{tsybakov2009}, for $n$ large enough that $M^{d^*}\ge8$,
\[
\inf_{\hat f}\max_{0\le j\le N}P_j\big(\|\hat f-f_{w^{(j)}}\|_{L^2}\ge s_M\big)\ge\frac{\sqrt N}{1+\sqrt N}\Big(1-2\alpha_0-\sqrt{\frac{2\alpha_0}{\log N}}\Big)\ge c_0>0,
\]
so $\inf_{\hat f}\sup_{P}\E_PR(\hat f)\ge c_0s_M^2$. Since $M\le2(c_*n)^{1/(2\beta+d^*)}$ for large $n$, $s_M^2\ge c\,n^{-2\beta/(2\beta+d^*)}$ with $c>0$ depending only on $(\beta,d^*,K,F,r)$. For the finitely many remaining $T$, note that the minimax risk is non-increasing in $T$, in this nested independent-regression subexperiment, because an estimator may ignore observations; it is therefore bounded below by its (positive) value at the smallest $T$ covered above, and the bound holds for all $n=ST$ after decreasing $c$. The constant does not depend on $S$, because only the number $n$ of independent pairs enters the argument. \qed

\section{Proof of Theorem~\ref{thm:pred}}\label{app:pred}

(a) By Lemma~\ref{lem:hell}, $\mathrm{TV}(p_{f(x)},p_{f_0(x)})\le\sqrt2H(p_{f(x)},p_{f_0(x)})\le(2c_2)^{1/2}|f(x)-f_0(x)|$, and $|P_f(C\mid x)-P_{f_0}(C\mid x)|\le\mathrm{TV}(p_{f(x)},p_{f_0(x)})$ for every set $C$, which gives \eqref{eq:cov1}. Averaging over the independent copy and using the Cauchy--Schwarz inequality, $n^{-1}\sum_{s,t}\E|f(X'_{s,t})-f_0(X'_{s,t})|\le R(f)^{1/2}$, which gives the averaged statement. For the prediction risk, since $\E[Y\mid X=x]=e^{f_0(x)}$ under Assumption~\ref{ass:model}, $\E[(Y-e^{f(x)})^2\mid X=x]-\E[(Y-e^{f_0(x)})^2\mid X=x]=(e^{f(x)}-e^{f_0(x)})^2\le e^{2F}\{f(x)-f_0(x)\}^2$ by the mean value theorem.

(b) Recall that $\mathbb P_f^{(h)}(\cdot\mid z,t)$ is the law of $(\mathbf Y_{t+1},\dots,\mathbf Y_{t+h})$ when $Z_t=z$ and the counts evolve according to the model with log-mean function $f$; under both laws the innovations $U_{t+2},\dots,U_{t+h}$ are drawn independently from $P_U$. Let $W_k=(\mathbf Y_{t+k},U_{t+k+1})$, so that $(W_1,\dots,W_h)$ determines $(\mathbf Y_{t+1},\dots,\mathbf Y_{t+h})$ and $\mathbf X_{t+k}$ is a function of $(z,W_1,\dots,W_{k-1})$. By the chain rule for the Kullback--Leibler divergence \citep[Theorem~2.5.3]{cover2006}, applied to the laws of $(W_1,\dots,W_h)$ under $f_0$ and under $f$, and because the conditional law of $W_k$ given $(W_1,\dots,W_{k-1})$ is $\bigotimes_sp_{f_0(X_{s,t+k})}\otimes P_U$ under $f_0$ and $\bigotimes_sp_{f(X_{s,t+k})}\otimes P_U$ under $f$,
\[
\mathrm{KL}\big(\mathbb P_{f_0}^{(h)}(\cdot\mid z,t)\,\big\|\,\mathbb P_{f}^{(h)}(\cdot\mid z,t)\big)\le\sum_{k=1}^{h}\E_{f_0}\Big[\sum_{s=1}^{S}\mathrm{KL}\big(p_{f_0(X_{s,t+k})}\,\big\|\,p_{f(X_{s,t+k})}\big)\,\Big|\,Z_t=z\Big],
\]
where the inequality, an equality for the laws of $(W_k)$, accounts for the marginalization to the counts, and the divergence between product measures is the sum of the divergences of the factors. By Lemma~\ref{lem:hell} each term is at most $\tfrac12V_{\max}\{f(X_{s,t+k})-f_0(X_{s,t+k})\}^2$, and Pinsker's inequality $\mathrm{TV}\le(\mathrm{KL}/2)^{1/2}$ \citep[Lemma~2.5]{tsybakov2009} gives \eqref{eq:simlemma}. For the average, take $z=Z'_t$. The copy follows the model with $f_0$, so $\E\big[\E_{f_0}[\,\cdot\mid Z_t=Z'_t]\big]$ is the unconditional expectation under the copy, and Jensen's inequality for the concave square root gives
\[
\begin{aligned}
\frac{1}{T-h}\sum_{t=1}^{T-h}\E\,\mathrm{TV}_h(Z'_t,t)&\le\Big\{\frac{V_{\max}}{4(T-h)}\sum_{t=1}^{T-h}\sum_{k=1}^{h}\sum_{s=1}^{S}\E\{f(X'_{s,t+k})-f_0(X'_{s,t+k})\}^2\Big\}^{1/2}\\
&\le\Big\{\frac{V_{\max}\,h\,nR(f)}{4(T-h)}\Big\}^{1/2},
\end{aligned}
\]
because every $u\in\{1,\dots,T\}$ equals $t+k$ for at most $h$ pairs $(t,k)$; this is \eqref{eq:simavg}, since $n=ST$. Marginalizing to any coordinate or function of $(\mathbf Y_{t+1},\dots,\mathbf Y_{t+h})$ does not increase total variation, which gives the statement for prediction regions.

(c) By the Dvoretzky--Kiefer--Wolfowitz inequality with the constant of \citet{massart1990}, $P\{\sup_q|G_B(q)-G(q)|>\varepsilon\}\le2e^{-2B\varepsilon^2}$, so with probability at least $1-\eta$ we have $\sup_q|G_B(q)-G(q)|\le\varepsilon_B=\{\log(2/\eta)/(2B)\}^{1/2}$, and then also $\sup_q|G_B(q-)-G(q-)|\le\varepsilon_B$ by taking limits. On this event the fitted-model probability of $[\hat q_-,\hat q_+]$ is
\[
G(\hat q_+)-G(\hat q_--)\ge G_B(\hat q_+)-G_B(\hat q_--)-2\varepsilon_B\ge1-\alpha-2\varkappa_B-2\varepsilon_B .
\]
For $\hat q_\pm=G_B^{-1}(p_\pm)=\inf\{q:G_B(q)\ge p_\pm\}$ with $p_+=1-\alpha/2$ and $p_-=\alpha/2$, $G_B(\hat q_+)\ge p_+$ by right continuity and $G_B(\hat q_--)<p_-$ because $G_B(q)<p_-$ for $q<\hat q_-$; thus $\varkappa_B=0$. Let $Y_{(1)}\le\dots\le Y_{(B)}$ be the ordered draws, which are integers. Linear interpolation returns, for level $p$, a value between $Y_{(j)}$ and $Y_{(j+1)}$ with $j=\lfloor(B-1)p\rfloor+1$ (equal to $Y_{(j)}$ when $(B-1)p$ is an integer), and rounding to an integer keeps it in $[Y_{(j)},Y_{(j+1)}]$. Hence $G_B(\hat q_+)\ge j/B>(B-1)p_+/B\ge p_+-1/B$, and $G_B(\hat q_--)\le j/B\le\{(B-1)p_-+1\}/B\le p_-+1/B$, because at most $j$ of the draws are smaller than $Y_{(j+1)}$; thus $\varkappa_B=1/B$. \qed

\noindent\textit{Proof of part (d).} Fix the fitted model and origin. Enlarge the probability space by drawing independent terminal counts $W_b$ conditionally on the independent histories $V_b$, with conditional distribution functions $K_b$. Then $W_1,\ldots,W_B$ are independent with common distribution function $G$, and $\widetilde G_B$ is the conditional expectation of their empirical distribution function $G_B$ given $V_1,\ldots,V_B$. Conditional Jensen's inequality for the supremum norm gives
\[
\E\|\widetilde G_B-G\|_\infty\le\E\|G_B-G\|_\infty
\le\int_0^\infty2e^{-2Bu^2}\,du=\sqrt{\frac{\pi}{2B}},
\]
where the second inequality integrates the Dvoretzky--Kiefer--Wolfowitz tail bound; using the untruncated upper bound only enlarges the integral. Replacing one history changes every value of $\widetilde G_B$ by at most $1/B$, and therefore changes $\|\widetilde G_B-G\|_\infty$ by at most $1/B$. The bounded-differences inequality \citep[Chapter~6]{boucheron2013} yields
\[
P\{\|\widetilde G_B-G\|_\infty>\E\|\widetilde G_B-G\|_\infty+u\}\le e^{-2Bu^2}.
\]
Set $u=\{\log(1/\eta)/(2B)\}^{1/2}$ and apply the endpoint argument of (c), including left limits. This proves (d). These statements are for one fixed origin, site and horizon; simultaneous guarantees require allocating the failure probability across the requested intervals. At horizon one each $K_b=G$, so the error is identically zero. \qed

\section{Proof of Theorem~\ref{thm:unknown}}\label{app:unknown}

Throughout, $\Theta=[-F,F]\times[r_{\rm lo},r_{\rm hi}]$, $\vartheta=(f,r)\in\Theta$ is a pair of a log-mean value and a size, $p_\vartheta=p_{f,r}$ is the $\NB(e^f,r)$ mass function and $|\cdot|$ is the Euclidean norm on $\mathbb R^2$. The constants are
\begin{equation}\label{eq:constants2a}
\begin{gathered}
\kappa=a_0+a_1e^F,\qquad a_1=\frac14+\frac1{4r_{\rm lo}}+\frac1{r_{\rm lo}^2},\qquad a_0=\frac{r_{\rm hi}}4+\frac{e^F}{4r_{\rm lo}}+\frac2{r_{\rm lo}}+\frac{1+e^F}{r_{\rm lo}^2},\\
b_1=1+\frac1{r_{\rm lo}},\qquad b_0=r_{\rm hi}+\frac{e^F}{r_{\rm lo}},\\
c_f=\frac{e^{-2F}}{8m_2},\qquad c_r=\Big[2r_{\rm hi}^2e^{2F}\big\{(2m_4)^{1/2}+2e^F(1+r_{\rm lo}^{-1})(2m_2)^{1/2}\big\}\Big]^{-2},\\
m_2=e^F+e^{2F}(1+r_{\rm lo}^{-1}),\qquad m_4=e^{4F}(1+3r_{\rm lo}^{-1})^3+4e^{3F}(1+2r_{\rm lo}^{-1})^2+2e^{2F}(1+r_{\rm lo}^{-1}),
\end{gathered}
\end{equation}
\begin{equation}\label{eq:constants2}
\begin{gathered}
\kappa_\gamma=\kappa\max(1,r_{\rm hi}^2),\qquad c_\gamma=c_rr_{\rm lo}^2,\qquad\rho=\frac{r_{\rm hi}}{r_{\rm lo}},\qquad c_0=\max(r_{\rm hi},e^F),\\
s_0=\frac12\log\Big(1+\frac{r_{\rm lo}}{2e^F}\Big),\qquad\bar\mu_\gamma=\frac{r_{\rm lo}\log2}{2s_0},\qquad M_0=\frac{4}{s_0^2}\,e^{2s_0c_0}\,2^{r_{\rm lo}},\\
\bar c_\gamma=\kappa_\gamma+2\rho^2M_0,\qquad\lambda_\gamma=\min\Big\{1,\frac{s_0}{2\rho a_{\max}}\Big\},\qquad a_{\max}=\max(2F,\gamma_{\rm hi}-\gamma_{\rm lo}).
\end{gathered}
\end{equation}
Here $\Theta=[-F,F]\times[r_{\rm lo},r_{\rm hi}]$ with $r_{\rm lo}=e^{\gamma_{\rm lo}}$ and $r_{\rm hi}=e^{\gamma_{\rm hi}}$.

\begin{lemma}[The two-parameter family]\label{lem:two}
For $\vartheta=(f,r)$ and $\vartheta'=(f',r')$ in $\Theta$ and $y\in\mathbb N_0$:
\begin{enumerate}
\item[(a)] $|\log p_{\vartheta}(y)-\log p_{\vartheta'}(y)|\le(y+r_{\rm hi})|f-f'|+(y+e^F)r_{\rm lo}^{-1}|r-r'|$;
\item[(b)] $\mathrm{KL}(p_\vartheta\,\|\,p_{\vartheta'})\le\tfrac12\kappa|\vartheta-\vartheta'|^2$, hence $H^2(p_\vartheta,p_{\vartheta'})\le\tfrac14\kappa|\vartheta-\vartheta'|^2$ and $\mathrm{TV}(p_\vartheta,p_{\vartheta'})\le\tfrac12\kappa^{1/2}|\vartheta-\vartheta'|$;
\item[(c)] $c_f(f-f')^2+c_r(r-r')^2\le2H^2(p_\vartheta,p_{\vartheta'})$.
\end{enumerate}
\end{lemma}
\begin{proof}
Write $\mu=e^f$ and $\psi$ for the digamma function, and recall $\psi(y+r)-\psi(r)=\sum_{j=0}^{y-1}(r+j)^{-1}\in[0,y/r]$ and $0\le\psi'(r)-\psi'(y+r)\le\psi'(r)\le1/r+1/r^2$. Differentiating $\log p_{f,r}(y)=\log\Gamma(y+r)-\log\Gamma(r)-\log y!+r\log r+y f-(y+r)\log(r+\mu)$,
\[
\begin{gathered}
\partial_f\log p=\frac{(y-\mu)r}{r+\mu},\qquad\partial_r\log p=\psi(y+r)-\psi(r)+\log\frac{r}{r+\mu}+\frac{\mu-y}{r+\mu},\\
-\partial_f^2\log p=\frac{r\mu(y+r)}{(r+\mu)^2},\qquad\partial_f\partial_r\log p=\frac{(y-\mu)\mu}{(r+\mu)^2},\\
-\partial_r^2\log p=\psi'(r)-\psi'(y+r)-\frac{\mu}{r(r+\mu)}+\frac{\mu-y}{(r+\mu)^2}.
\end{gathered}
\]
(a) $|\partial_f\log p|\le y+r\le y+r_{\rm hi}$, and since $\log\{r/(r+\mu)\}\in[-\mu/r,0]$ and $(\mu-y)/(r+\mu)\in[-y/r,\mu/r]$, $|\partial_r\log p|\le(y+\mu)/r\le(y+e^F)/r_{\rm lo}$ on $\Theta$. The mean value theorem along the segment from $\vartheta'$ to $\vartheta$, which lies in the convex set $\Theta$, gives (a).

(b) Fix $\vartheta$ and let $g(\vartheta')=\mathrm{KL}(p_\vartheta\|p_{\vartheta'})=\E_\vartheta\{\log p_\vartheta(Y)-\log p_{\vartheta'}(Y)\}$, where $\E_\vartheta$ is the expectation under $p_\vartheta$. All moments of $Y$ are finite, so differentiation under the expectation is allowed; $g(\vartheta)=0$, $\nabla g(\vartheta)=0$ and $\nabla^2g(\vartheta'')=\E_\vartheta\{-\nabla^2\log p_{\vartheta''}(Y)\}$. By Taylor's theorem, $g(\vartheta')\le\tfrac12\sup_{\vartheta''\in\Theta}v^\top\nabla^2g(\vartheta'')v$ with $v=\vartheta'-\vartheta$. For a symmetric $2\times2$ matrix with entries $A,B,C$, $v^\top\binom{A\ B}{B\ C}v\le(|A|+|B|+|C|)|v|^2$. With $\mu_\vartheta=\E_\vartheta Y\le e^F$ and the bounds $r\mu/(r+\mu)^2\le1/4$, $\mu/(r+\mu)^2\le1/(4r)$ and $\mu/\{r(r+\mu)\}\le1/r$, the entries of $\E_\vartheta\{-\nabla^2\log p_{\vartheta''}(Y)\}$ satisfy
\[
|A|\le\frac{\mu_\vartheta+r_{\rm hi}}{4},\qquad|B|\le\frac{\mu_\vartheta+e^F}{4r_{\rm lo}},\qquad|C|\le\frac1{r_{\rm lo}}+\frac1{r_{\rm lo}^2}+\frac1{r_{\rm lo}}+\frac{e^F+\mu_\vartheta}{r_{\rm lo}^2},
\]
so $|A|+|B|+|C|\le a_0+a_1\mu_\vartheta\le\kappa$. This gives the bound on $\mathrm{KL}$. The bound on $H^2$ follows from $\mathrm{KL}\ge-2\log(1-H^2)\ge2H^2$, which is Jensen's inequality for $\E_\vartheta\log(p_{\vartheta'}/p_\vartheta)^{1/2}$, and the bound on $\mathrm{TV}$ from Pinsker's inequality.

(c) For a function $g$ on $\mathbb N_0$ and mass functions $p,q$, the Cauchy--Schwarz inequality and $(\sqrt p+\sqrt q)^2\le2(p+q)$ give, with the argument $y$ omitted,
\[
\Big|\sum_yg\,(p-q)\Big|=\Big|\sum_yg\,(\sqrt p-\sqrt q)(\sqrt p+\sqrt q)\Big|\le2H(p,q)\Big\{\sum_yg^2(p+q)\Big\}^{1/2}.
\]
The factorial moments of $\NB(\mu,r)$ are $\E\{Y(Y-1)\cdots(Y-k+1)\}=\mu^k\prod_{j=1}^{k-1}(1+j/r)$, and $y^2=y(y-1)+y$ and $y^2(y-1)^2=y(y-1)(y-2)(y-3)+4y(y-1)(y-2)+2y(y-1)$, so $\E_\vartheta Y^2\le m_2$ and $\E_\vartheta\{Y(Y-1)\}^2\le m_4$ on $\Theta$. With $g(y)=y$ this gives $|\mu-\mu'|\le2H(2m_2)^{1/2}$, where $\mu'=e^{f'}$ and $H=H(p_\vartheta,p_{\vartheta'})$; since $|\mu-\mu'|\ge e^{-F}|f-f'|$, $H^2\ge c_f(f-f')^2$. With $g(y)=y(y-1)$, $|\mu^2(1+1/r)-\mu'^2(1+1/r')|\le2H(2m_4)^{1/2}$, and
\[
\mu'^2\Big|\frac1r-\frac1{r'}\Big|\le\big|\mu^2(1+r^{-1})-\mu'^2(1+r'^{-1})\big|+(1+r_{\rm lo}^{-1})|\mu-\mu'|(\mu+\mu').
\]
Using $\mu'^2\ge e^{-2F}$, $rr'\le r_{\rm hi}^2$ and $\mu+\mu'\le2e^F$, $|r-r'|\le2r_{\rm hi}^2e^{2F}H\{(2m_4)^{1/2}+2e^F(1+r_{\rm lo}^{-1})(2m_2)^{1/2}\}$, that is, $H^2\ge c_r(r-r')^2$. Adding the two lower bounds gives (c).
\end{proof}

\begin{lemma}[Log-size parametrization and moments]\label{lem:gamma}
For $(f,\gamma)$ and $(f',\gamma')$ in $[-F,F]\times[\gamma_{\rm lo},\gamma_{\rm hi}]$, write $p=p_{f,e^\gamma}$, $p'=p_{f',e^{\gamma'}}$ and $\Delta=\max(|f-f'|,|\gamma-\gamma'|)$. Then
\begin{enumerate}
\item[(a)] $|\log p(y)-\log p'(y)|\le\rho(y+c_0)(|f-f'|+|\gamma-\gamma'|)\le2\rho(y+c_0)\Delta$ for $y\in\mathbb N_0$;
\item[(b)] $\mathrm{KL}(p\,\|\,p')\le\tfrac12\kappa_\gamma\{(f-f')^2+(\gamma-\gamma')^2\}$ and $H^2(p,p')\le\tfrac14\kappa_\gamma\{(f-f')^2+(\gamma-\gamma')^2\}$;
\item[(c)] $c_f(f-f')^2+c_\gamma(\gamma-\gamma')^2\le2H^2(p,p')$;
\item[(d)] if $Y\sim\NB(\mu,r)$ with $\mu\le e^F$ and $r\in[r_{\rm lo},r_{\rm hi}]$, then $\E e^{2s_0Y}\le2^{r_{\rm lo}}$.
\end{enumerate}
\end{lemma}
\begin{proof}
For $r=e^\gamma$ and $r'=e^{\gamma'}$ in $[r_{\rm lo},r_{\rm hi}]$, the mean value theorem gives $r_{\rm lo}|\gamma-\gamma'|\le|r-r'|\le r_{\rm hi}|\gamma-\gamma'|$. Part~(a) follows from Lemma~\ref{lem:two}(a), because $y+r_{\rm hi}\le\rho(y+c_0)$ and $(y+e^F)r_{\rm hi}/r_{\rm lo}\le\rho(y+c_0)$. Parts~(b) and~(c) follow from Lemma~\ref{lem:two}(b,c) and the two inequalities between $|r-r'|$ and $|\gamma-\gamma'|$. For~(d), the moment generating function of the negative binomial law is $\E e^{uY}=\{1-(\mu/r)(e^u-1)\}^{-r}$ when $(\mu/r)(e^u-1)<1$; for $u=2s_0$, $e^u-1=r_{\rm lo}/(2e^F)$, so $\E e^{2s_0Y}\le(1-a/r)^{-r}$ with $a=r_{\rm lo}/2$. Because $-\log(1-x)\le x/(1-x)$ for $0\le x<1$, the function $r\mapsto-r\log(1-a/r)$ has a nonpositive derivative on $r\ge r_{\rm lo}$, where it equals $r_{\rm lo}\log2$; hence $\E e^{2s_0Y}\le2^{r_{\rm lo}}$.
\end{proof}

\begin{proof}\textbf{of Theorem~\ref{thm:unknown}.}
(a) We follow the five steps of the proof of Theorem~\ref{thm:oracle}, indexing the $n$ site-weeks as in Appendix~\ref{app:oracle}, with $\vartheta_i(f,g)=(f(X_i),g(X_i))$, $p_{i}(f,g)=p_{f(X_i),\exp g(X_i)}$ and $H_i^2(f,g)=H^2(p_i(f,g),p_i(f_0,g_0))$. Under \eqref{eq:modelgen} the conditional law of $Y_i$ given $\mathcal H_{i-1}$ is $p_i(f_0,g_0)$, and its mean and size satisfy the conditions of Lemma~\ref{lem:gamma}(d).

\emph{Steps~1 and~2.} Step~1 uses only that $p_i(f_0,g_0)$ is the conditional law of $Y_i$; so for every fixed pair, with probability at least $1-\delta$, $\sum_iH_i^2(f,g)\le\tfrac12\{\ell_n(f_0,g_0)-\ell_n(f,g)\}+\log(1/\delta)$. With an internal $\varepsilon$-net $(f_j,g_j)_{j\le N}$ of $\mathcal P_n$, $N=N_n(\varepsilon)$, a union bound gives an event $\Omega_1$ of probability at least $1-\delta$ on which this holds for all $j$ with $\log(N/\delta)$ in place of $\log(1/\delta)$.

\emph{Step~3 (approximation term).} Let $D_i=\log\{p_i(f_0,g_0)(Y_i)/p_i(f^*,g^*)(Y_i)\}$. By Lemma~\ref{lem:gamma}(a), $|D_i|\le2\rho a(Y_i+c_0)$, and by Lemma~\ref{lem:gamma}(b), $\E[D_i\mid\mathcal H_{i-1}]=\mathrm{KL}\le\kappa_\gamma a^2$. For $0<\lambda\le\lambda_\gamma$ we have $2\lambda\rho a\le s_0$ because $a\le a_{\max}$, and the inequality $e^x\le1+x+\tfrac12x^2e^{|x|}$ gives
\[
\E[e^{\lambda D_i}\mid\mathcal H_{i-1}]\le1+\lambda\kappa_\gamma a^2+2\lambda^2\rho^2a^2\,\E\big[(Y_i+c_0)^2e^{s_0(Y_i+c_0)}\,\big|\,\mathcal H_{i-1}\big].
\]
Since $x^2\le4s_0^{-2}e^{s_0x}$ for $x\ge0$, Lemma~\ref{lem:gamma}(d) bounds the last expectation by $4s_0^{-2}e^{2s_0c_0}\E[e^{2s_0Y_i}\mid\mathcal H_{i-1}]\le M_0$. With $\log(1+x)\le x$ and $\lambda\le1$, $\log\E[e^{\lambda D_i}\mid\mathcal H_{i-1}]\le\lambda a^2(\kappa_\gamma+2\rho^2M_0)=\lambda\bar c_\gamma a^2$, and Lemma~\ref{lem:mart} with $\xi_i=\lambda_\gamma(D_i-\bar c_\gamma a^2)$ gives an event $\Omega_2$ of probability at least $1-\delta$ on which $\ell_n(f_0,g_0)-\ell_n(f^*,g^*)\le n\bar c_\gamma a^2+\lambda_\gamma^{-1}\log(1/\delta)$.

\emph{Step~4 (sum of the counts).} By Lemma~\ref{lem:gamma}(d) and Jensen's inequality, $\E[e^{s_0Y_i}\mid\mathcal H_{i-1}]\le2^{r_{\rm lo}/2}$, so Lemma~\ref{lem:mart} with $\xi_i=s_0Y_i-\tfrac12r_{\rm lo}\log2$ gives an event $\Omega_3$ of probability at least $1-\delta$ on which $\sum_iY_i\le n\bar\mu_\gamma+s_0^{-1}\log(1/\delta)$.

\emph{Step~5 (assembly).} On $\Omega_1\cap\Omega_2\cap\Omega_3$, choose $j$ with $\max(\|\hat f-f_j\|_\infty,\|\hat g-g_j\|_\infty)\le\varepsilon$. By Lemma~\ref{lem:gamma}(b) and the triangle inequality for $H$, $H_i^2(\hat f,\hat g)\le2H_i^2(f_j,g_j)+\kappa_\gamma\varepsilon^2$; by Lemma~\ref{lem:gamma}(a), $\ell_n(\hat f,\hat g)-\ell_n(f_j,g_j)\le2\rho\varepsilon\sum_i(Y_i+c_0)$; and since $(f^*,g^*)\in\mathcal P_n$, $\ell_n(f^*,g^*)-\ell_n(\hat f,\hat g)\le n\Delta_n$. Hence
\[
\begin{aligned}
\sum_iH_i^2(\hat f,\hat g)&\le\{\ell_n(f_0,g_0)-\ell_n(f_j,g_j)\}+2\log(N/\delta)+n\kappa_\gamma\varepsilon^2\\
&\le n\bar c_\gamma a^2+\lambda_\gamma^{-1}\log\tfrac1\delta+n\Delta_n+2\rho\varepsilon\Big(n\bar\mu_\gamma+\frac{\log(1/\delta)}{s_0}+nc_0\Big)+2\log\frac N\delta+n\kappa_\gamma\varepsilon^2,
\end{aligned}
\]
which is the bound in (a) after division by $n$. The last claim of (a) is Lemma~\ref{lem:gamma}(c), averaged over $i$.

(b) A network with two outputs that approximates $(f_0,g_0)$ is obtained by placing the two approximating networks of Lemma~\ref{lem:sh} in parallel, padding the shallower one with identity layers ($\sigma(x)-\sigma(-x)=x$); this at most doubles the width and adds the sparsities, and conditions (i)--(iv) of Theorem~\ref{thm:rate} are stated up to constants, so $a^2\lesssim\phi_n$. The covering bound of Lemma~\ref{lem:sh} holds for networks with any number of outputs, with the supremum norm taken over outputs as well, and truncation of each output is 1-Lipschitz. Taking $\varepsilon=1/n$ in (a), the trajectory bound follows as Theorem~\ref{thm:rate} follows from Theorem~\ref{thm:oracle}. For the out-of-sample bound, Proposition~\ref{prop:mixing} with the constant $\epsilon_0$ of (c) and Lemma~\ref{lem:transfer} concern only the true process, and the proof of Theorem~\ref{thm:oos} goes through with $g_{f,g}(\mathbf x)=S^{-1}\sum_sH^2(p_{f(x_s),e^{g(x_s)}},p_{f_0(x_s),e^{g_0(x_s)}})$, the net $(f_j,g_j)$ and $2c_2\varepsilon^2$ replaced by $\kappa_\gamma\varepsilon^2$; Lemma~\ref{lem:gamma}(c) converts the bound on $R^H(\hat f,\hat g)$ into the bound on $R(\hat f)+R_g(\hat g)$.

(c) In the proof of Proposition~\ref{prop:mixing}, the probability of a zero count is now $\{r/(r+\mu)\}^r$ with $\mu\le e^F$ and $r\le r_{\rm hi}$; it is decreasing in $\mu$ and, because $(1+\mu/r)^r$ is increasing in $r$, decreasing in $r$, so it is at least $\{r_{\rm hi}/(r_{\rm hi}+e^F)\}^{r_{\rm hi}}$, and the rest of the proof is unchanged. In the proof of Theorem~\ref{thm:pred}, replace Lemma~\ref{lem:hell} by Lemma~\ref{lem:gamma}(b): the one-step total variation distance is at most $(\mathrm{KL}/2)^{1/2}\le\tfrac12\kappa_\gamma^{1/2}[\{f(x)-f_0(x)\}^2+\{g(x)-g_0(x)\}^2]^{1/2}$, and each Kullback--Leibler term of the chain rule is at most $\tfrac12\kappa_\gamma[\{f-f_0\}^2+\{g-g_0\}^2](X_{s,t+k})$. The rest of the proof is unchanged.

(d) Pointwise, $(\bar f-f_0)^2\le n_{\rm fit}^{-1}\sum_k(\hat f_k-f_0)^2$ and $(\bar g-g_0)^2\le n_{\rm fit}^{-1}\sum_k(\hat g_k-g_0)^2$ by convexity. By Lemma~\ref{lem:gamma}(b) and~(c), at every covariate value,
\[
H^2(p_{\bar f,e^{\bar g}},p_{f_0,e^{g_0}})\le\frac{\kappa_\gamma}4\cdot\frac1{n_{\rm fit}}\sum_k\{(\hat f_k-f_0)^2+(\hat g_k-g_0)^2\}\le\frac{\kappa_\gamma}{2c_{\min}}\cdot\frac1{n_{\rm fit}}\sum_kH^2(p_{\hat f_k,e^{\hat g_k}},p_{f_0,e^{g_0}}),
\]
and averaging over site-weeks, or taking expectations over the copy, gives (d).

(e) The bound follows from (b), since $\|\hat f-f_0\|_n+\|\hat g-g_0\|_n\le\{2(\|\hat f-f_0\|_n^2+\|\hat g-g_0\|_n^2)\}^{1/2}$. For the average of $e^f$ over a set $A$ of $wn$ site-weeks, $|e^a-e^b|\le e^F|a-b|$ on $[-F,F]$ and the Cauchy--Schwarz inequality give $(wn)^{-1}\sum_{i\in A}|e^{f(X_i)}-e^{f_0(X_i)}|\le e^F\{(wn)^{-1}\sum_{i\in A}(f-f_0)^2(X_i)\}^{1/2}\le e^Fw^{-1/2}\|f-f_0\|_n$, and likewise for $e^g$, with $e^{\gamma_{\rm hi}}$ in place of $e^F$, and for predictive probabilities, by Lemma~\ref{lem:gamma}(b) and Pinsker's inequality. \qed
\end{proof}

\section{Proof of Theorem~\ref{thm:dep}}\label{app:dep}

The proof of Theorem~\ref{thm:unknown}(a) is modified in its three probabilistic steps, with the filtration $(\mathcal I_t)_{t\ge0}$ indexed by periods and the generalized H\"older inequality: for non-negative $W_1,\dots,W_S$, $\E[\prod_sW_s\mid\mathcal I_{t-1}]\le\prod_s\E[W_s^S\mid\mathcal I_{t-1}]^{1/S}$. Under Assumption~\ref{ass:marginal} the conditional law of $Y_{s,t}$ given $\mathcal I_{t-1}$ is the negative binomial law $p_{s,t}(f_0,g_0)$ of the site.

\emph{Step~1.} Fix a pair $(f,g)$ with values in the ranges, let $\Lambda_{s,t}=p_{s,t}(f,g)(Y_{s,t})/p_{s,t}(f_0,g_0)(Y_{s,t})$ and let $H_{s,t}^2$ be the squared Hellinger distance between the two laws. Since $\E[\Lambda_{s,t}^{1/2}\mid\mathcal I_{t-1}]=1-H_{s,t}^2$, the inequality with $W_s=\Lambda_{s,t}^{1/(2S)}$ gives $\E[\prod_s\Lambda_{s,t}^{1/(2S)}\mid\mathcal I_{t-1}]\le\prod_s(1-H_{s,t}^2)^{1/S}$. Hence $\zeta_t=(2S)^{-1}\sum_s\log\Lambda_{s,t}-S^{-1}\sum_s\log(1-H_{s,t}^2)$ satisfies $\E[e^{\zeta_t}\mid\mathcal I_{t-1}]\le1$, and Lemma~\ref{lem:mart}, applied to the $T$ periods, gives with probability at least $1-\delta$ that $\sum_t\zeta_t\le\log(1/\delta)$, that is, because $-\log(1-x)\ge x$,
\[
\sum_{s,t}H_{s,t}^2(f,g)\le\tfrac12\{\ell_n(f_0,g_0)-\ell_n(f,g)\}+S\log(1/\delta).
\]
A union bound over an internal $\varepsilon$-net gives the event $\Omega_1$ with $S\log(N/\delta)$ in place of $\log(N/\delta)$.

\emph{Steps~3 and~4.} With $D_{s,t}$ as in Step~3 of the proof of Theorem~\ref{thm:unknown}, that step shows $\E[\exp\{\lambda_\gamma(D_{s,t}-\bar c_\gamma a^2)\}\mid\mathcal I_{t-1}]\le1$ for every site, using only the conditional law of $Y_{s,t}$. By the H\"older inequality with $W_s=\exp\{(\lambda_\gamma/S)(D_{s,t}-\bar c_\gamma a^2)\}$, the variable $\xi_t=(\lambda_\gamma/S)\sum_s(D_{s,t}-\bar c_\gamma a^2)$ satisfies $\E[e^{\xi_t}\mid\mathcal I_{t-1}]\le1$, so $\ell_n(f_0,g_0)-\ell_n(f^*,g^*)\le n\bar c_\gamma a^2+S\lambda_\gamma^{-1}\log(1/\delta)$ on an event $\Omega_2$ of probability at least $1-\delta$. In the same way, $\xi_t=S^{-1}\sum_s(s_0Y_{s,t}-\tfrac12r_{\rm lo}\log2)$ gives $\sum_{s,t}Y_{s,t}\le n\bar\mu_\gamma+Ss_0^{-1}\log(1/\delta)$ on an event $\Omega_3$ of probability at least $1-\delta$.

\emph{Assembly.} Step~5 of the proof of Theorem~\ref{thm:unknown} is deterministic given the three events. Every term $\log(N/\delta)$ or $\log(1/\delta)$ now carries the factor $S$, and dividing by $n=ST$ gives (a); the comparison with the squared norms is Lemma~\ref{lem:gamma}(c), which is pointwise.

(b) Apply (a) with $\varepsilon=1/T$ and with $(f^*,g^*)$ such that $a^2\le2C_{\rm SH}\phi_T$. By Lemma~\ref{lem:sh} with $T$ in place of $n$, applied to each output as in the proof of Theorem~\ref{thm:unknown}(b), $\log N_n(1/T)\le CT\phi_TL\log^2T$, and the remaining terms are of order $1/T\le\phi_T$. Parts (d) and (e) of Theorem~\ref{thm:unknown} and (c) rest on pointwise inequalities and do not involve the joint law of the sites.

(d) Under the stated laws the conditional law of $Y_{s,t}$ given $\mathcal I_{t-1}$ is that of $Y_{1,t}$, the negative binomial law with log-mean $f_0(X_{1,t})=f_0(X_{s,t})$, so Assumption~\ref{ass:marginal} holds. The data are a function of $(X_{1,t},Y_{1,t})_{t\le T}$, every estimator is therefore an estimator in the experiment with one site and $T$ observations, and $R(\hat f)$ is the same in both experiments because the sites are identical. Theorem~\ref{thm:lower}, whose constant does not depend on the number of sites, gives the bound. \qed

\vskip 0.2in
\bibliography{refs}

\clearpage
\includepdf[pages=-]{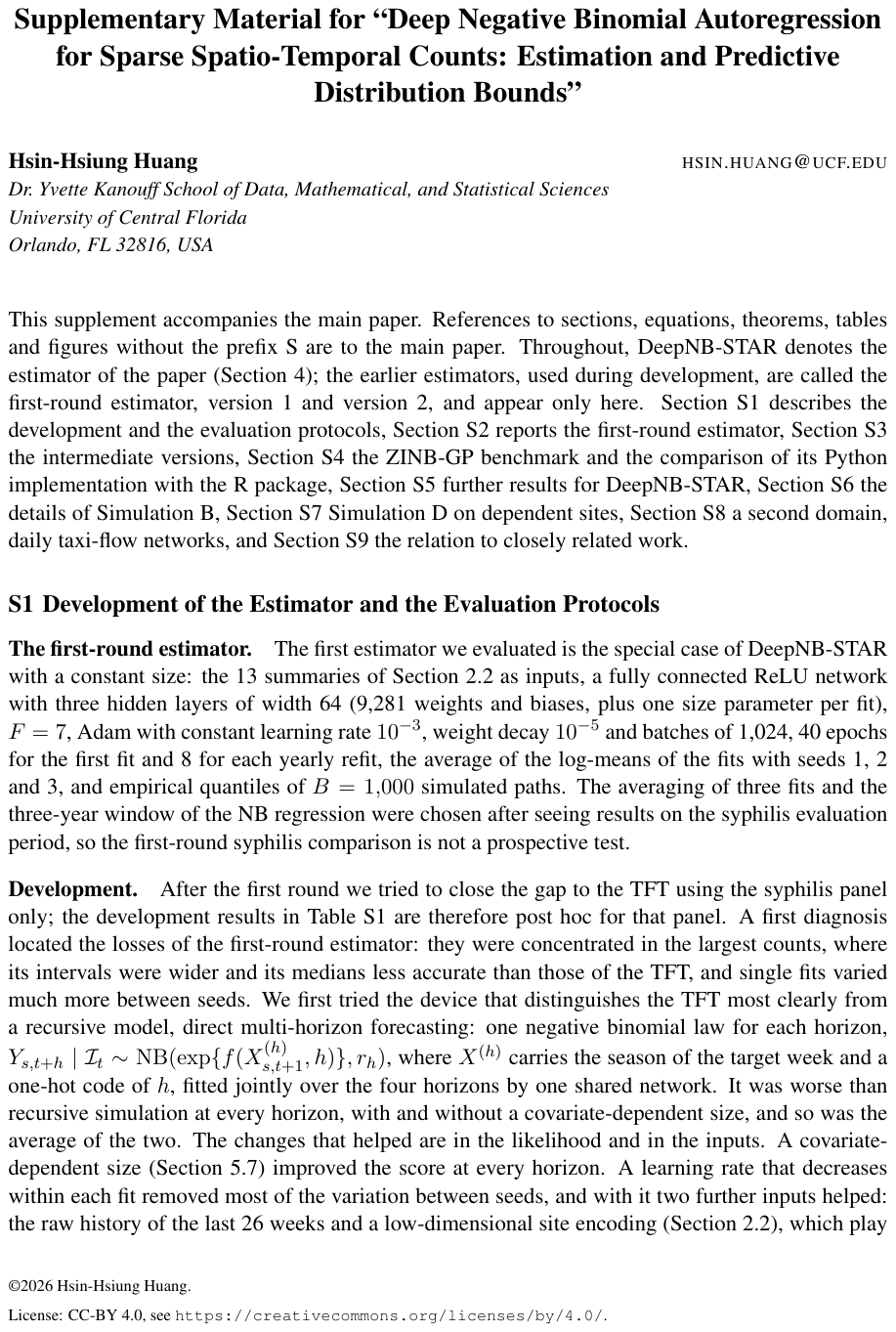}

\end{document}